\documentclass[aps,prb,reprint,groupedaddress]{revtex4-1}
\usepackage{graphicx}
\usepackage{amsmath}
\usepackage{color}

\begin{document}


\title{Fractional periodicity of persistent current in coupled quantum rings}


\author{M. Ij\"{a}s}
\email[]{mari.ijas@aalto.fi}
\author{A. Harju}
\affiliation{Department of Applied Physics and Helsinki Institute of Physics, Aalto University School of Science, Espoo, Finland}


\date{\today}

\begin{abstract}

We study the transmission properties of a few-site Hubbard rings with up to second-nearest neighbor coupling embedded to a ring-shaped lead using exact diagonalization. The approach captures all the correlation effects and enables us to include interactions both in the ring and in the ring-shaped lead, and study  on an equal footing weak and strong coupling between the ring and the lead as well as asymmetry. In the weakly coupled case, we find fractional periodicity at all electron fillings at sufficiently high Hubbard $U$, similar to isolated rings. For strongly coupled rings, on the contrary, fractional periodicity is only observed at sufficiently large negative gate voltages and high interaction strengths. This is explained by the formation of a bound correlated state in the ring that is effectively weakly coupled to the lead. 
\end{abstract}


\maketitle


\section{Introduction}

Systems having a connected geometry exhibit interesting properties in the presence of a magnetic flux. In the Aharonov-Bohm effect, transmission through a ring pierced by a magnetic flux shows interference effects as electrons traversing through the two different paths acquire different phase factors. On the other hand, a magnetic field induces a non-dissipative current called the persistent current in isolated phase-coherent rings or ring ensembles due to the single-valuedness of the electron wave function.\cite{Kotlyar-DasSarma} Both before and after the experimental observations of the persistent currents,\cite{Bleszynski-Jayich, Bluhm, Jariwala, Deblock, Chandrasekhar, Fuhrer, Mailly,Morpurgo, Pedersen, Giebers}  intense theoretical study has been devoted to the different factors affecting the periodicity of the current, such as the effects of electron-electron interaction, disorder and finite temperature.\cite{Buttiker-Imry-Landauer, Buttiker, Cheung-Gefen, Gogolin-Prokofev, Yu-Fowler, Loss-Goldbart, Fye, Kusmartsev_1991, Weisz-Kishore-Kusmartsev, Kusmartsev-Weisz, Kusmartsev, Kusmartsev_1997, Kusmartsev_1999} The persistent currents in more complex systems, such as multichannel rings or cylinders \cite{Maiti_PhysicaE, Kusmartsev_1999} and two-dimensional quantum dot arrays,\cite{Kotlyar-DasSarma} have also been the topic of theoretical considerations. 

The understanding of these phenomena is not only interesting for fundamental science but also from an applicational point of view.  Quantum rings could serve as components for future nanoelectronics. For instance, there have been suggestions to use ring molecules as quantum interference effect transistors.\cite{Cardamone, Ke, Solomon} Aharonov-Bohm phenomena are also relevant for carbon nanotubes in which band gap oscillations as a function of external magnetic flux have been observed.\cite{Charlier, Sangalli-Marini} 

Metallic rings with a large number of charge carriers are well described by non-interacting theories and by now the properties of the persistent current in them, such as the magnitude, periodicity, and direction, are well understood.\cite{Bleszynski-Jayich}   In these systems, the measured periodicity of the persistent current is a flux quantum, $\Phi_0 =ch/e$, or half of it, $\Phi_0/2$.\cite{Bleszynski-Jayich, Deblock, Bluhm, Jariwala, Chandrasekhar, Levy}. Theoretically, the two different periods are related to parity effects with respect to number of electrons in the ring.\cite{Loss-Goldbart, Kusmartsev-Weisz} In the experiment, the half-flux periodicity is observed only in measurements over an ensemble of metallic rings.\cite{Jariwala, Deblock, Levy} This has been explained in terms of ensemble averaging of the different parities of the individual rings.\cite{Bouchiat-Montambaux}

In semiconducting rings with a small number of electrons, electron-electron correlations are important. With a strong enough electron-electron interaction, the appearance of fractional periodicity $\Phi_0/N_{el}$, where $N_{el}$ is the number of electrons in the ring, has been theoretically predicted.\cite{Kusmartsev_1991, Yu-Fowler} Fractional periodicity has also been shown to occur at low values of $U$ in dilute systems when $N_{el}/NU$ is small, associated with the disappearance of parity effects.\cite{Kusmartsev-Weisz}  However, only Keyser~\emph{et al.}\cite{Keyser} have been able to show this experimentally by studying a semiconducting ring with less than ten electrons.   Very recently, Hernandez~\emph{et al.}\cite{Hernandez-Gusev} reported measurements on a 20-40--electron ring but they were only able to observe the $\Phi_0$ and $\Phi_0/2$ oscillations. As the effect seems to be elusive in experiments, it is necessary to study which additional factors could complicate the experimental observation of fractional periodicity.

In the experiments, either isolated rings or rings connected to external leads are studied. A superconducting quantum interference device (SQUID)-type setup allows the measurement of the persistent current from the induced magnetic moment in isolated rings or ring ensembles.\cite{Mailly,Bluhm, Levy, Chandrasekhar, Jariwala} Also resonator-based methods have been used in the literature to study isolated ring systems.\cite{Bleszynski-Jayich, Deblock} These methods provide access to both the magnitude of the persistent current and its flux periodicity. On the other hand, the magnetoconductance of the ring can be measured in an Aharonov-Bohm -type setup with external leads.\cite{Giebers, Keyser, Morpurgo, Fuhrer, Fuhrer_prl, Pedersen} In these measurements, a bias voltage is applied, and the strength of the coupling between the ring and the leads can be controlled by gate voltages. These measurements yield only the magnetic flux periodicity through the measured conductance but provide no direct information on the magnitude of the persistent current.

Theoretical study of conductance through interacting systems is far from trivial. A Landauer-B\"uttiker\cite{Buttiker_PRL} -type formalism is available also for the interacting case, although the practical implementation is cumbersome.\cite{Meir-Wingreen-Lee} Flux-pierced interacting rings have been widely studied in the literature using perturbative approaches with respect to the coupling between the ring and the leads,\cite{Jagla-Balseiro,Hallberg-Aligia,Rincon-Aligia-Hallberg} using a variety of theoretical models to describe the ring part of the system: a Luttinger liquid,\cite{Friederich-Meden} spinless interacting fermions,\cite{Meden-Schollwock_rings} the $t-J$ model,\cite{Rincon-Aligia-Hallberg} and also the Hubbard model.\cite{Jagla-Balseiro} The strongly coupled case has, to the best of our knowledge, not yet been addressed for interacting systems, and in general only non-interacting leads have this far been considered. Moreover, we are not aware of any studies dealing with the conductance or transmission properties of Hubbard rings with second-nearest neighbor coupling. 

It has been proposed that the conductance properties of interacting nanostructures can be studied using the so-called embedding approach,\cite{Rejec-Ramsak, Rejec-Ramsak_AB} in which the interacting system is embedded into a non-interacting lead. Instead of using semi-infinite leads and introducing a bias voltage between them, as in the widely-used Landauer-B\"uttiker approach,\cite{Buttiker_PRL} the lead is formed into a ring that is pierced by a magnetic flux.  As a consequence, a persistent current is induced in the ring and its magnitude in the presence and in the absence of the nanostructure can be related to the transmission coefficient of the nanostructure, and thus to conductance. In the limit of an infinitely long lead, the approach yields linear response conductance. If electron-electron interaction is introduced to the lead part of the system, no formal relationship between the persistent current and the conductance has been found in the literature. However, the magnitude of the lead current can still be thought to probe the transmission properties of the nanostructure. The advantage of the embedding approach is that all coupling strengths can be treated on equal footing, and that the ground-state energy of the composite system is, in principle, all that is needed to extract the conductance properties.

In this paper, we study a fully interacting system consisting of a flux-pierced Aharonov-Bohm ring embedded in an interacting ring-shaped lead. By studying a Hubbard ring with second-nearest neighbor couplings,  we show that in the weakly coupled case, the fractional periodicity of the persistent current can be observed for all electron fillings. For strong coupling between the ring and the lead, on the contrary, fractional periodicity is only observable at higher electron fillings, stronger interaction and larger gate voltages, where a correlated bound state that is weakly coupled to the lead is formed. We also discuss the effect of asymmetry, both in the coupling strengths and lead positions as well as in the interaction strength in the ring and lead parts of the system. Our results provide insight into why the fractional periodicity is difficult to observe experimentally. 

\begin{figure}
 \includegraphics[width=0.6\columnwidth]{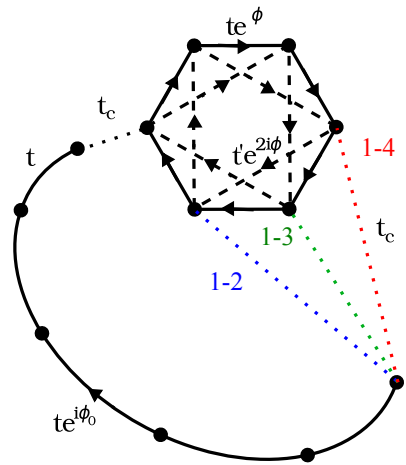}
 \caption{\label{fig:structure} (Color online) The geometry of the double ring system. Dotted lines: alternative connections between the ring and the lead (the three colors/shades of gray show the different positions) Dashed lines: second-nearest neighbor hoppings in the ring. For symbols, see Section~\ref{sec:model}.}
\end{figure}

\section{Model and method \label{sec:model}}

We use the Hubbard model to describe our system consisting of an interacting ring (hereafter ring) and an equally interacting ring-shaped lead (hereafter lead) connected to it, schematically shown in Fig.~\ref{fig:structure}. We note that a related multiring or multiarm setup has recently been studied within the non-interacting tight-binding formalism.\cite{Maiti-EPJB} We take the hopping amplitude between the sites within the ring and within the lead as a constant, $t_0=-t$, and measure all energies in the units of $t$. For the ring part of the system, we consider rings with the next-nearest neighbour coupling (2NN) that is given by the hopping amplitude $t'$. 

In 1NN Hubbard rings, fractional $\Phi_0/N_{el}$ periodicity appears in the $U\rightarrow \infty$ limit.\cite{Viefers} Isolated 2NN rings have been shown to map to continuum rings, and the inclusion of 2NN hoppings has been shown to lower the required value of the on-site electron-electron interaction $U$ for the onset of fractional periodicity.\cite{Hancock} The presence of the 2NN hopping makes thus the ring effectively quasi-1D as electrons are allowed to pass each other even at very high interaction strengths. We denote the onset value by $U_c$. We use the value $t' = 0.2t$, following a previous study on isolated rings,\cite{Hancock} for which this value was found to lead to a relatively low value for $U_c$  ($U_c = 5.3 t$ for an eight-site  ring at quarter-filling).  The ring is pierced by an Aharonov-Bohm (AB) flux $\Phi$ and the system properties as a function of this flux are studied. $N_{\mathrm{ring}}$ gives the number of sites in the ring and $N_l$ the number of sites in the lead.

In order to drive a current through the ring to probe the transmission properties, another AB flux that is fixed to the value $\phi_0$ pierces the lead ring. Throughout this paper, we set $\hbar = c=e=1$ and the fluxes are thus represented in units of $\Phi_0 = ch/e = 2\pi$. In the limit of an infinitely long noninteracting lead, the transmission through the nanostructure can be expressed using the current in the lead part in the presence and in the absence of the nanostructure, when the boundary condition in the lead part is given by a flux $\phi_0 = 0.25\Phi_0 = \pi/2$, also called a "twisted boundary condition".\cite{Sushkov, Rejec-Ramsak, Rejec-Ramsak_AB} We choose this value of $\phi_0$ also for the interacting leads, although we can not analytically link the value of the current and the transmission coefficient. The current in the lead is thus driven by a hopping $t_0e^{i\phi_0}$ between one pair of sites in the lead, the rest of the hoppings between the nearest neighbors being $t_0$. 

The Hubbard Hamiltonian is given by 
\begin{equation} H =  H_{\mathrm{kin}} + H_U + H_g, \end{equation}
where the kinetic energy contribution is given by 
\begin{equation} \label{eq:Hkin} H_{\mathrm{kin}} = \sum_{ i, j, \sigma}  (t_{ij}c_{i\sigma}^{\dagger}c_{j\sigma} + t_{ij}^*c_{j\sigma}^{\dagger}c_{i\sigma}),  \end{equation}
the contribution due to the on-site interaction by
\begin{equation} 
 H_U = \sum_i U_i n_{i\downarrow}n_{i\uparrow}, \end{equation}
and the gate voltage applied upon the ring by
\begin{equation} 
H_g = \sum_{i\sigma} V^g_{i\sigma}n_{i\sigma}. \end{equation}
Here, $t_{ij}$ is the spin-independent hopping amplitude between sites $i$ and $j$, $U_i$ is the interaction strength at site $i$, and $V^g_{i\sigma}$ is the gate voltage applied onto the ring sites, modelled by an on-site energy. Other notation is standard second quantization.  We denote by $U_r$ and $U_l$ the interaction strength in the ring and in the lead, respectively, if necessary. The hopping amplitude within the ring is given by $t_0e^{i\phi}$ between nearest neighbours, and by $t'e^{2i\phi}$ between second-nearest neighbours for clockwise rotation.\cite{Hancock} Due to hermiticity of the Hamiltonian, $t_{ji} = t_{ij}^*$ naturally applies. The total flux piercing the ring is given by $\Phi = N_{\mathrm{ring}}\phi$.  The coupling between the lead and the ring is given by $t_c<0$.    We assume there is no magnetic field present in the electron paths, and thus neglet the spin-orbit interaction and Zeeman splitting. As for a given total spin $S$, the states with different $S_z$ values are degenerate in the absence of Zeeman splitting, we may choose $S_z = 0$ ($S_z = 1/2$) for even(odd)-electron systems without losing information on states with a higher total spin. 

We exactly diagonalize the Schr{\"{o}}dinger equation, $H\Psi = E\Psi$, presented in the many-body basis formed from the many-electron states for both spin species. The diagonalization is performed in a block of the Hamiltonian with a fixed number of electrons in the system, $N_{el} = N_{\uparrow}+N_{\downarrow}$, and a fixed $S_z = N_{\uparrow}-N_{\downarrow}$, as the Hubbard interaction does not couple sectors of the Hamiltonian with different $S_z$ or $N_{el}$.  To obtain the lowest-energy eigenstate, we apply the Lanczos diagonalization algorithm \cite{Dagotto}, converging the lowest eigenenergy to within $10^{-12}t$. The computational effort associated with the exact diagonalization method limits the number of sites in our system to below 30, and also the number of electrons in the system to around six, as the size of the Hilbert space grows factorially as $\binom{N}{N_{\sigma}} \binom{N}{N_{-\sigma}}$. We mostly focus on a six-site ring to allow sufficient variation in the lead length for the consideration of finite-size effects. We also consider different coupling positions between the ring and the lead, shown by the dotted lines with different colors/shades of gray in Fig.~\ref{fig:structure}.

From the wave function of lowest-energy eigenstate, we calculate the observables of interest, such as the current between any two sites $i$ and $j$, by calculating the expectation value $J_{ij}~=~\langle \Psi |\hat{J}_{ij}|\Psi \rangle$, where the current operator is given by ($\hbar = c = e =1$) \cite{Koskinen-Manninen} 
\begin{equation} \label{eq:J} \hat{J}_{ij}= i\sum_{\sigma}(t_{ij}c_{i\sigma}^{\dagger}c_{j\sigma} - t_{ij}^*c_{j\sigma}^{\dagger}c_{i\sigma}).\end{equation}
We define $J_{\mathrm{ring}}$ and $J_{\mathrm{lead}}$ such that the sites $i$ and $j$ in Eq.~(\ref{eq:J}) both belong to the ring or lead part of the system, respectively. Alternatively, we could calculate the persistent current in the ring from the variation of the ground-state energy $E_0$ with respect to a change in the flux, \cite{Buttiker-Imry-Landauer}
\begin{equation} \label{eq:J2} J = -\frac{\partial E_0}{\partial \Phi}. \end{equation} 
We prefer, however, the current operator approach as it allows us to easily calculate the current in the different branches of the system by choosing the sites $i$ and $j$ accordingly. Similarly, we calculate the occupation of a given site ($\hat{n}_i = \sum_{\sigma} \hat{n}_{i\sigma} = \sum_{\sigma} c_{i\sigma}^{\dagger}c_{i\sigma}$) or the total spin of our system [$\langle \hat{S}^2 \rangle = S_{tot}(S_{tot}+1)$]  by calculating the expectation value of the corresponding operator.  

\section{Results}

\subsection{Weakly coupled rings \label{sec:weak}}

To begin with, we consider weak coupling between the ring and the lead. We expect the behavior of the fractional periodicity to be close to that of isolated rings, and also to observe single-electron tunneling. For the time being, we restrict ourselves to the symmetric ring-lead coupling (Fig.~\ref{fig:structure}, 1-4 connection). 

\begin{figure*}
\includegraphics[width = 0.95\columnwidth]{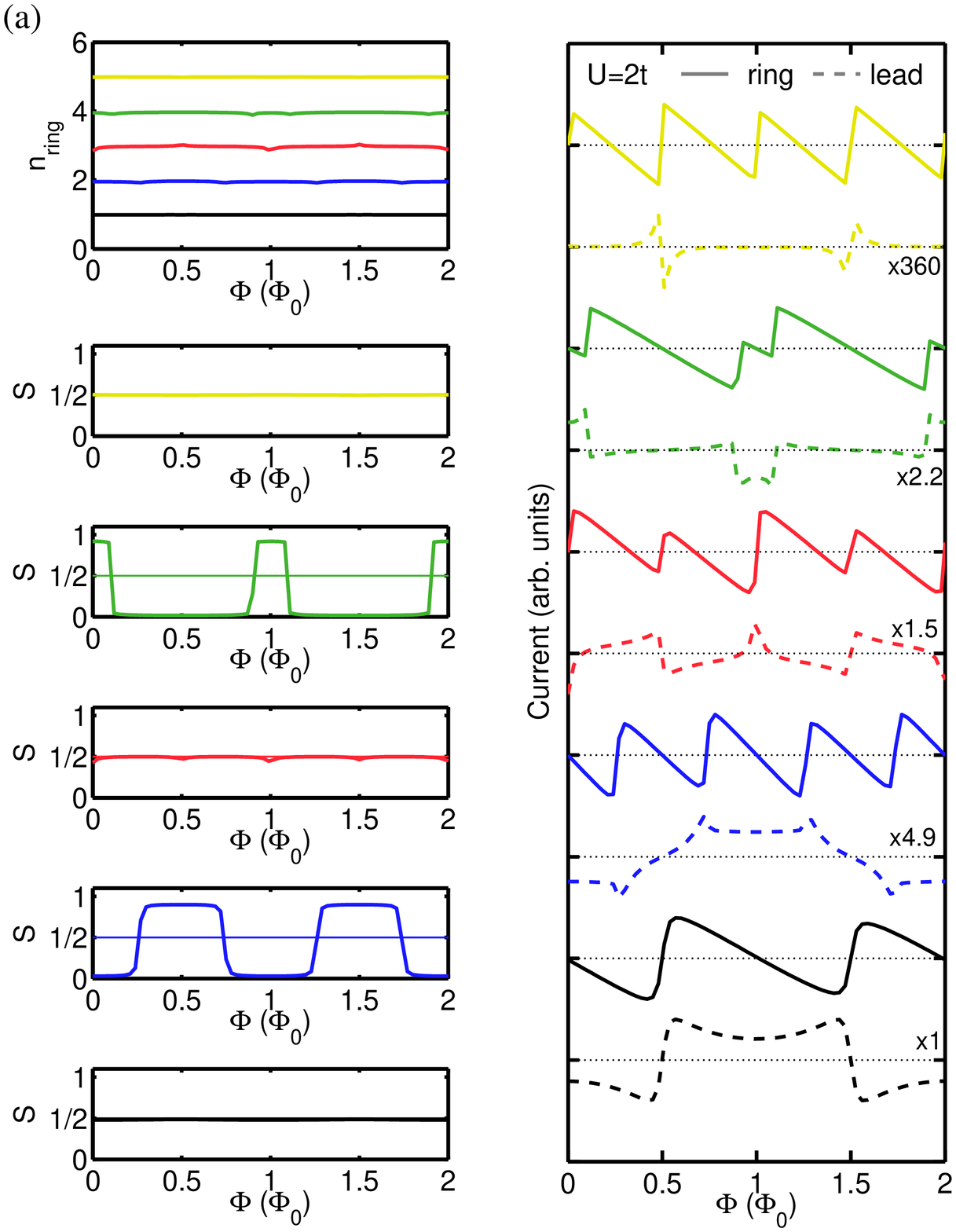}
\includegraphics[width = 0.95\columnwidth]{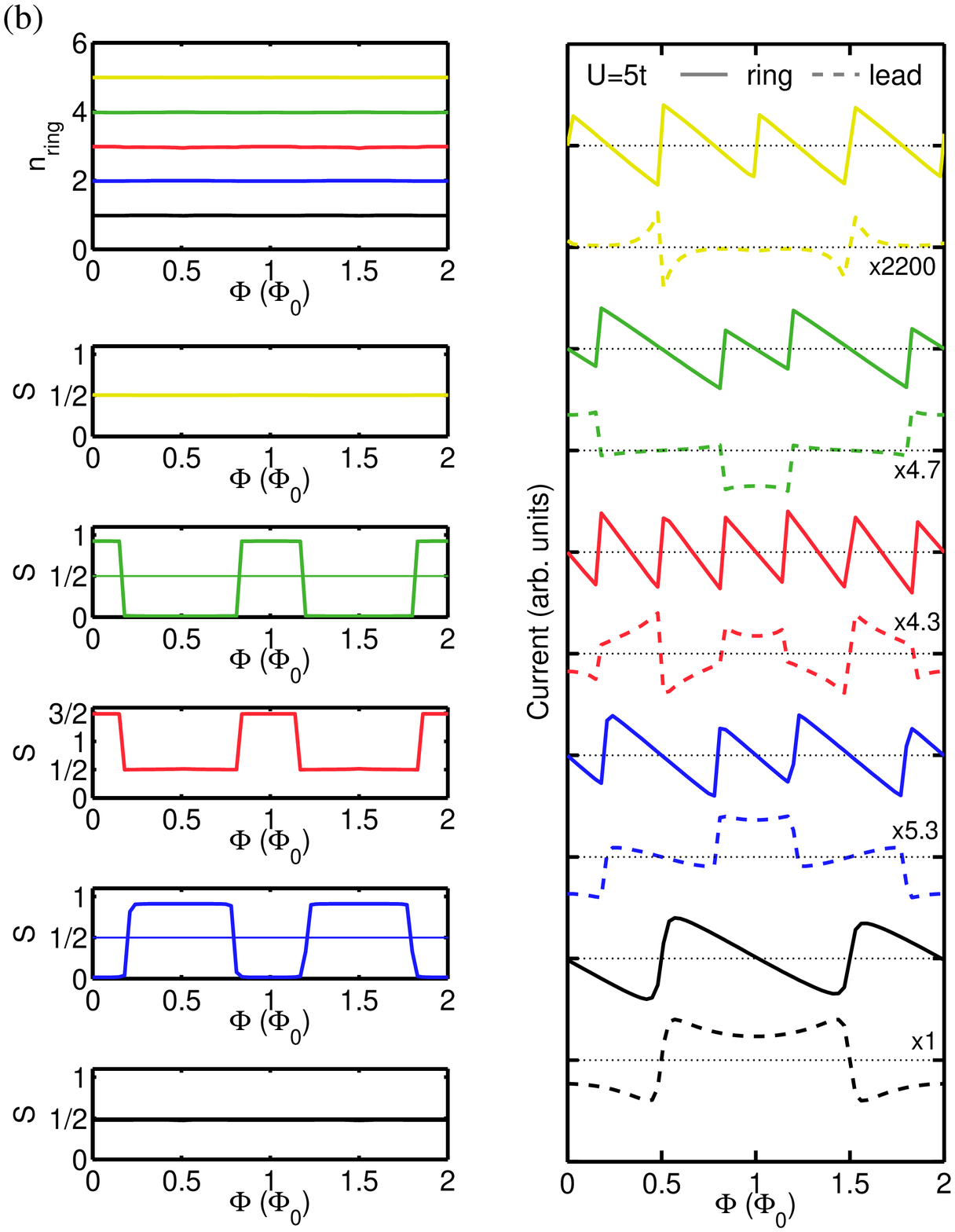}\\
\includegraphics[width = 0.95\columnwidth]{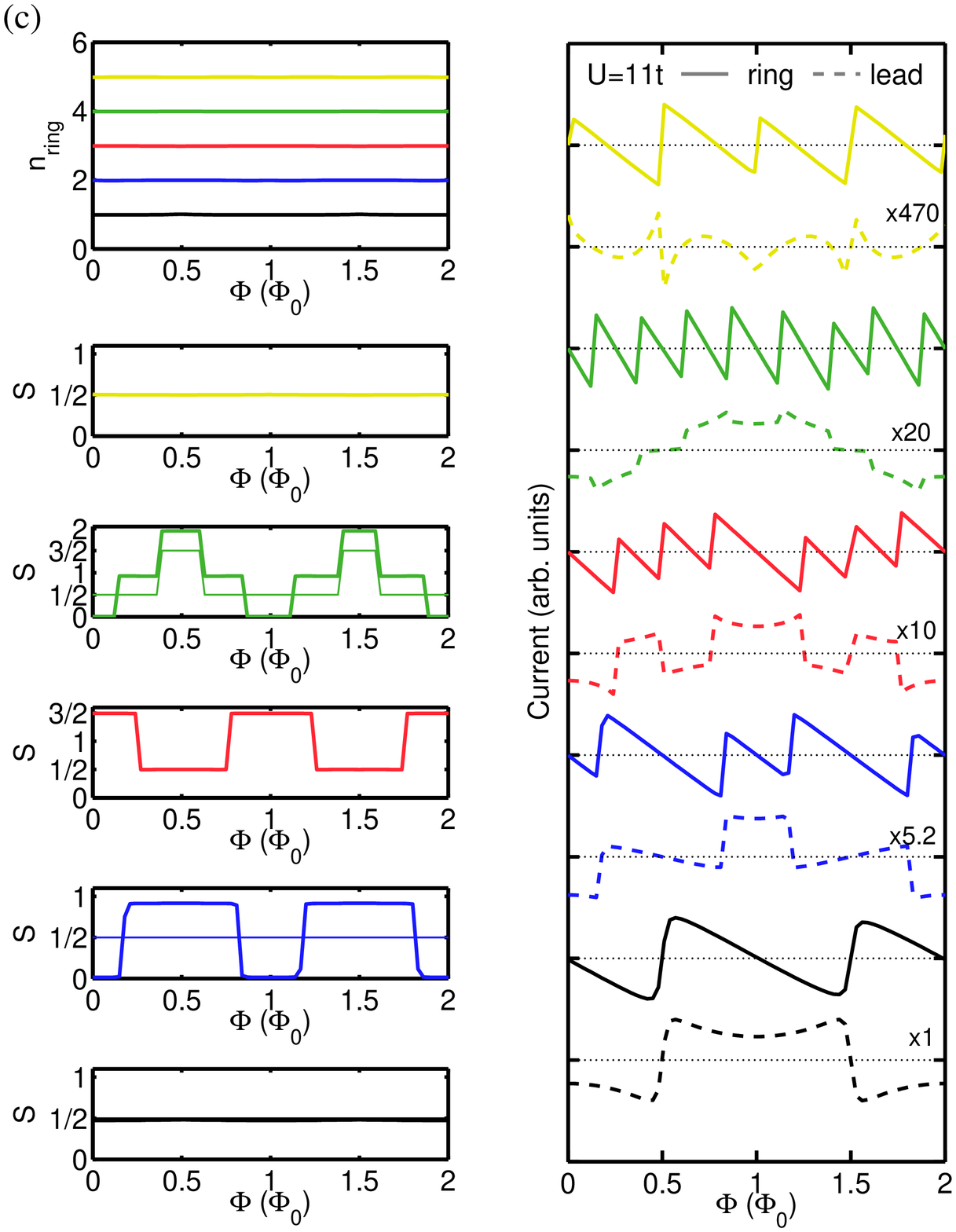}
\includegraphics[width = 0.95\columnwidth]{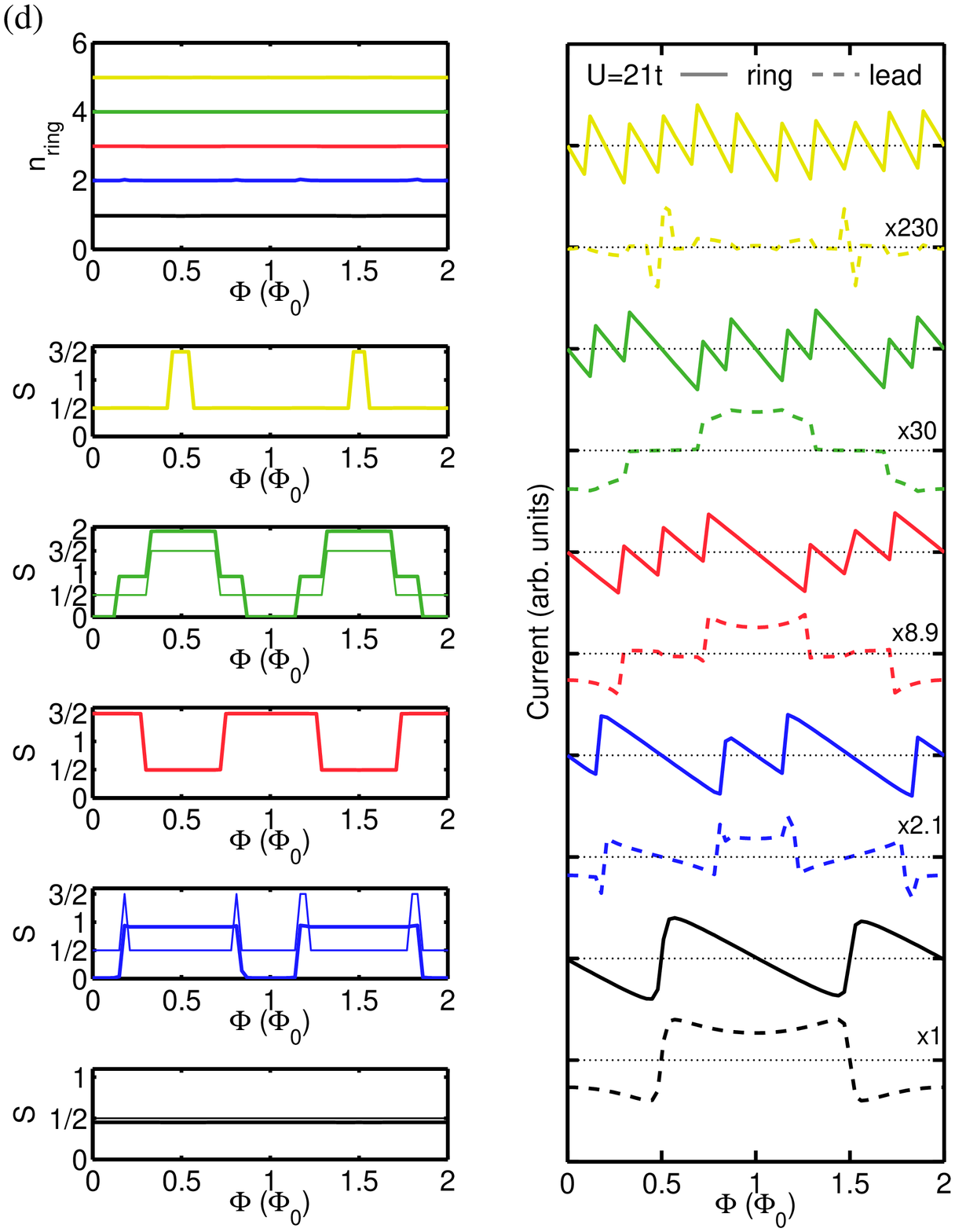}

 \caption{\label{fig:murtoluku} (Color online) The properties of a six-site 2NN ring weakly connected ($t_c = -0.2t$) to a 14-site lead  as a function of the total flux $\Phi$ piercing the ring, with the ring occupation $n_{\mathrm{ring}}$ at integer values (top left panels). Bottom left panels show the total spin (thin line) and its projection onto the ring part of the system (thick line) and the right panels the ring current (solid lines) and the lead current (dashed lines). The dotted lines show the zero level, and the curves have been normalized to facilitate comparison. For lead currents, the scaling factors rounded to two significant digits are indicated. The values of $U$ are chosen as to show fractional periodicity at each integer $n_{\mathrm{ring}}$, and the gate voltage $V^g$ to keep the occupation constant as a function of the flux $\Phi$. The five curves illustrate the different ring occupations ($n_{\mathrm{ring}}\approx 1,2,3,4,5$: black, blue, red, green, yellow in color, black to light gray in grayscale). (a) $U = 2t$ (b) $U = 5t$ (c) $U = 11t$ (d) $U = 21t$.}
\end{figure*}

We use a 14-site lead and leave the discussion on finite-size effects to Section~\ref{sec:fin size}. We occupy the system with an odd number of electrons, $n_{\uparrow}=2$ and $n_{\downarrow}=3$. The ring and the lead are weakly coupled,  $t_c = -0.2t$.  We calculate the current in the ring and lead parts of the system, as well as the total spin state of the composite system and its contribution from the ring, at $V^g$ values chosen as to keep the ring occupation $n_{\mathrm{ring}}$ near an integer value, as shown in the top left panels in Fig.~Ê\ref{fig:murtoluku} at $U = 2t, 5t, 11t$, and 21$t$. The values of $U$ have been chosen to best illustrate the fractional peridicity $\Phi_0/N_{el}$ for each of the ring fillings $N_{el} = 2,3,4$ and 5. To restrict the computational effort, however, only integer values for $U$ were considered. In the right panels of Fig.~\ref{fig:murtoluku}(a)-(d), the current in the ring, $J_{\mathrm{ring}}$, is given by the solid lines and the one in the lead, $J_{\mathrm{lead}}$, by the dotted lines. The current have been shifted and normalized to better illustrate the shape and periodicity of the ring current. 

\begin{figure}
\includegraphics[width = 0.95\columnwidth]{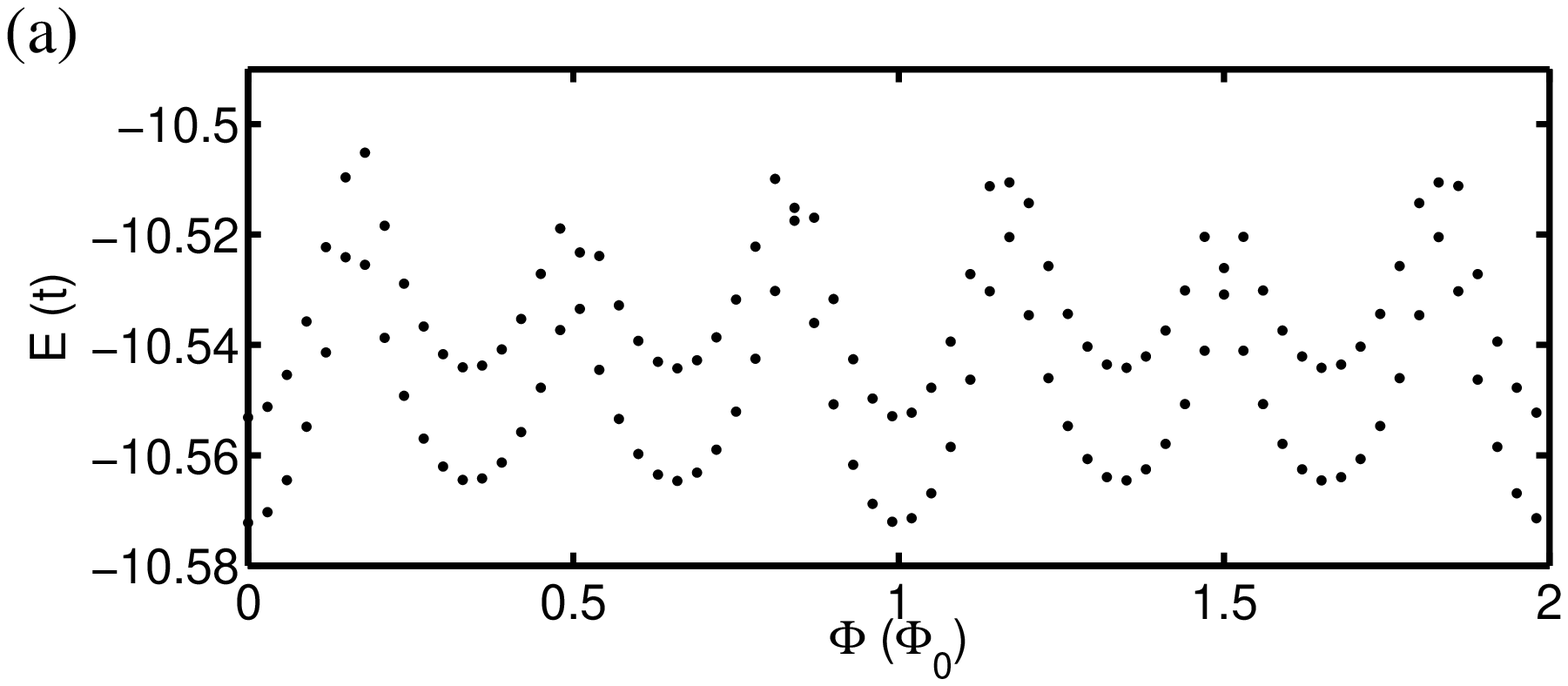}
\includegraphics[width = 0.95\columnwidth]{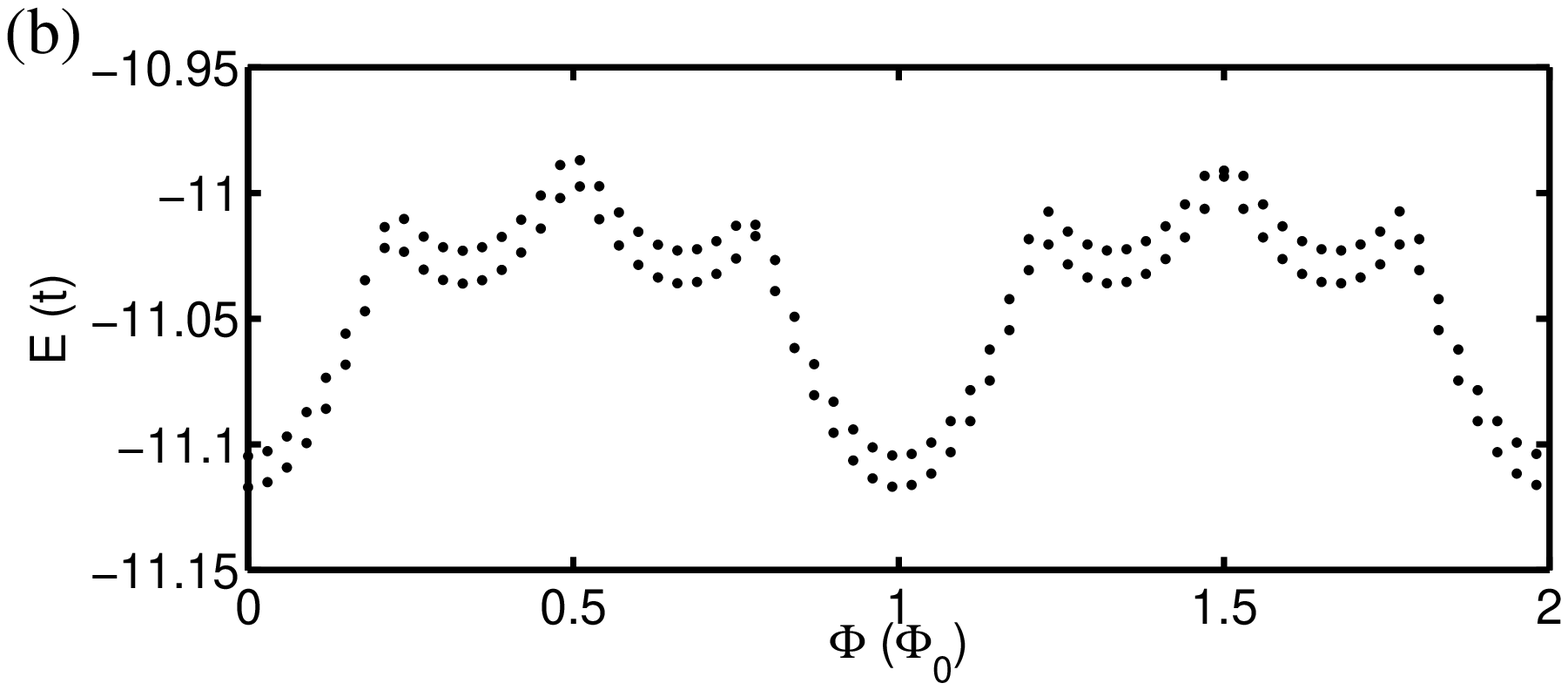}
\includegraphics[width = 0.95\columnwidth]{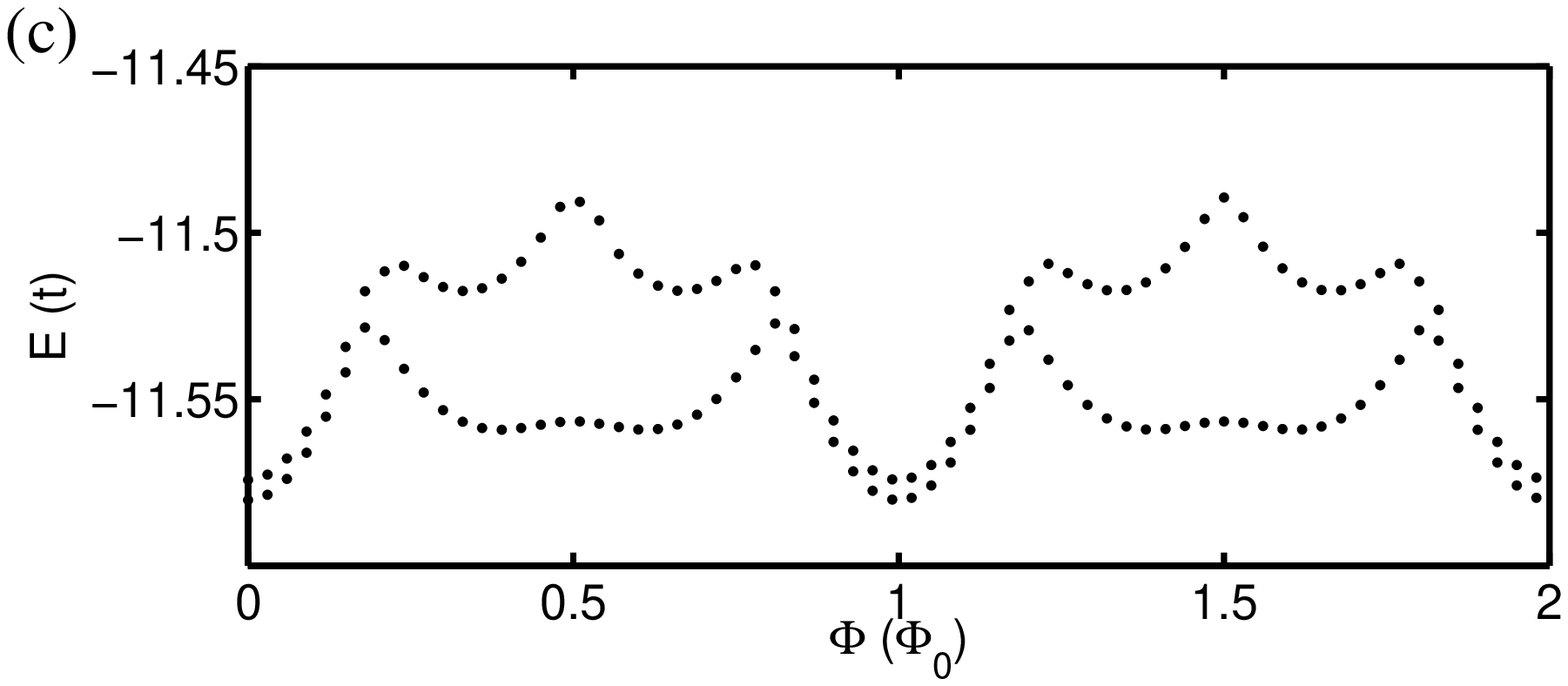}
\caption{\label{fig:enephi} The two lowest energy levels  as a function of the flux $\Phi$ for a six-site ring with $N_l = 14$ as given by the Lanczos diagonalization in which the lowest-energy state has been converged. The ring occupation $n_{\mathrm{ring}}$ is approximately three electrons. (a) Weakly coupled system at fractional periodicity ($U=5t$, $t_c = -0.2t$, $V^g  = -0.9t$). (b) Weakly coupled system for $U > U_c$. ($U=8t$, $t_c = -0.2t$,$V^g = -1.1t$) (c) Strongly coupled system near fractional periodicity ($U=8t$, $t_c = -t$, $V^g = -1.1t$). Other parameters $t_t = 0.2t$, $N_{el} = 5$. }
\end{figure}

In Fig.~\ref{fig:murtoluku} (right panels, solid lines), the kinks in the ring current appear due to crossings between the energy parabola that determine the ground state of the system. Fig.~\ref{fig:enephi} illustrates this, showing the energy of the two lowest-energy states as a function of $\Phi$ as obtained from the Lanczos diagonalization in which the lowest-energy state has been converged. The energy parabolas are easily identified.    At the crossings, the slope of the ground-state energy changes sign, leading also to an abrupt sign change in the persistent current [see Eq. (\ref{eq:J2})]. For $n_{\mathrm{ring}} = 1$, the periodicity of the ring current is always $\Phi_0$. For higher occupations, in the absence of interactions a period of $\Phi_0$ ($\Phi_0/2$) would be expected for even (odd) occupation numbers in the ring.\cite{Loss-Goldbart, Yu-Fowler} Interaction, however, shifts the energy parabola, leading to additional crossings and a pseudo-$\Phi_0/2$ period for even electrons below $U_c$\cite{Loss-Goldbart, Yu-Fowler} as parabola corresponding to states with a higher total spin are less shifted. This is also reflected in the total spin projected onto the ring, shown in left panels in Fig.~\ref{fig:murtoluku} by thicker lines. When $U$ is increased, the regions where the ring spin is high or even maximal ($S = n_{\mathrm{ring}}/2$) are widened.
  
At the onset of the fractional periodicity ($U = U_c$), the period of the ring current is $\Phi_0/n_{\mathrm{ring}}$ and segments between the kinks are of equal length in $\Phi$.This is also illustrated in Fig.~\ref{fig:enephi}(a) that shows the ground-state energy as  a function of $\Phi$ at $U=5t$ corresponding to $n_{\mathrm{ring}}=3$ with fractional periodicity. Notably, even though two of the ring sites are perturbed through the presence of the leads, the fractional periodicity is not affected.  Above $U_c$, the parabola continue to shift in energy and the kinks in the ring current are no longer of equal length, shown in Fig.~\ref{fig:enephi}(b) at $U=8t$. It is readily seen that the parabola around $\Phi=\Phi_0/2+n\Phi_0$ have shifted relative to the ones at $\Phi = n\Phi_0$. The periodicity, however, remains pseudo-$\Phi_0/n_{\mathrm{ring}}$. 

The different periodicities of the persistent current in the ring are also reflected in the lead current, shown in dashed lines in the right panels of Fig.~\ref{fig:murtoluku}. Apart from the $n_{\mathrm{ring}} = 5$ case, in which the magnitude of the ring current is so small that it is significantly affected by numerical instabilities, the kinks in the ring current are associated with jumps or peaks in the lead current. The overall periodicity of the lead current is twice that of the ring current. This is understood by considering the double-ring geometry (Fig.~\ref{fig:structure}). In the present configuration (1-4 connection, see Fig.~\ref{fig:structure}), the ring is pierced by the flux 6$\phi$. The lead and some of the ring sites can be thought to form another ring that is pierced by the flux $-3\phi+\phi_0$.  Thus, the periodicity in the lead current is twice that of the ring current. The effect of the position of the coupling sites  will be discussed in more detail in Section~\ref{sec:asymmetry}. 

In Fig.~\ref{fig:murtoluku}, the lower left panels show the total spin of the ring-lead system (thin line) and its projection onto the ring part (thick lines). The total spin is restricted to half-integer values by the electron number ($n_{\uparrow}$ = 2, $n_{\downarrow}$ = 3), and at low interaction values it is 1/2 for all $n_{\mathrm{ring}}$.  At higher $U$, some regions of $S = 3/2$ appear. The total spin projected onto the ring, on the contrary, reflects the changes in the ground-state energy parabola. With an increasing interaction strength,  the regions with a higher total spin value widen in $\Phi$, and for $n_{\mathrm{ring}}$ = 4 even regions of maximal spin $S=2$ appear at and above $U_c$. This effect is similar than in isolated rings \cite{Hancock} in which a maximally spin polarized state ($S = N_{el}/2$) was found above $U_c$ at a fixed value of $\Phi=0.5\Phi_0$ for a quarter-filled eight-site ring.  Due to the coupling between the lead and the ring, the values of the total spin projected onto the ring are, however, not strictly integer or half-integer. 

The onset value of the fractional periodicity scales approximately as $U_c \propto N_{el}^2$. Thus, in order to experimentally observe fractional periodicity, a large number of electrons in the ring requires very strong interaction that might not be experimentally realistic. In the weakly coupled case, the fractional periodicity is very similar to the isolated case and we thus turn our attention to the case with a strong coupling between the lead and the ring. 

\subsection{Strongly coupled systems \label{sec:strong}}

\begin{figure*}
\includegraphics[width = 0.95\columnwidth]{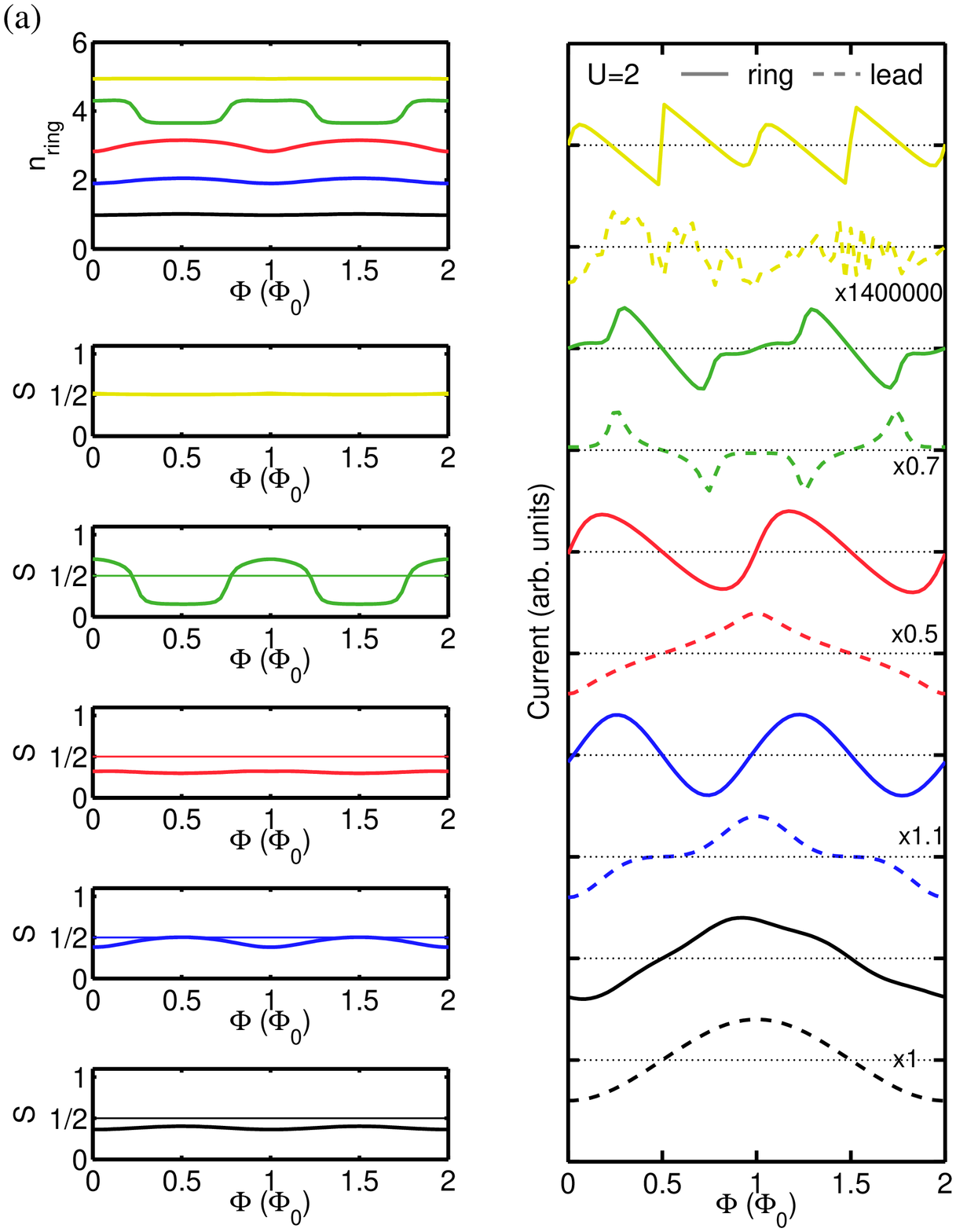} 
\includegraphics[width = 0.95\columnwidth]{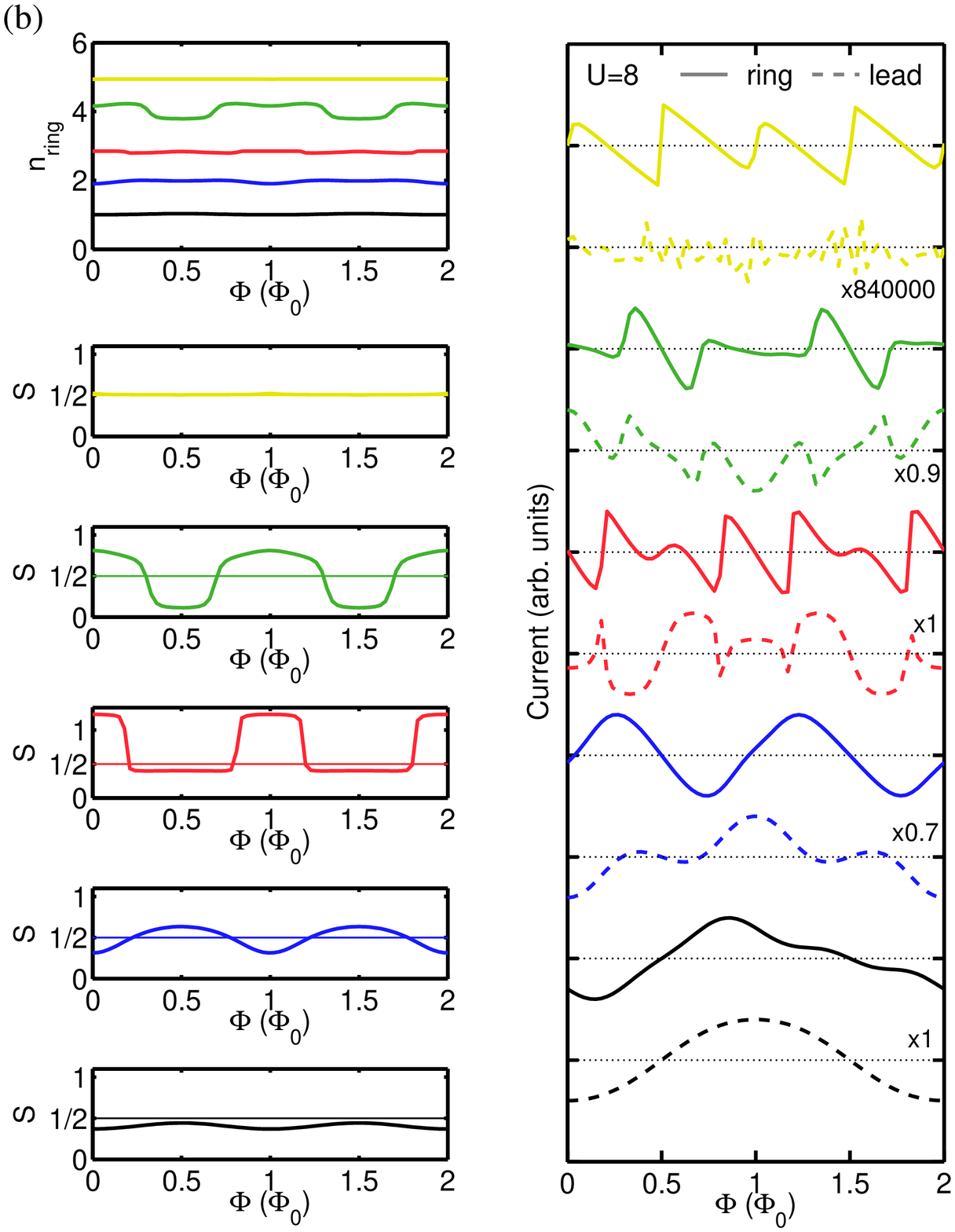} \\
\includegraphics[width = 0.95\columnwidth]{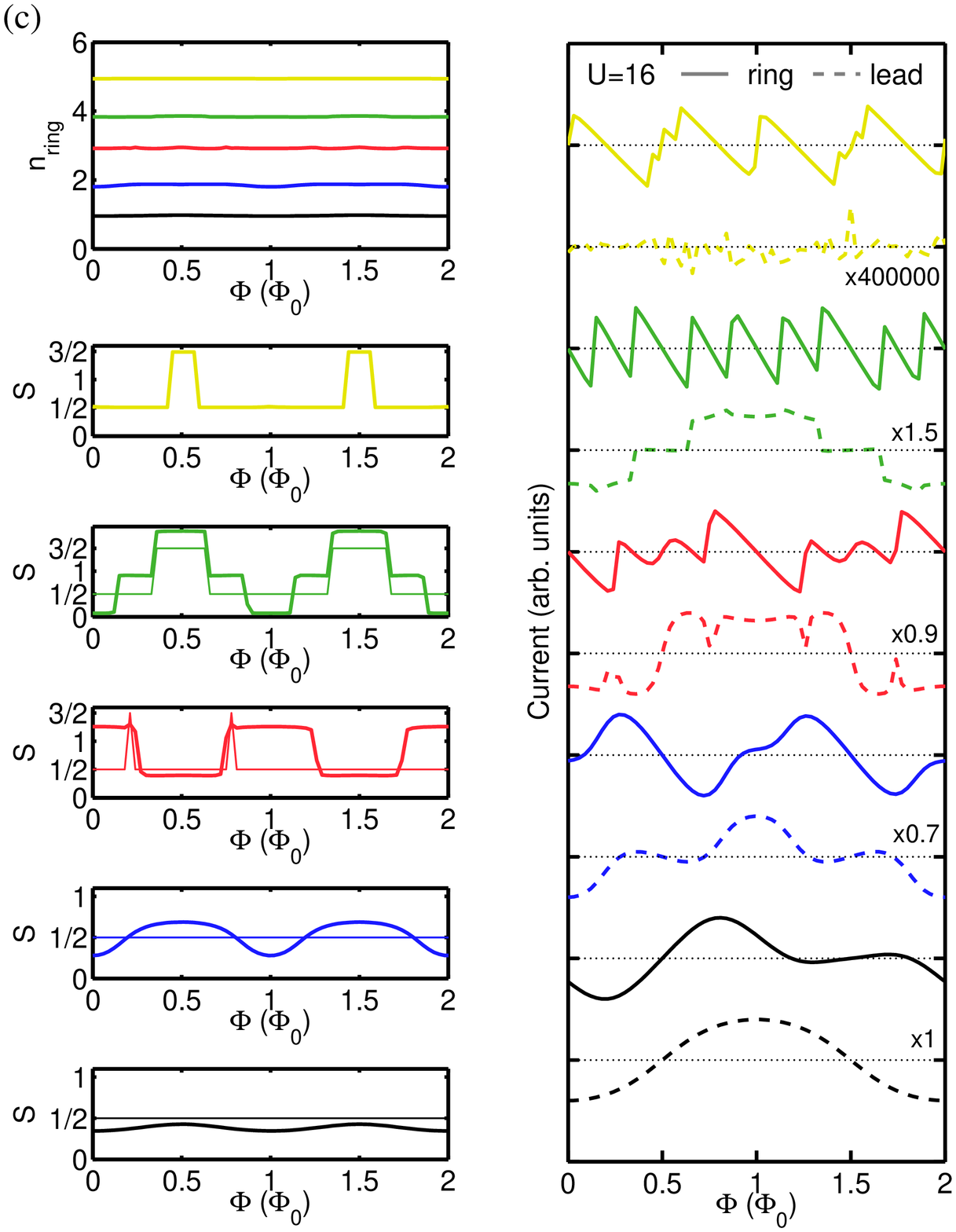} 
\includegraphics[width = 0.95\columnwidth]{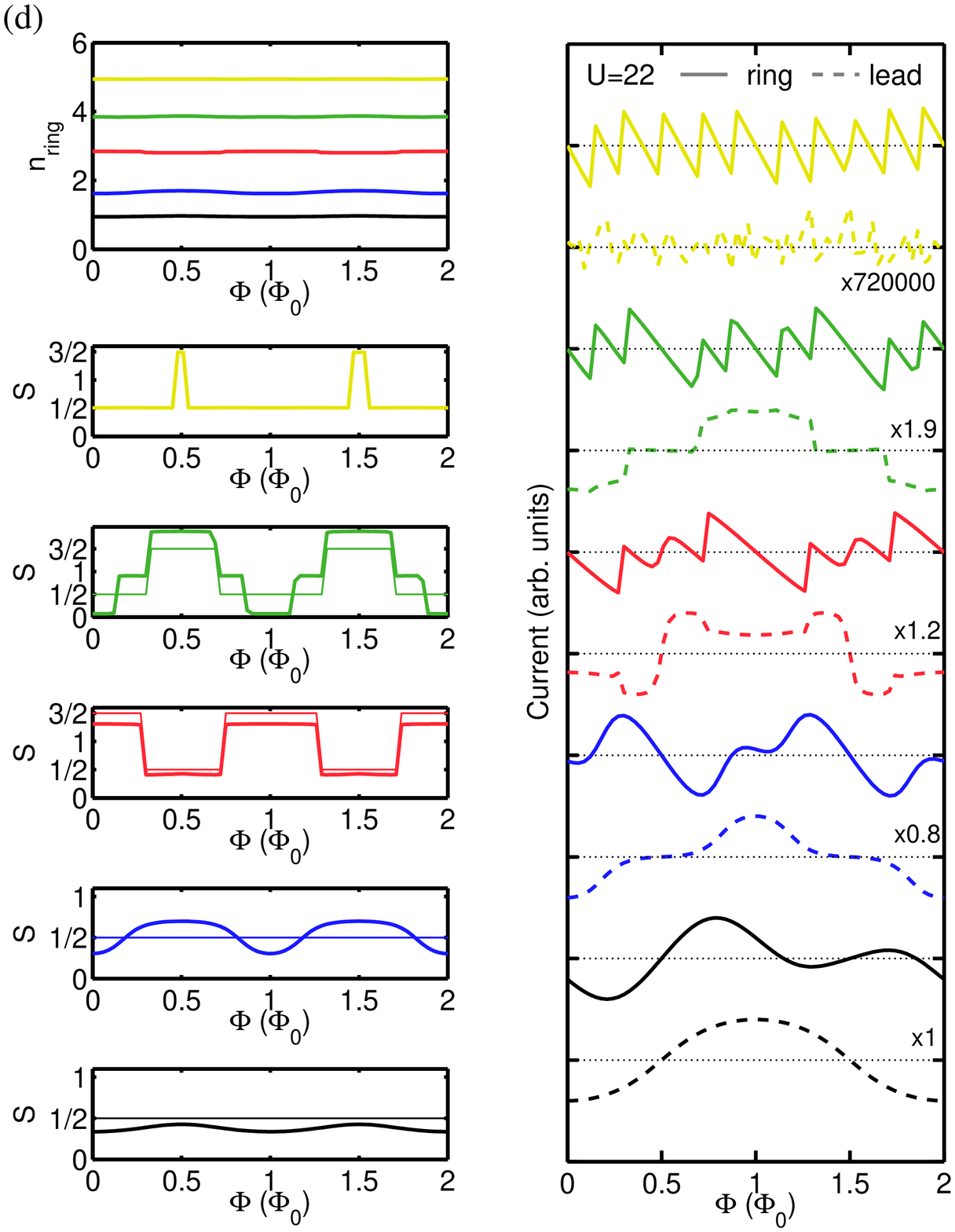} 

\caption{\label{fig:murtoluku_tc1} (Color online) The properties of a six-site 2NN ring strongly connected ($t_c = -t$) to a 14-site lead as a  function of the total flux $\Phi$ piercing the ring, with the ring occupation $n_{\mathrm{ring}}$ as close as possible to integer values (top left panels). Bottom left panels show the total spin (thin line) and its projection onto the ring part of the system (thick line), and the right panels the ring current (solid lines) and the lead current (dashed lines). The dotted lines show the zero level, and the curves have been normalized to facilitate comparison. For the lead currents, the scaling factor rounded to two significant digits is indicated. The values of $U$ chosen as to show fractional periodicity at each $n_{\mathrm{ring}}$, or as close as the system gets to it, and the gate voltage $V^g$ to keep the occupation constant as well as possible. Due to the strong coupling, large fluctuations in the electron number are present especially at low values of interaction. The five curves illustrate different ring occupations ($n_{\mathrm{ring}}\approx 1,2,3,4,5$: black, blue, red, green, yellow in color, black to light gray in grayscale). (a) $U=2t$ (b) $U=8t$ (c) $U=16t$ (d) $U=22t$.}
\end{figure*}

Similar to Fig.~\ref{fig:murtoluku} for weak coupling, Fig.~\ref{fig:murtoluku_tc1} shows the ring and lead currents, along with the total spin and the ring occupation in the strongly coupled systems. Again, the values of $U$ are chosen as to best show the fractional periodicity at each $n_{\mathrm{ring}}$ if possible. The top left panels in Fig.~\ref{fig:murtoluku_tc1} show the ring occupation as a function of $\Phi$. Unlike for weak coupling, the charge state of the ring can not be fixed to integer values for all electron numbers and interaction strengths. For $n_{\mathrm{ring}}=2$, the value $U=2t$ was chosen to facilitate comparison with the weakly coupled case (Fig.~\ref{fig:murtoluku}), and for $n_{\mathrm{ring}} = 3$, $U=8t$ to show the closest the system gets to fractional periodicity.  
 
The ring and lead currents (Fig.~\ref{fig:murtoluku_tc1}, right panels) for low occupations appear irregular and in the $n_{\mathrm{ring}} = 1$ case even $2\Phi_0$ periodic.  This is, again, related to the doubled period in the lead part of the system. In the strongly coupled case, the magnitudes of the lead and ring current are comparable and due to the current conservation at the coupling points, the lead current alters the ring current.  In contrast to the weakly coupled case, we only observe fractional periodicity for higher ring occupations, $n_{\mathrm{ring}} \geq 4$.  For electron fillings $n_{\mathrm{ring}} < 3$, the saw-tooth shape of the persistent current is completely smoothened at low interaction strengths, and for $n_{\mathrm{ring}}=3$ a persistent current resembling fractional periodicity with a partly smoothened shape appears. This is related to entanglement with the lead states and the resulting fluctuation in the occupation of the ring as a function of $\Phi$.\cite{Cheung-Gefen, Maiti_PhysicaE} On the other hand, from the viewpoint of the ring, the strong coupling to the lead can be interpreted as a form of disorder at the coupling sites. Disorder has also been shown to lead to smoothening of the kinks.\cite{Maiti_PhysicaE} 
In the $n_{\mathrm{ring}} = 3$ case, however, we observe a ring current characteristic to $U > U_c$ in the weakly coupled case at strong interaction, even though the actual fractional periodicity with sharp saw-tooth features is missing.

At higher ring occupations,  $n_{\mathrm{ring}} = 4$ or 5, we see fractional periodicity at sufficient $U$, and (pseudo-)$\Phi_0/2$ periodicity below it. Thus, it appears that at sufficiently low $V^g$ values corresponding to high $n_{\mathrm{ring}}$, the ring becomes effectively weakly coupled to the lead.  Some of the kinks are slightly smoothened due to avoided crossings of the energy parabola, fluctuations in the ring occupation, and large magnitude of the lead current. Fig.~\ref{fig:enephi}(c) shows the ground-state energy as a function of $\Phi$ at $U=8t$ for $n_{\mathrm{ring}} \approx 3$. As opposed to the weakly coupled case of Fig.~Ê\ref{fig:enephi}(b), some of the sharp crossings between the parabola disappear in the strongly coupled case. For occupations and interaction strengths with fractional periodicity, the behavior of the lead current, as well as the behavior above $U_c$, is similar to that at weak coupling. For values of $V^g$ for which the ring occupation strongly fluctuates as a function of $\Phi$, also other periodicities are possible. For instance, in a six-site ring strongly coupled to a 14-site lead with five electrons,  at $V^g \approx -1.5t$  and $U=2t$ the occupation fluctuates between three and five electrons, and the periodicity of the ring current is pseudo-$\Phi_0/3$ (not shown). 

In general, in the strongly coupled systems the peaks and kinks in the lead current do not reflect the kinks in the ring current as accurately as for weak coupling. At fractional periodicity, the behavior of the ring and lead currents are similar irrespective of the coupling strength, namely, with abrupt magnitude and sign changes that coincide with the jumps in the ring current. At low $U$ and lower ring occupations we mainly see a single peak in the 2$\Phi_0$-periodic lead current, as if the ring were just a disordered site in the lead ring. For $n_{\mathrm{ring}} = 5$, the lead current shown in Fig.~\ref{fig:murtoluku_tc1} is essentially numerical noise due to minuscule magnitude of the current. Thus, in the strongly coupled case, the observation of fractional periodicity from conductance or transmission is much more difficult than at weak coupling. 

The evolution of the total spin with increasing interaction is similar to the weakly coupled case, although the spin projection onto the ring is smoothened out at low $n_{\mathrm{ring}}$ due to the strong coupling. At fractional periodicity, however, the crossings between different-$S$ parabola are clearly seen as abrupt changes in the spin state in the upper right panels of Fig.~\ref{fig:murtoluku_tc1}. Again, with increasing $U$, states with a higher total spin become the ground state for larger regions in $\Phi$.  

\begin{figure}
 \includegraphics[width = 0.95\columnwidth]{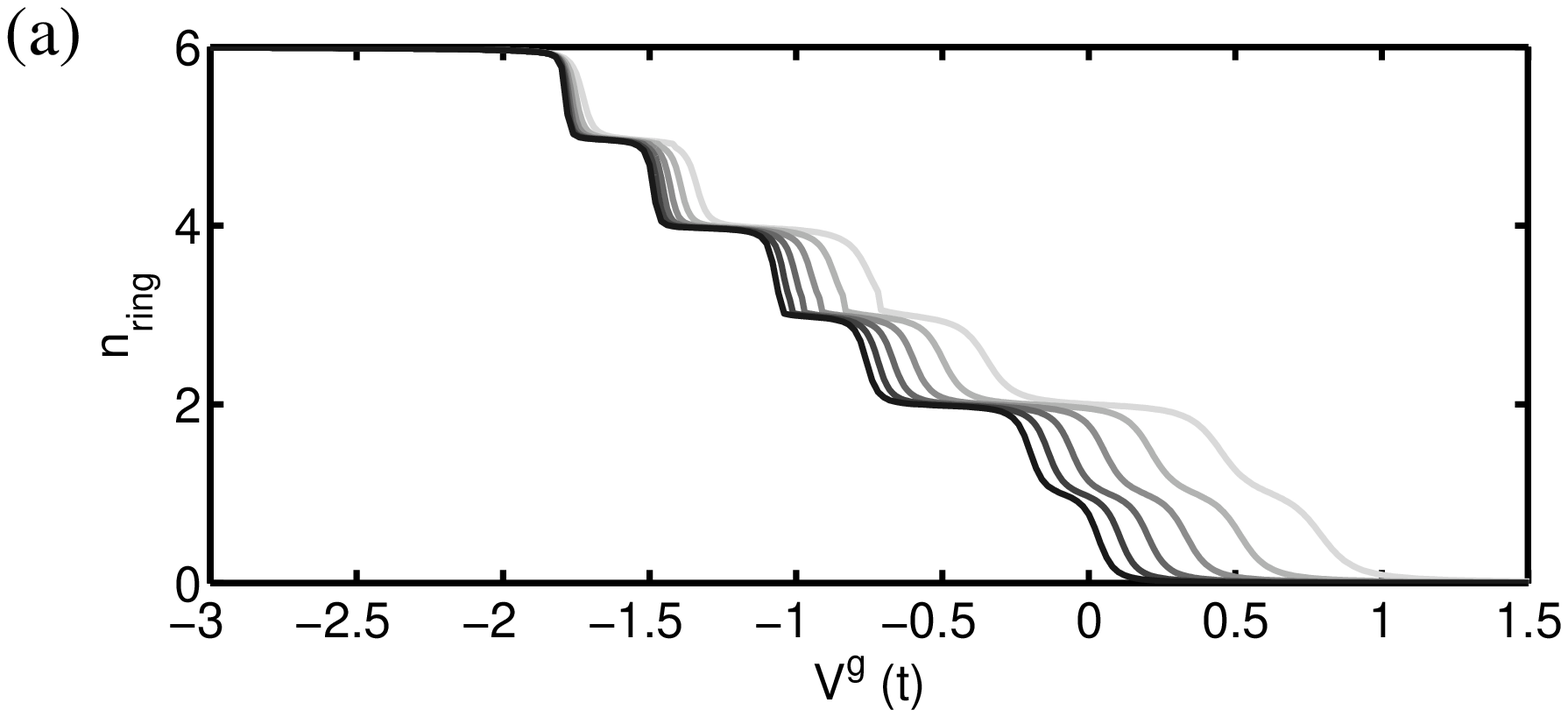}\\
 \includegraphics[width=0.95\columnwidth]{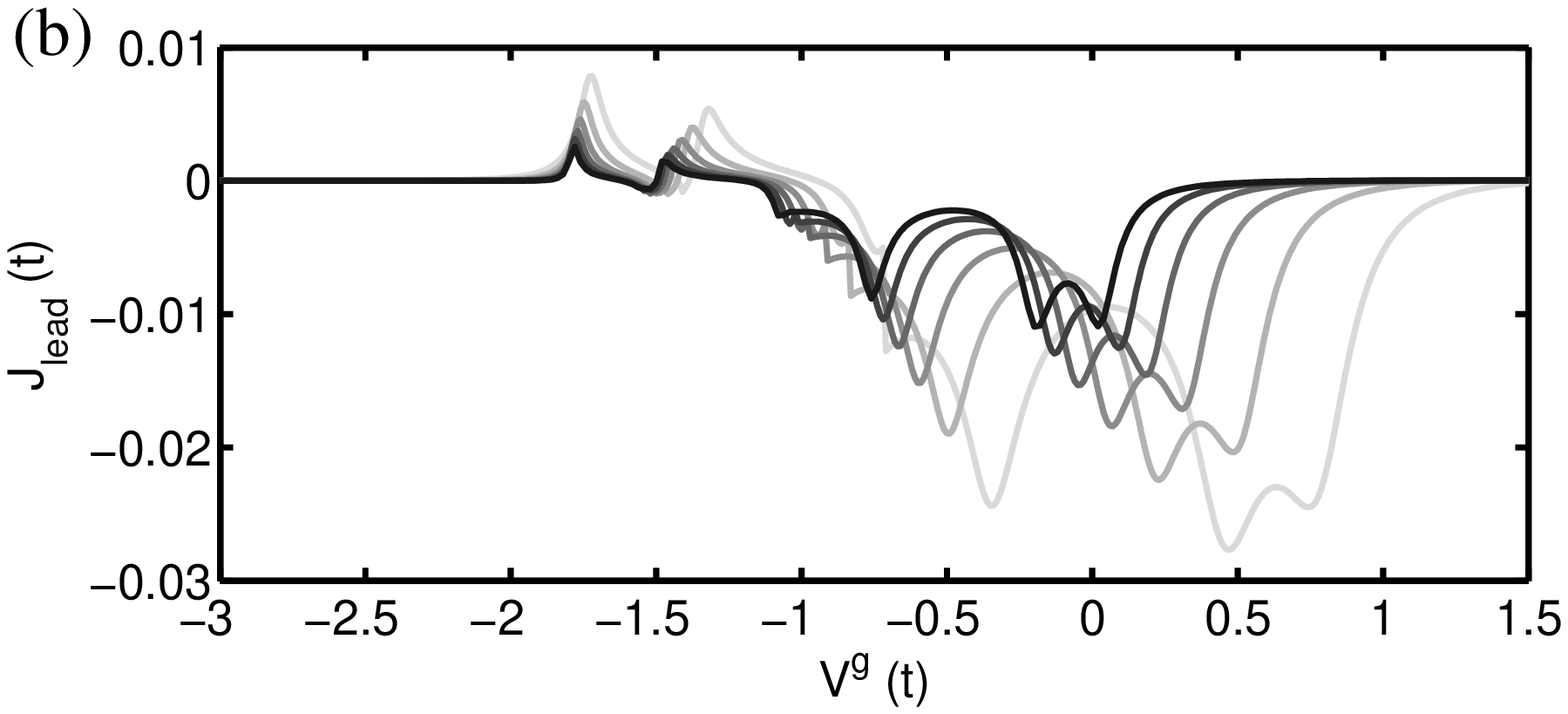}\\
 \includegraphics[width = 0.95\columnwidth]{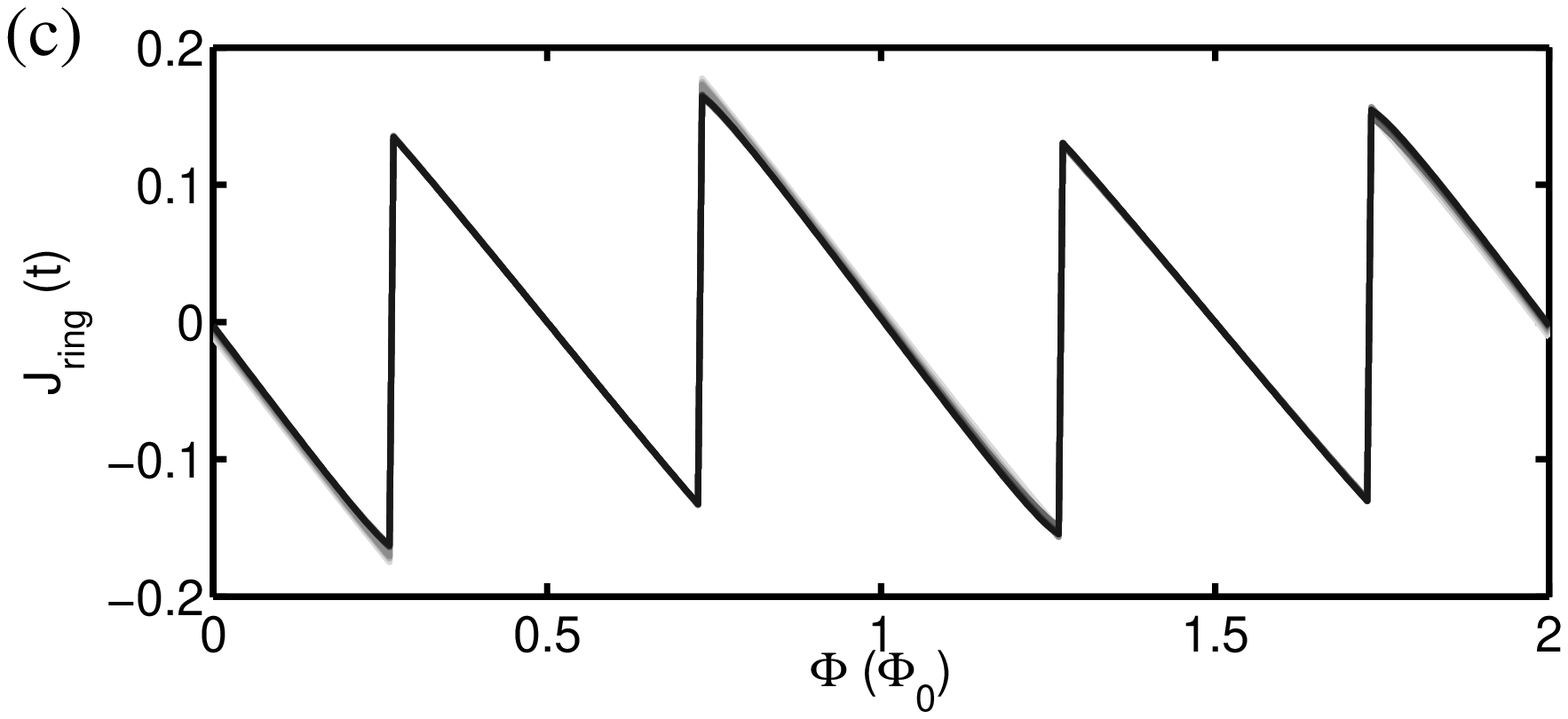} \\
 \includegraphics[width = 0.95\columnwidth]{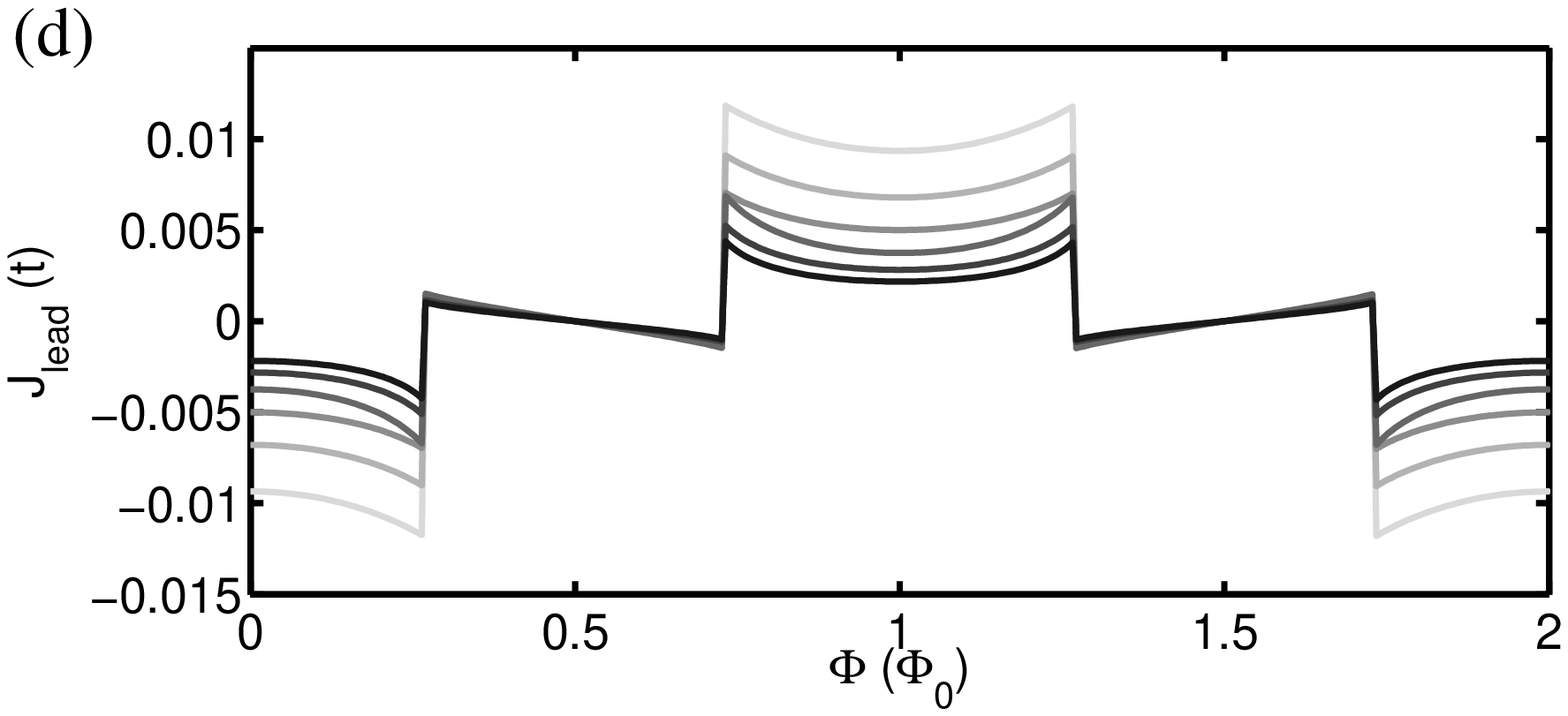}  \\
 \caption{\label{fig:finsize} The effect of lead length at weak coupling ($t_c = -0.2t$) with a fixed flux $\phi_i = 0.01\Phi_0$ on (a) the occupation of the ring (b) the lead current as a function of the gate voltage $V^g$. (c) The persistent current in the ring (d) persistent current in the lead as a function of the flux piercing the ring, $\Phi$, at $V_g$ chosen such that $n_{\mathrm{ring}}=2$. The number of sites in the lead, $N_l$, increases from 10 to 20 in steps of two with the darkening shade of gray. Other parameters $N_{el}$= 6, $U = 2t$ and $t' = 0.2t$.}
\end{figure}

In Figs.~\ref{fig:murtoluku} and \ref{fig:murtoluku_tc1}, the current curves have been normalized as to facilitate qualitative comparison. It is, however, instructive to also consider the magnitudes of the current. For the lead currents, the scaling factors relative to the $n_{\mathrm{ring}}=1$ current are given in the figures, rounded to two significant digits.  In general, the magnitude of the ring current is only slightly smaller in the strongly coupled case than in the weakly coupled case, the difference increasing with increasing interaction strength. The lead current, on the other hand,  is  an order of magnitude larger in the strongly coupled case than in the weakly coupled case. In particular, for the strongly coupled $n_{\mathrm{ring}} = 1$ case, the lead and ring current are of the same order of magnitude. The strong coupling disturbs the ring part, decreasing the ring current as the hybridization with the lead states is enhanced. The increased magnitude of the lead current in the strongly coupled case can be understood as increased transmission through the ring.

\begin{figure}
 \includegraphics[width=1.00\columnwidth]{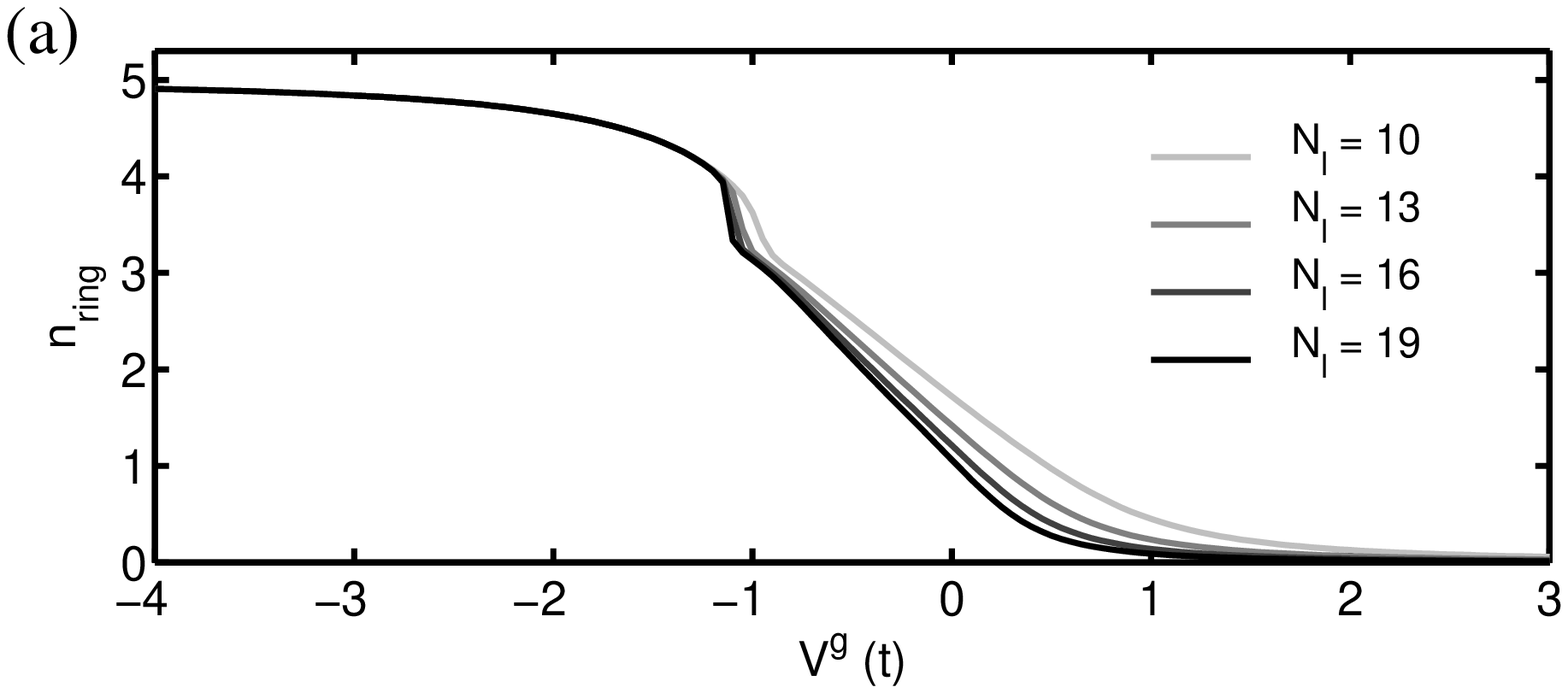}
  \includegraphics[width=1.00\columnwidth]{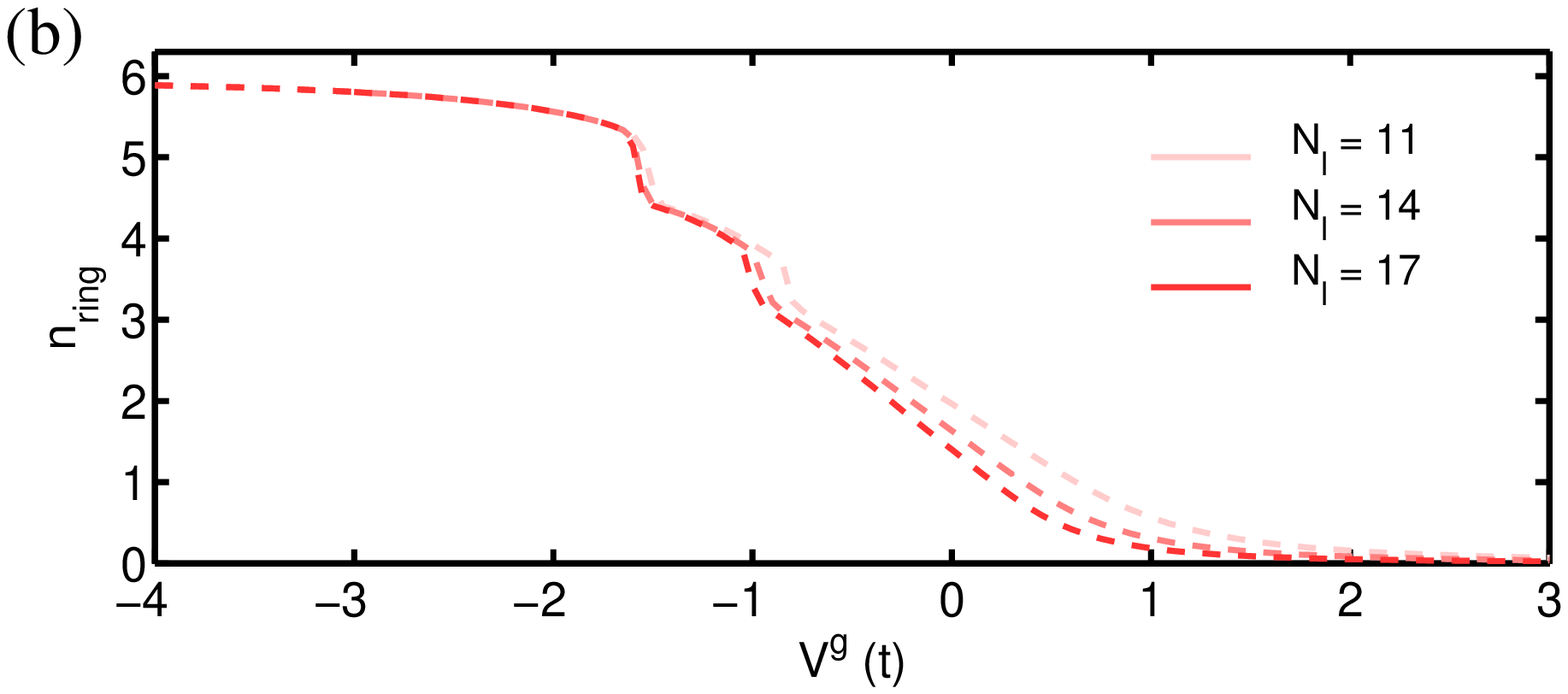}
   \includegraphics[width=1.00\columnwidth]{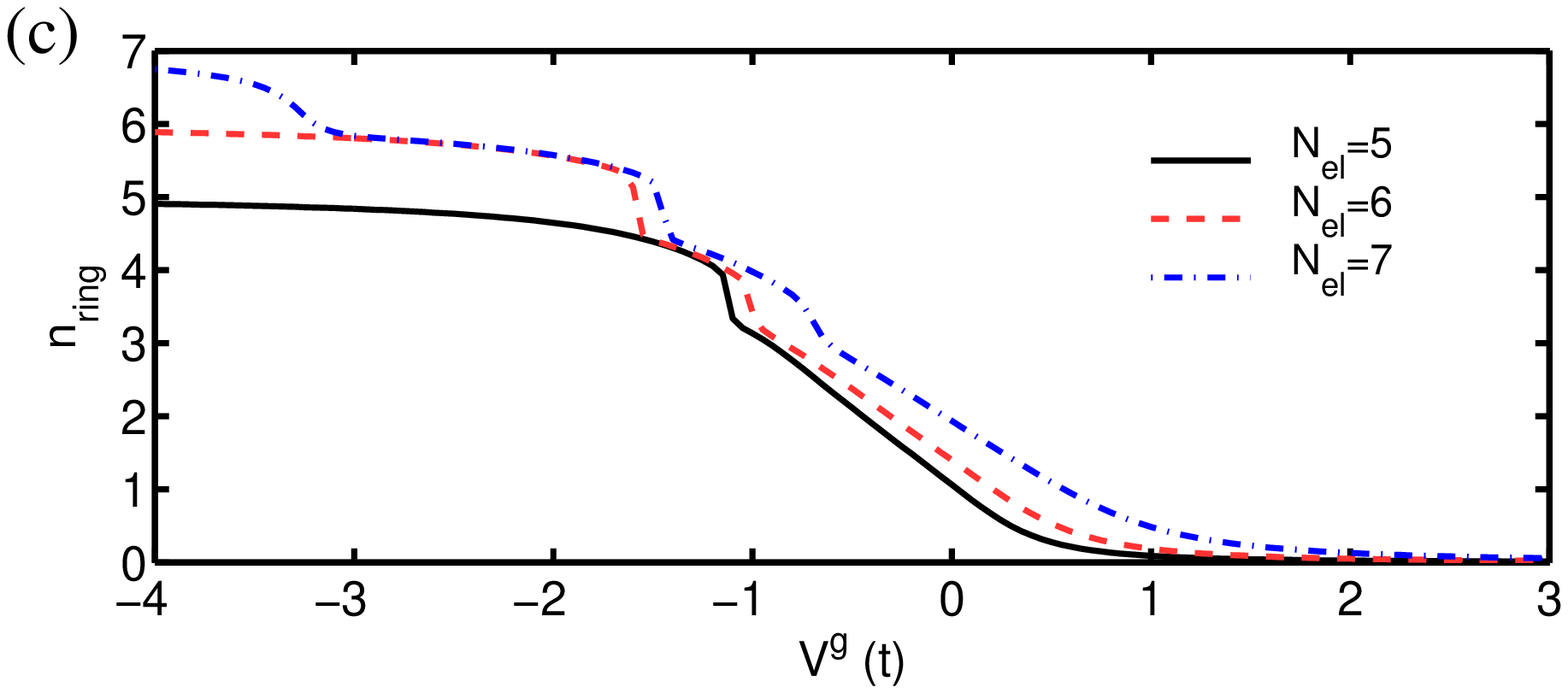}
    \caption{\label{fig:finsizetc} (Color online) The effect of an increasing lead length $N_l$ on the Coulomb staircase in a six-site ring strongly coupled ($t_c = -t$) with the lead. In (a),  $N_{el}=5$, the lead lengths 10, 13, 16, and 19 sites, and in (b) $N_{el}=6$, the lead lengths 11, 14, and 17 sites, the lead length increasing with darkening shade of gray/red. In (c), the Coulomb staircase for different electron numbers is compared ($N_{el} = 5$: $N_l = 19$, $N_{el} = 6$: $N_l = 17$, and $N_{el} = 7$: $N_l = 14$). In all calculations, $\phi_i = 0.01\Phi_0$ and $U=2t$.}
\end{figure}

\subsection{Finite-size effects \label{sec:fin size}}

\begin{figure}
 \includegraphics[width=1.00\columnwidth]{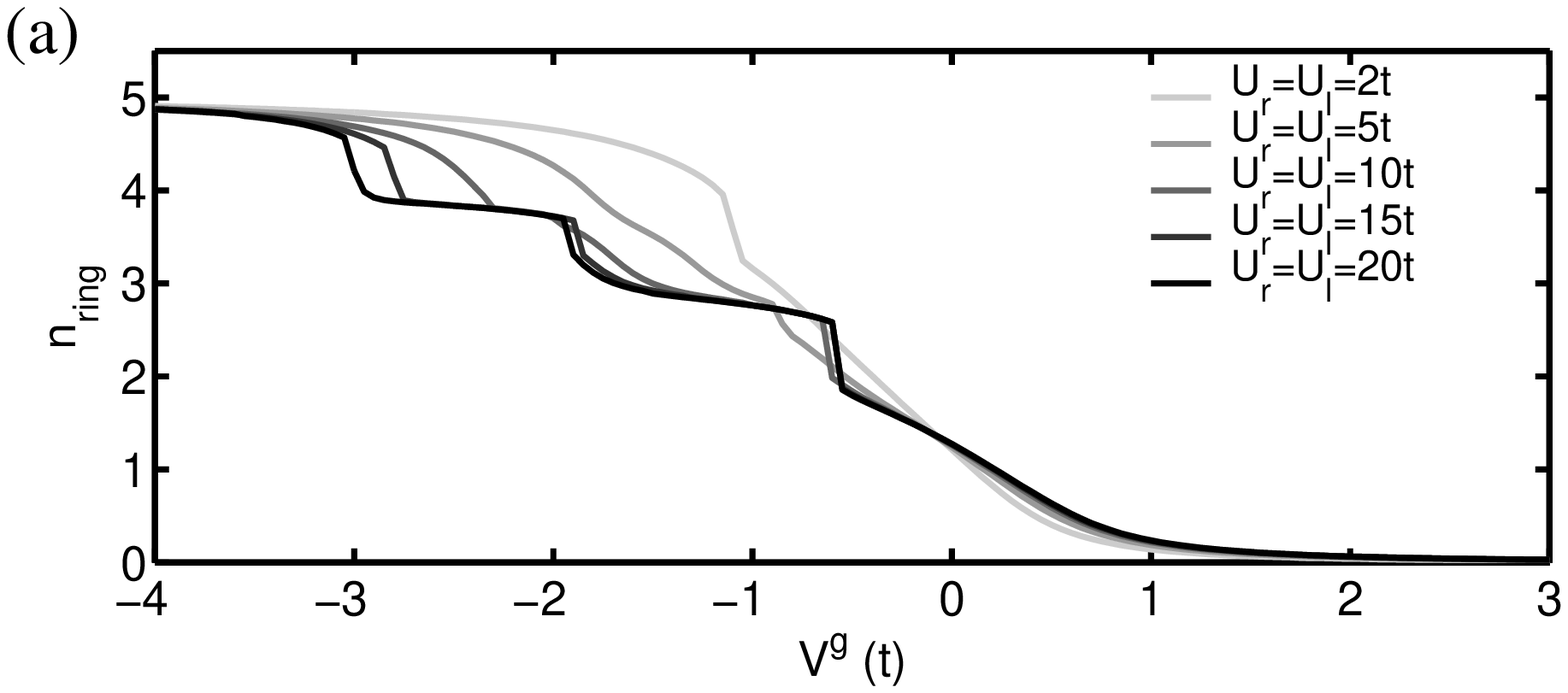}
    \includegraphics[width=1.00\columnwidth]{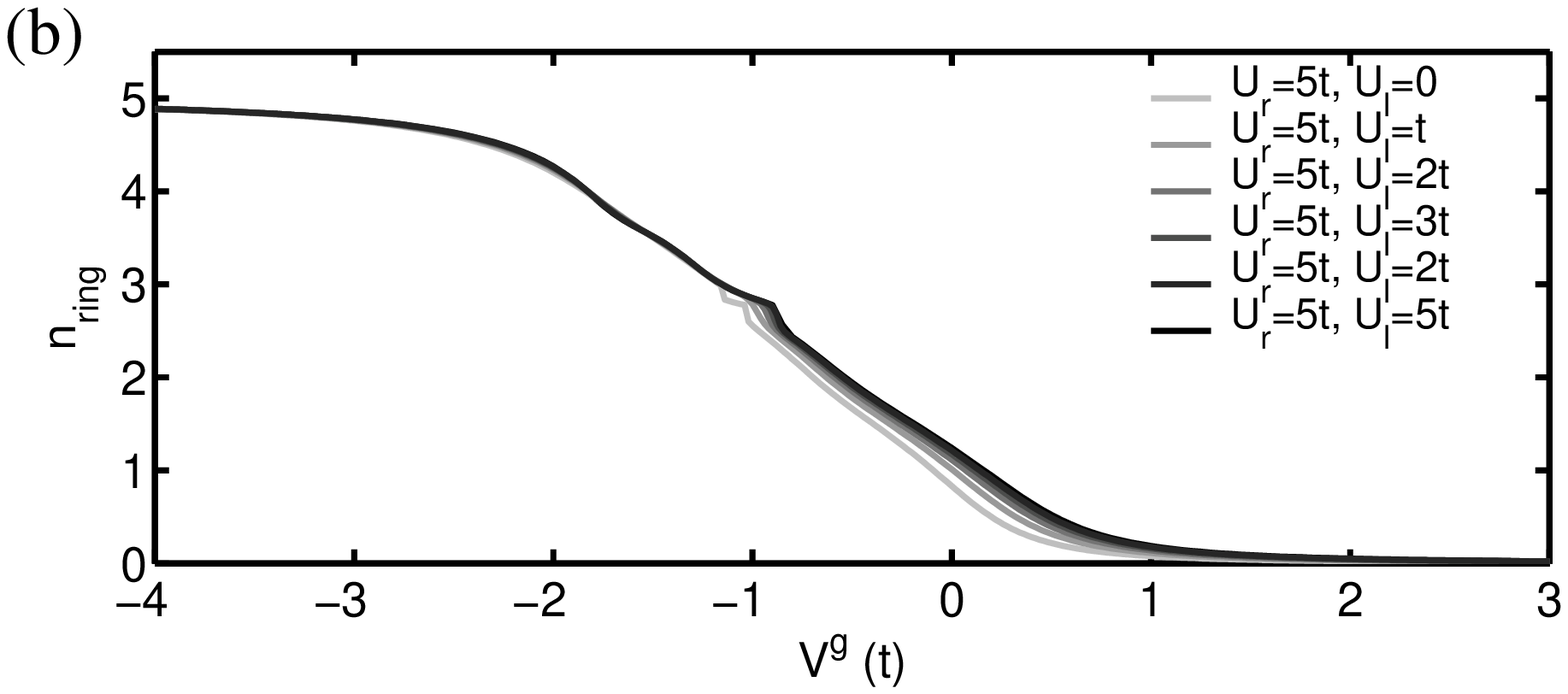}
      \includegraphics[width=1.00\columnwidth]{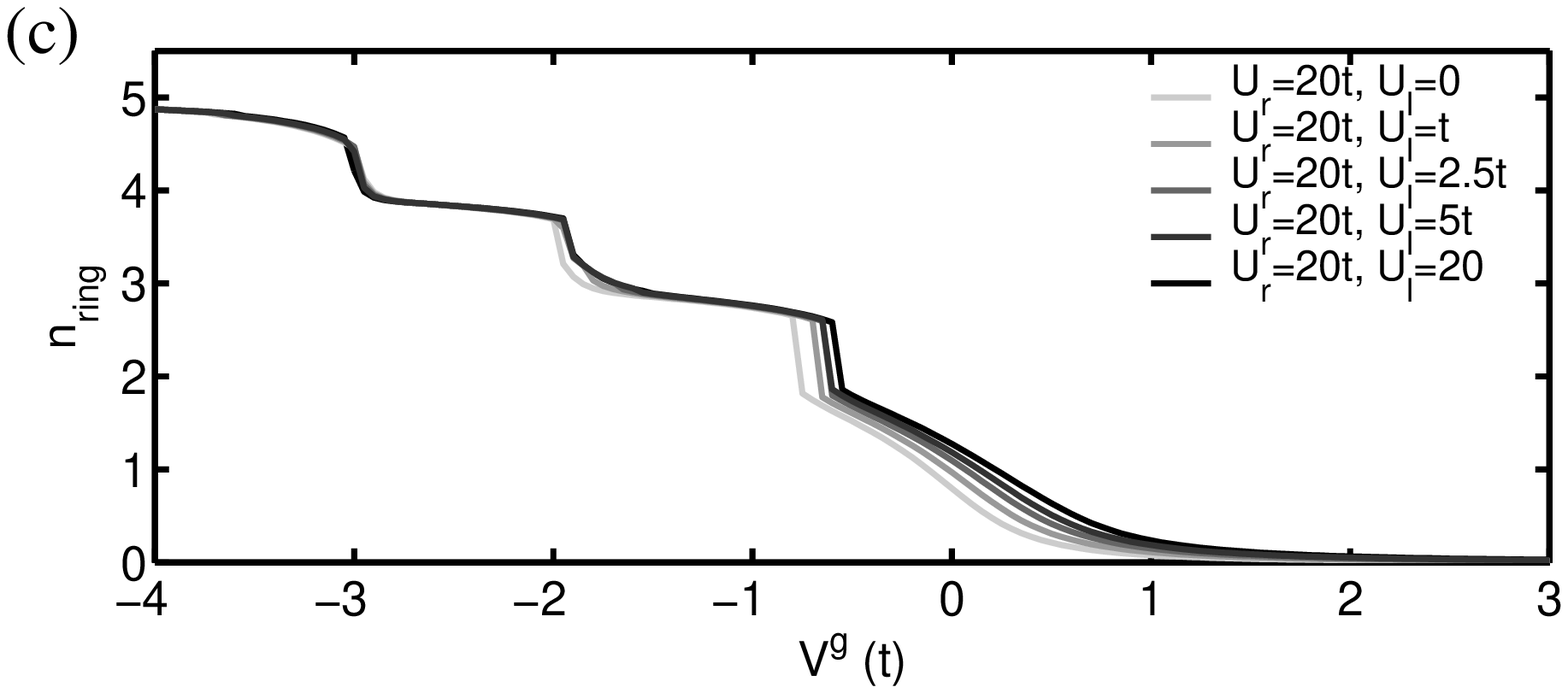}
  \caption{\label{fig:Uvar}  The effect of the interaction strength for $N_{el}=5$ for strong coupling between the ring and the lead. (a) The solid gray to black line: $U=2t$, 5$t$, 10$t$, 15$t$ and 20$t$.   (b),(c) The effect of the interaction strength in the lead, $U_l$,  with $U_r = 5t$ and $U_r = 20t$ in the ring part for all curves, respectively. $U_l$ increases with darkening shade of gray.  In all calculations, $N_{\mathrm{ring}} = 6$, $t_c = -t$, $N_l = 16$, and $\phi_i = 0.01\Phi_0$.}
\end{figure}

As we are dealing with relatively small systems, it is necessary to pay attention to finite size effects, namely, what is the effect of the length of the lead.  We are not interested in the diminishing magnitude of the lead current with increasing lead length\cite{Ferrari} but instead, the behavior of the ring persistent current and the gate voltages at which the charge state of the ring changes. Fig.~\ref{fig:finsize} shows the properties a six-site ring  with 2NN hopping ($t'=0.2t$) with the length of the lead increasing from 10 sites to 20 sites in steps of two sites. The ring is weakly coupled, $t_c = -0.1\,t$, and the number of spin up- and spin down- electrons are equal ($N_{el} = 6$). In (a) and (b), $\phi_i$ is fixed to 0.01$\Phi_0$ and in (c) and (d), $V^g$ in the ring is chosen to give $n_{\mathrm{ring}} \approx 2$. 

In Fig.~\ref{fig:finsize}(a), a clear Coulomb staircase characteristic for weakly coupled systems with well-defined charge states and transition regions associated with an electron tunneling from the lead to the ring between them is seen. As the lead length is increased, the steps shift toward more negative gate voltages, and the values of $V^g$ corresponding to the steps converge starting from the steps with highest $n_{\mathrm{ring}}$.  As the total number of electrons in the ring-lead system is fixed, a removal of an electron in the ring corresponds to an addition of an electron in the lead. For high ring occupations, the filling factor in the lead is low and thus the exact number of sites is not as important as for higher fillings.  The lead can be thought to act as a sort of a small reservoir but due to the system configuration, neither the electron number of the ring nor its chemical potential remain constant when the gate voltage or flux piercing the ring are changed. Comparing the lead current as a function of the gate voltage, shown in Fig.~\ref{fig:finsize}(b), we see that changes in the ring occupation are accompanied by peaks in the lead current due to charge transfer between the two subsystems.

Figs.~\ref{fig:finsize}(c) and \ref{fig:finsize}(d) show the persistent current in the ring and in the lead as a function of $\Phi$, respectively, at a gate voltage $V^g$ fixed such that there are approximately two electrons in the ring. The small differences between the curves in Fig.~\ref{fig:finsize}(c) are mainly due to small fluctuations in the ring occupation. We see that the lead length has little effect on the current within the ring, whereas with an increase in the lead length leads, the magnitude of the lead current is somewhat decreased [Fig.~\ref{fig:finsize}(d)]. The persistent current in a perfect, non-interacting ring without a nanostructure decays as $1/N_{\mathrm{lead}}$, and with a weakly coupled embedded Anderson impurity-type quantum dot as $1/N_{\mathrm{lead}}^{1/2}$.\cite{Ferrari} Again, qualitatively the curves are very similar, the changes in the  current occurring at same flux values with sign changes in the ring current.

\begin{figure}
 \includegraphics[width=0.95\columnwidth]{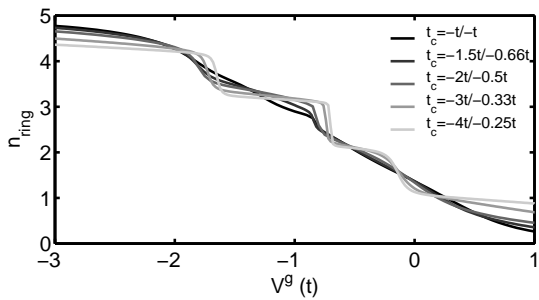}
 \caption{\label{fig:unequal_tc} The effect of asymmetry in the strength of the lead-ring coupling. The product of the couplings is fixed to one, $t_{c1}t_{c2} = 1$, but the ratio between them is varied. Parameters: Six-site ring, $N_l$ = 14, $U = 5t$, $N_{el} = 5$, $\phi_i = 0.01\Phi_0$, and $t'=0.2t$.}
\end{figure}

In the strongly coupled case, the Coulomb staircase is partly smoothened out, as illustrated in Fig.~\ref{fig:finsizetc}. In principle, the strong coupling between the lead and the ring should lead to broadening of the transmission resonances leading to the smoothening of the Coulomb steps. Fig.~\ref{fig:finsizetc}(a) and (b) show the ring occupation as a function of the gate voltage $V^g$ for a six-site 2NN ring at strong coupling ($t_c =-t$) and $U=2t$ for $N_{el} = 5$ and $N_{el} = 6$, respectively. In both subfigures, the darkening shade of the curves indicates the increasing number of sites, $N_l$, in the lead. We indeed observe a smoothening of the steps, to the extent that in the $N_{el}=5$  case only one step is observable, compared to all five seen at weak coupling. Fig.~\ref{fig:finsizetc}(c) compares the staircase for $N_{el}$ = 5, 6, and 7 with $N_l$ = 19, 17, and 14, respectively. The low-occupation steps remain smoothened out but increasing the total number of electrons in the system leads to the presence of additional high-occupation steps.  

\begin{figure*}
\begin{tabular}{cccc}
\includegraphics[width=0.6\columnwidth]{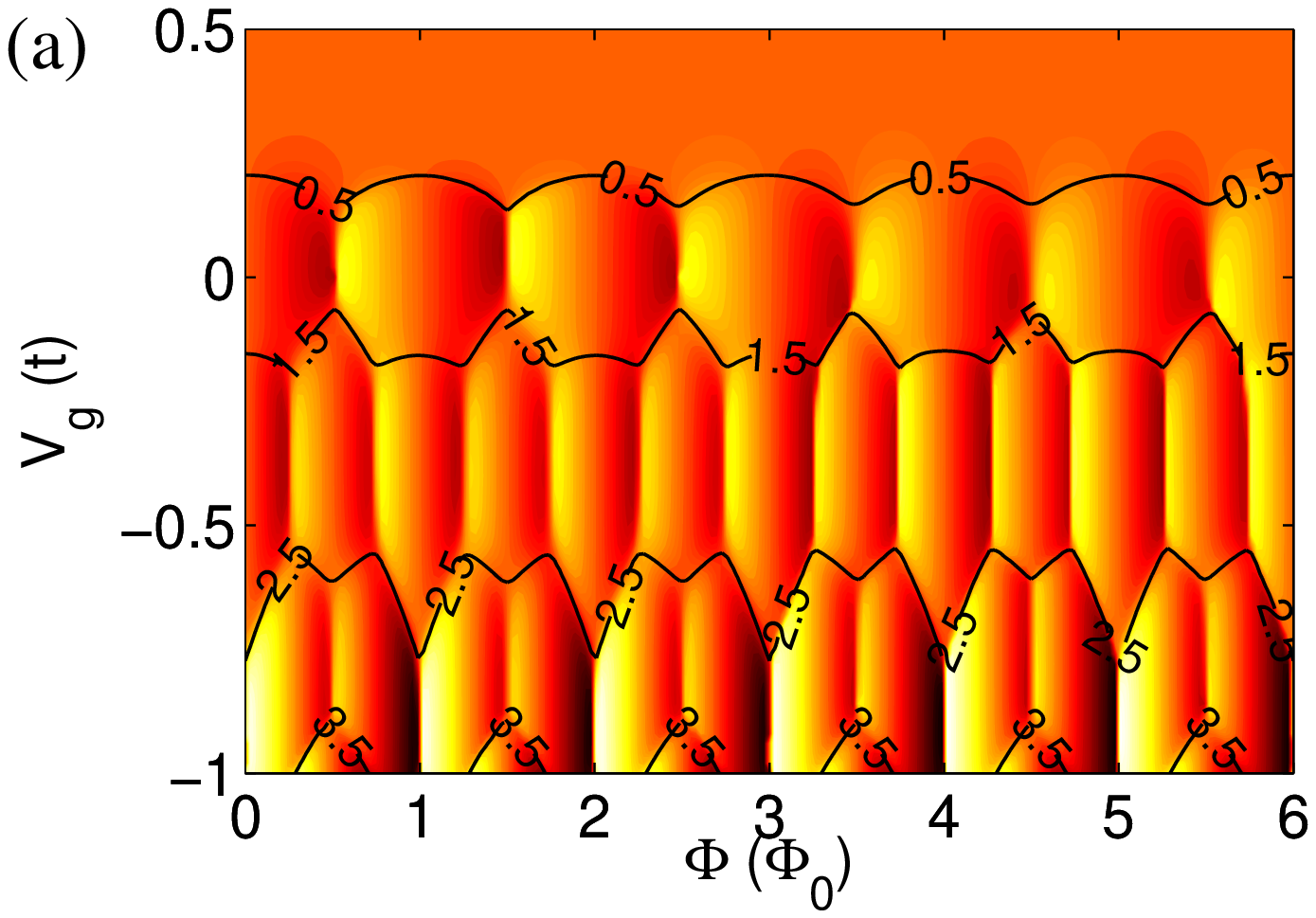}&
 \includegraphics[width=0.6\columnwidth]{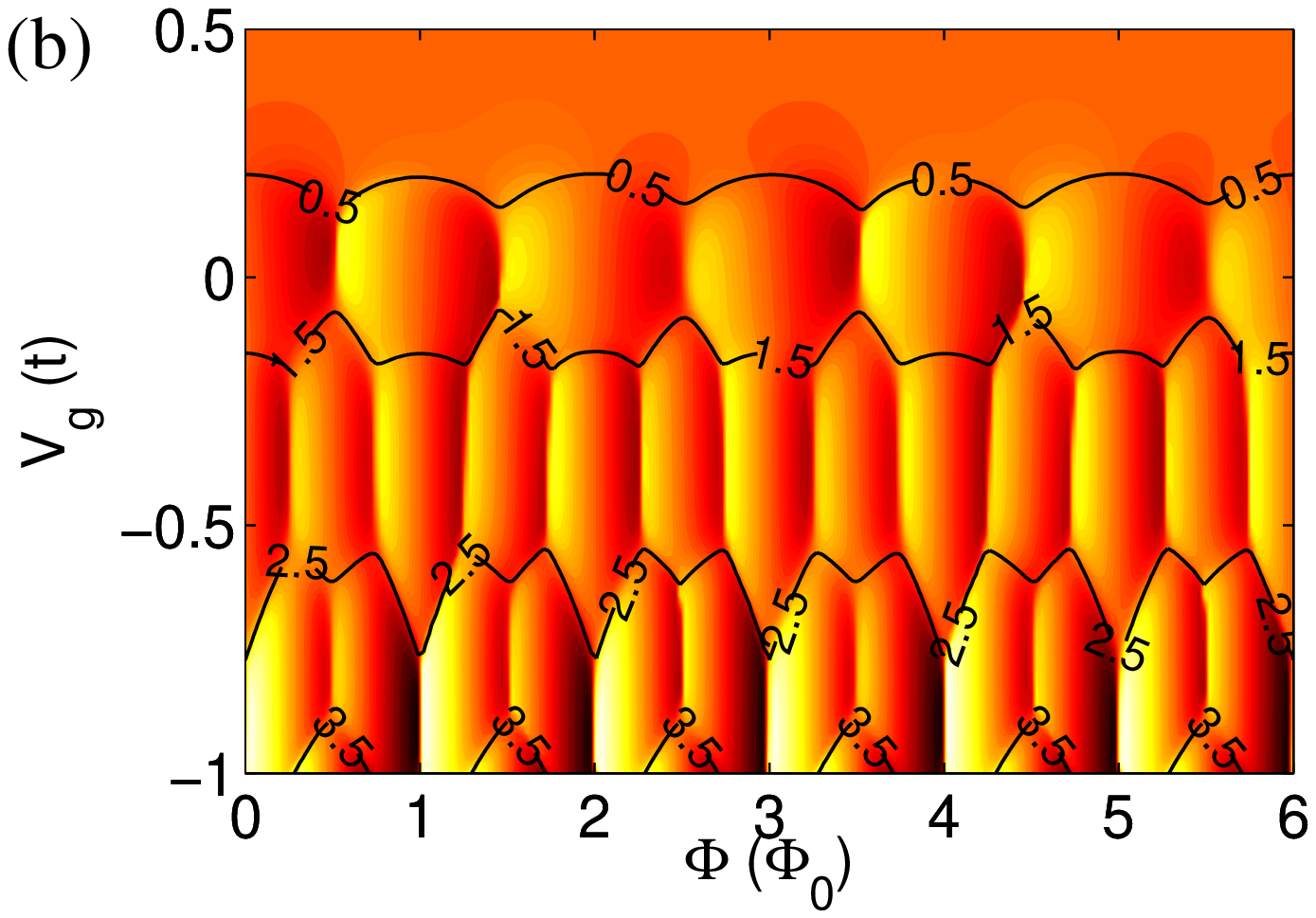}&
 \includegraphics[width=0.6\columnwidth]{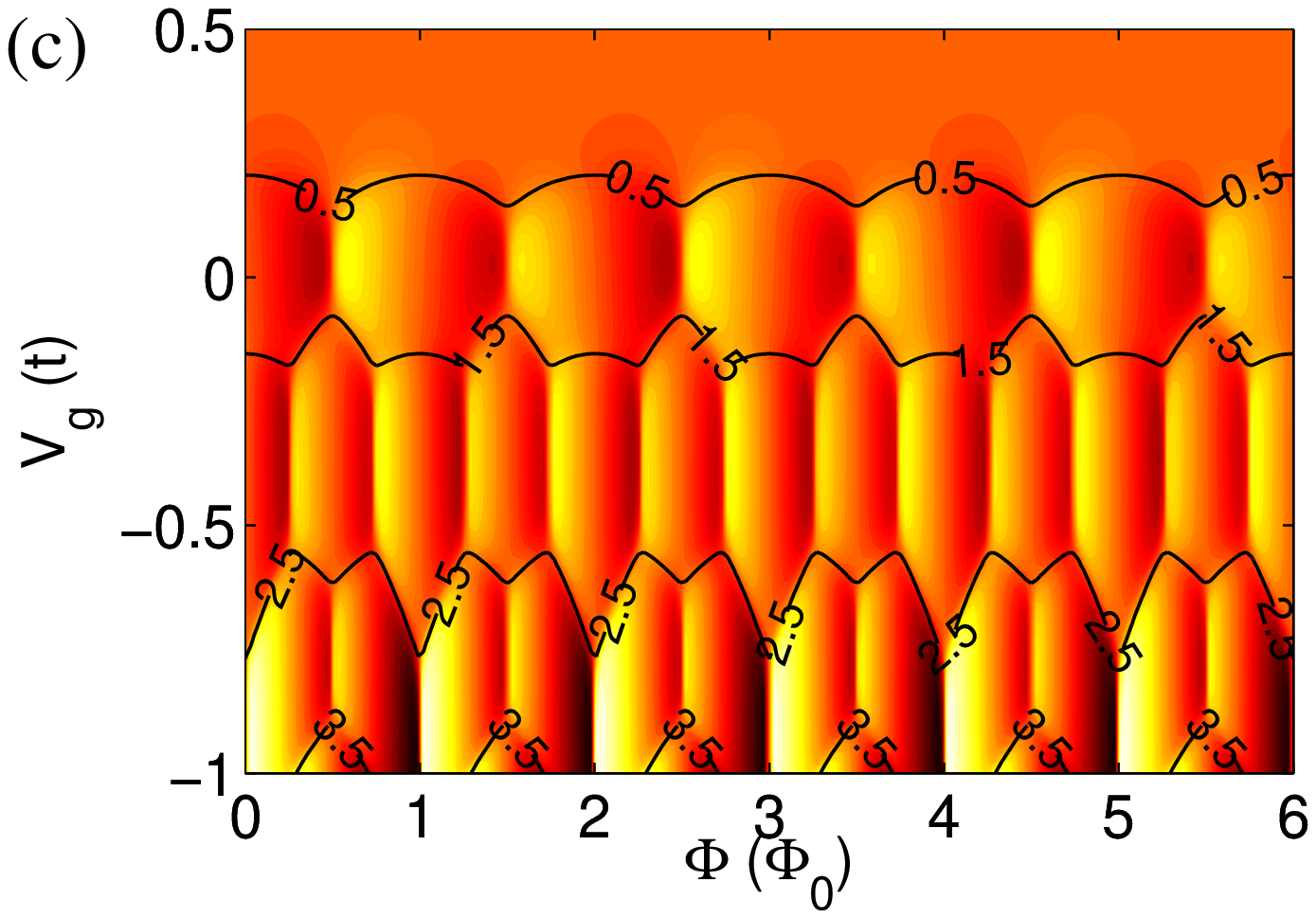}&

  \includegraphics[width=0.178\columnwidth]{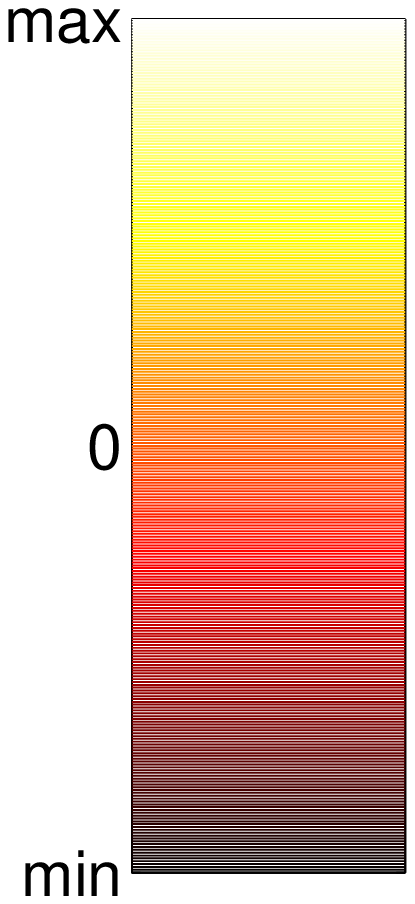}\\
\includegraphics[width=0.6\columnwidth]{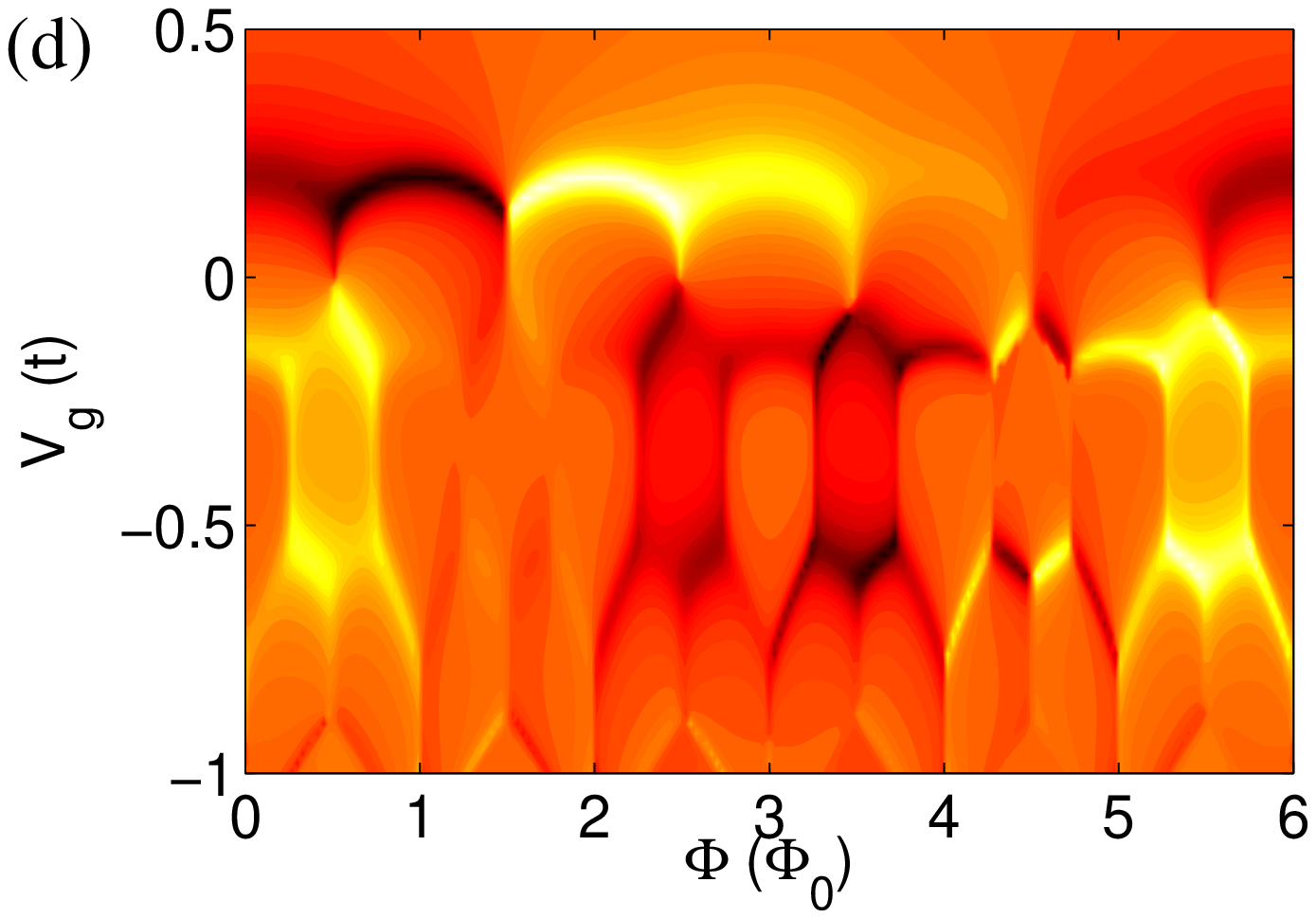}&
 \includegraphics[width=0.6\columnwidth]{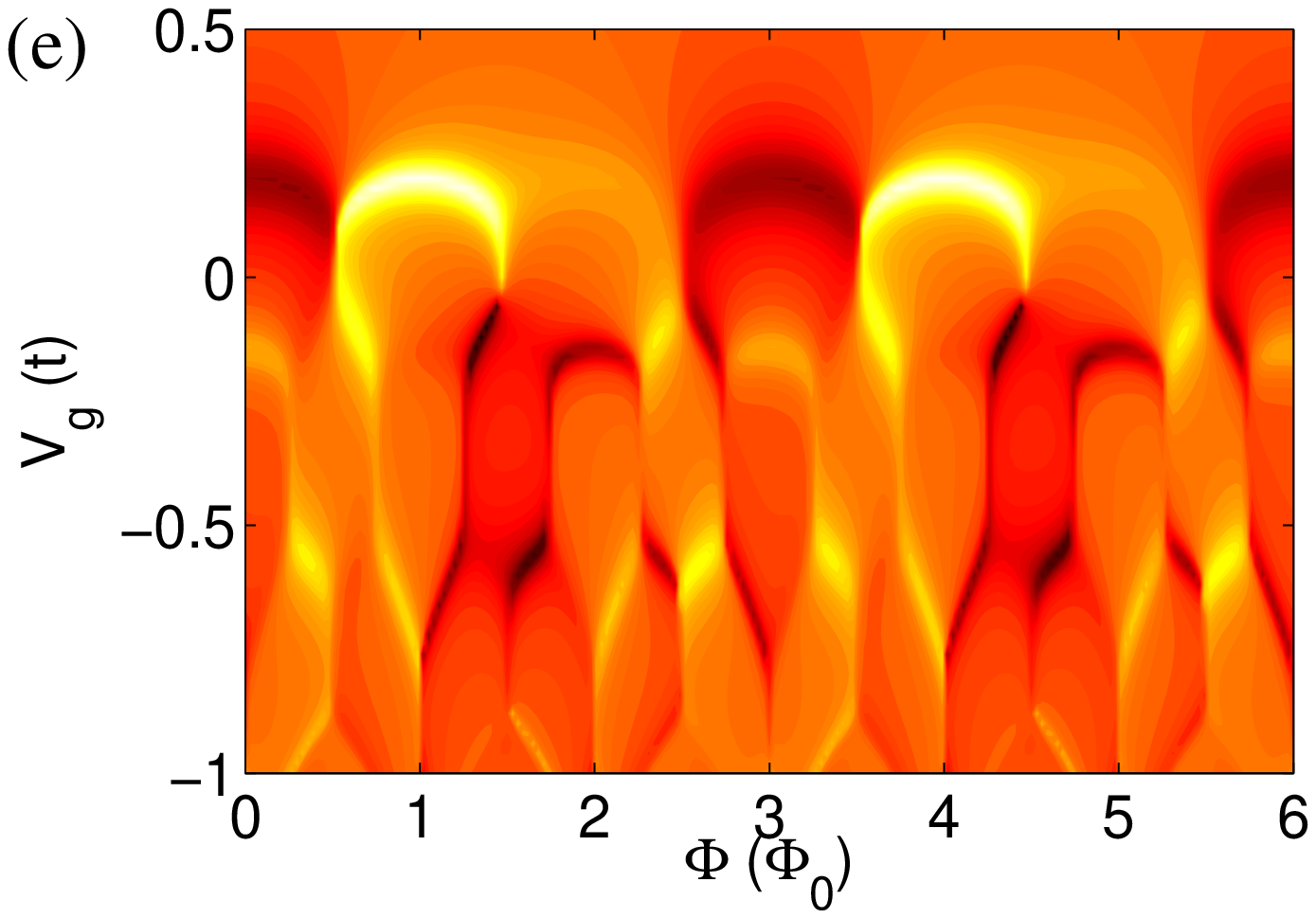}&
 \includegraphics[width=0.6\columnwidth]{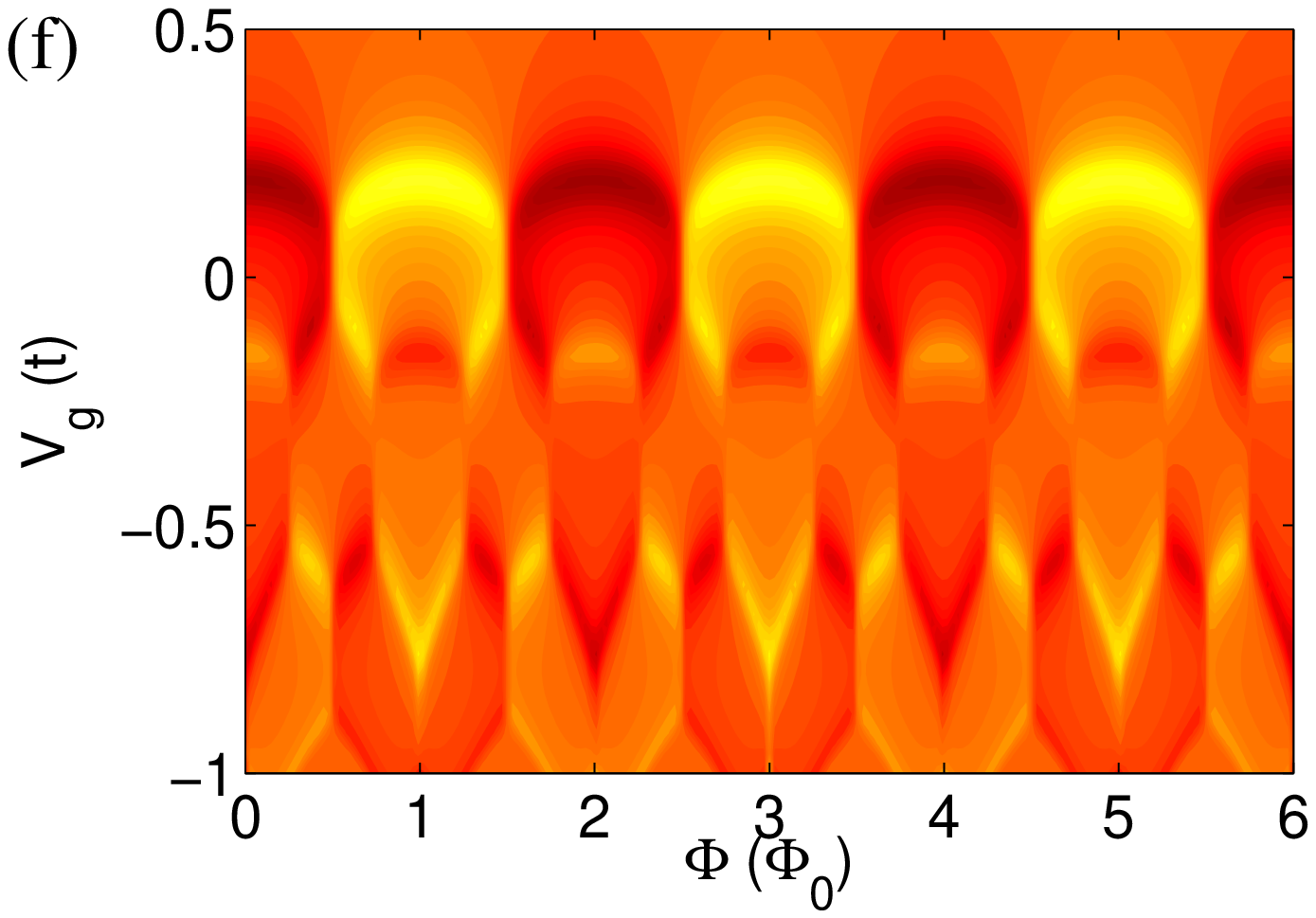}&
  \includegraphics[width=0.178\columnwidth]{cbar_J.eps}\\
 \includegraphics[width=0.6\columnwidth]{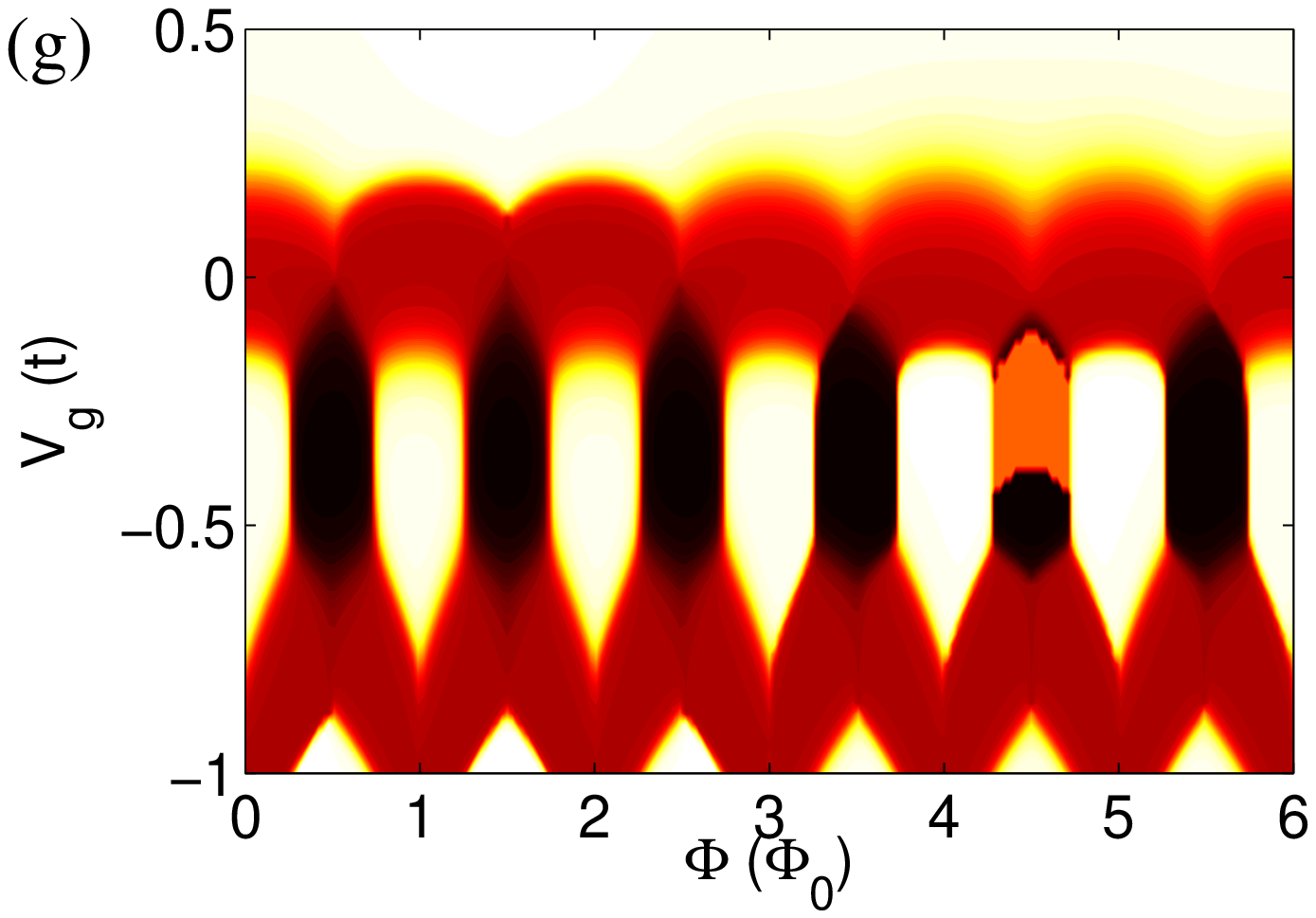} & 
 \includegraphics[width=0.6\columnwidth]{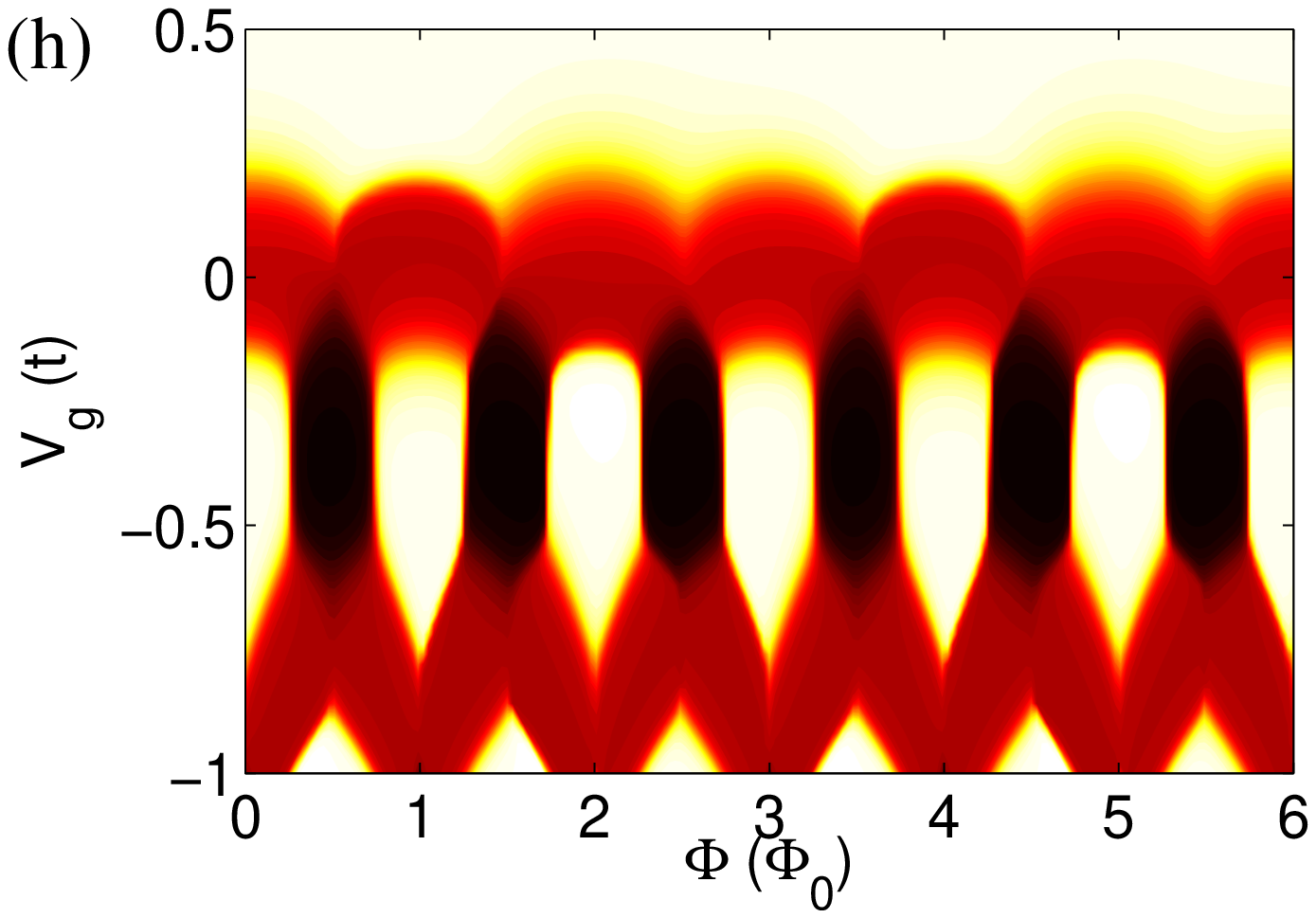}  & 
  \includegraphics[width=0.6\columnwidth]{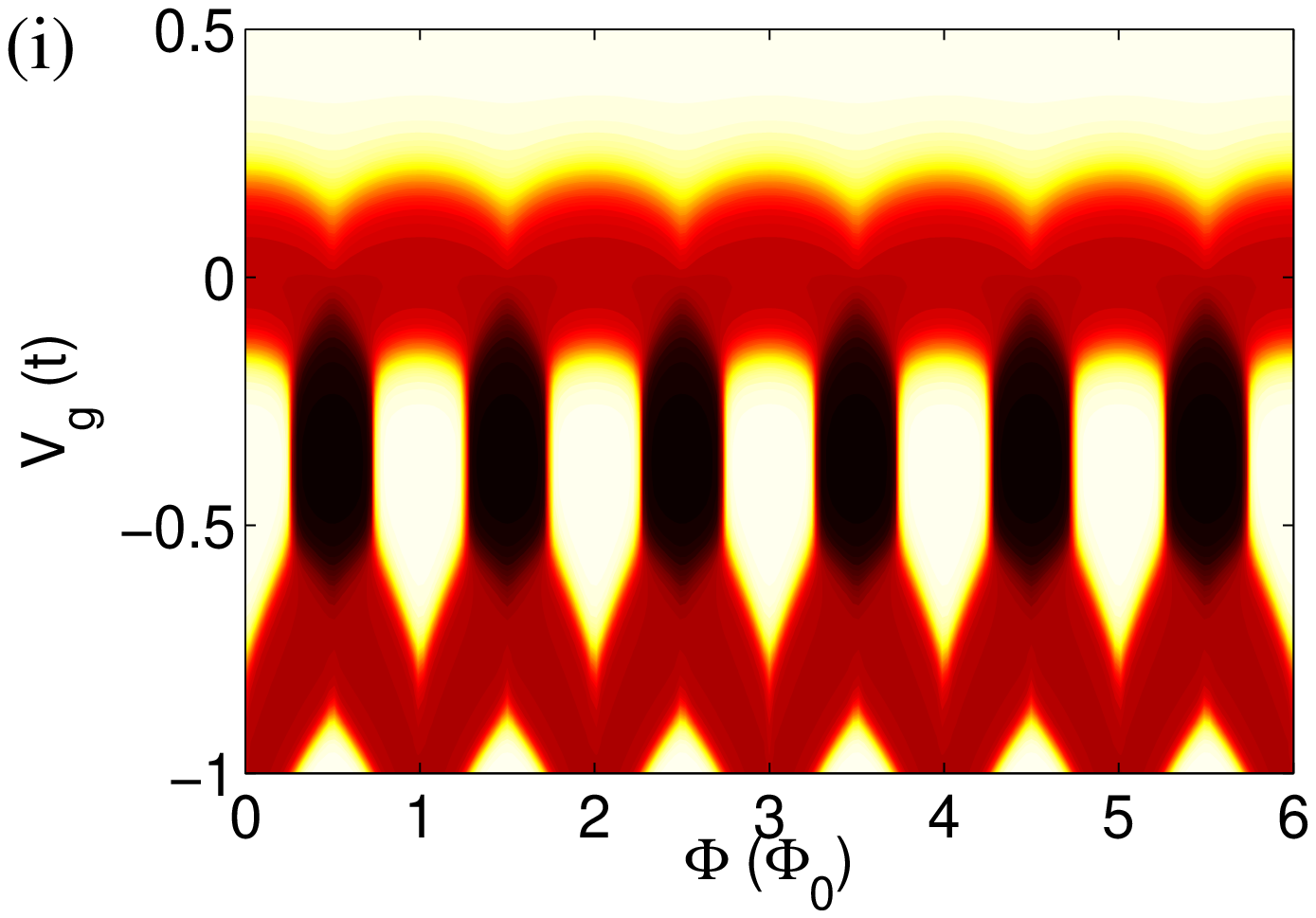} &
  \includegraphics[width=0.178\columnwidth]{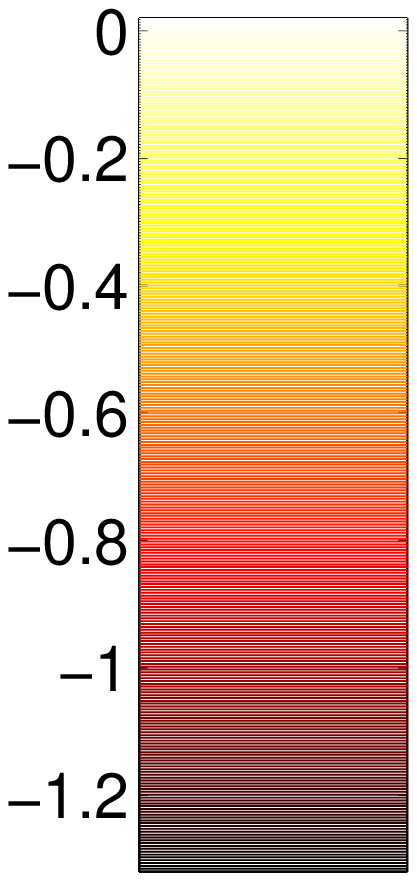}\\

\end{tabular}
 \caption{\label{fig:couplingsite} (Color online) The ring and lead currents [(a)-(c) and (d)-(f), respectively], as well as spin imbalance $n_{\uparrow}-n_{\downarrow}$ in the ring [(g)-(i)], in the ($\Phi$, $V^g$) space for the three different coupling sites for weak coupling between the lead and the ring.  (a),(d),(g): 1--2 coupling (b),(e),(h): 1--3 coupling (c),(f),(i): 1--4 coupling. In (a)-(c), the contours of half-integer ring occupation $n_{\mathrm{ring}}$ are additionally drawn, separating regions of approximately integer values of $n_{\mathrm{ring}}$.  Other parameters: Six-site ring, 14-site lead, $t_c = -0.2t$, $t' = 0.2t$, $N_{el}=5$, $U = 2t$. The color scales for each row are fixed to allow comparison in the magnitudes between different coupling positions.   } 
 \end{figure*}

The disappearance of the low-occupation steps can be understood if the competition between kinetic and interaction energy is considered. At large negative gate voltages, the localization of electrons in the ring is favored as the kinetic effects due to the strong coupling are small compared to the effect of the gate voltage. As the ring is close to half-filling, the density of doubly occupied sites is low and due to the presence of the 2NN hoppings, the electrons are allowed to move within the ring without having to doubly occupy any single site.  When the gate voltage is increased, the electrons are increasingly delocalized both in the ring and the lead. Charging effects associated with the addition of a single electrons can no longer be observed. Thus, at large negative $V_g$ the system is effectively weakly coupled to the gate voltage, and shows behavior similar to weak coupling. At higher $V_g$, on the contrary, neither well-defined charge states in the ring, nor fractional periodicity in the ring current are observed.

\begin{figure*}
\begin{tabular}{cccc}
\includegraphics[width=0.6\columnwidth]{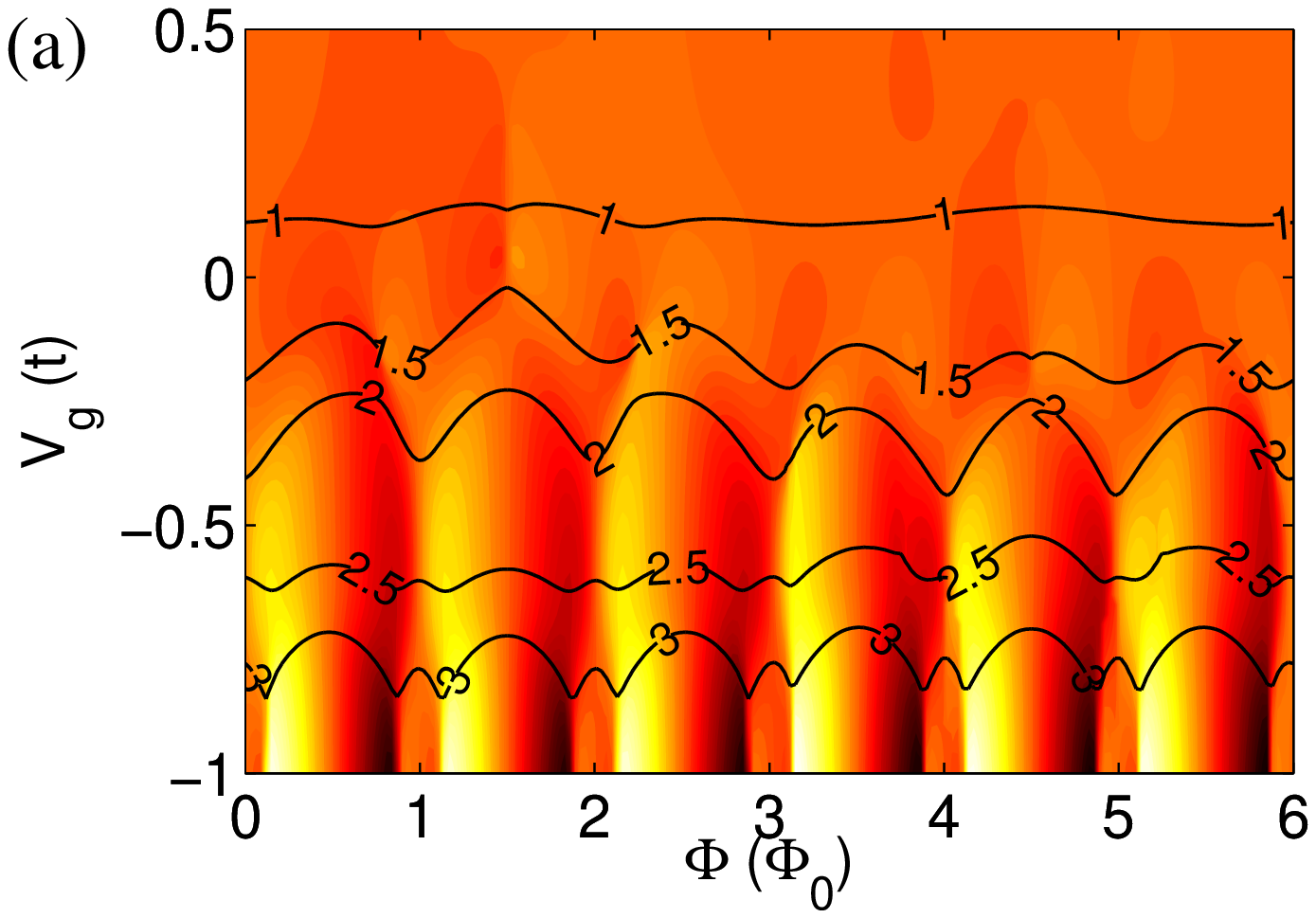}&
\includegraphics[width=0.6\columnwidth]{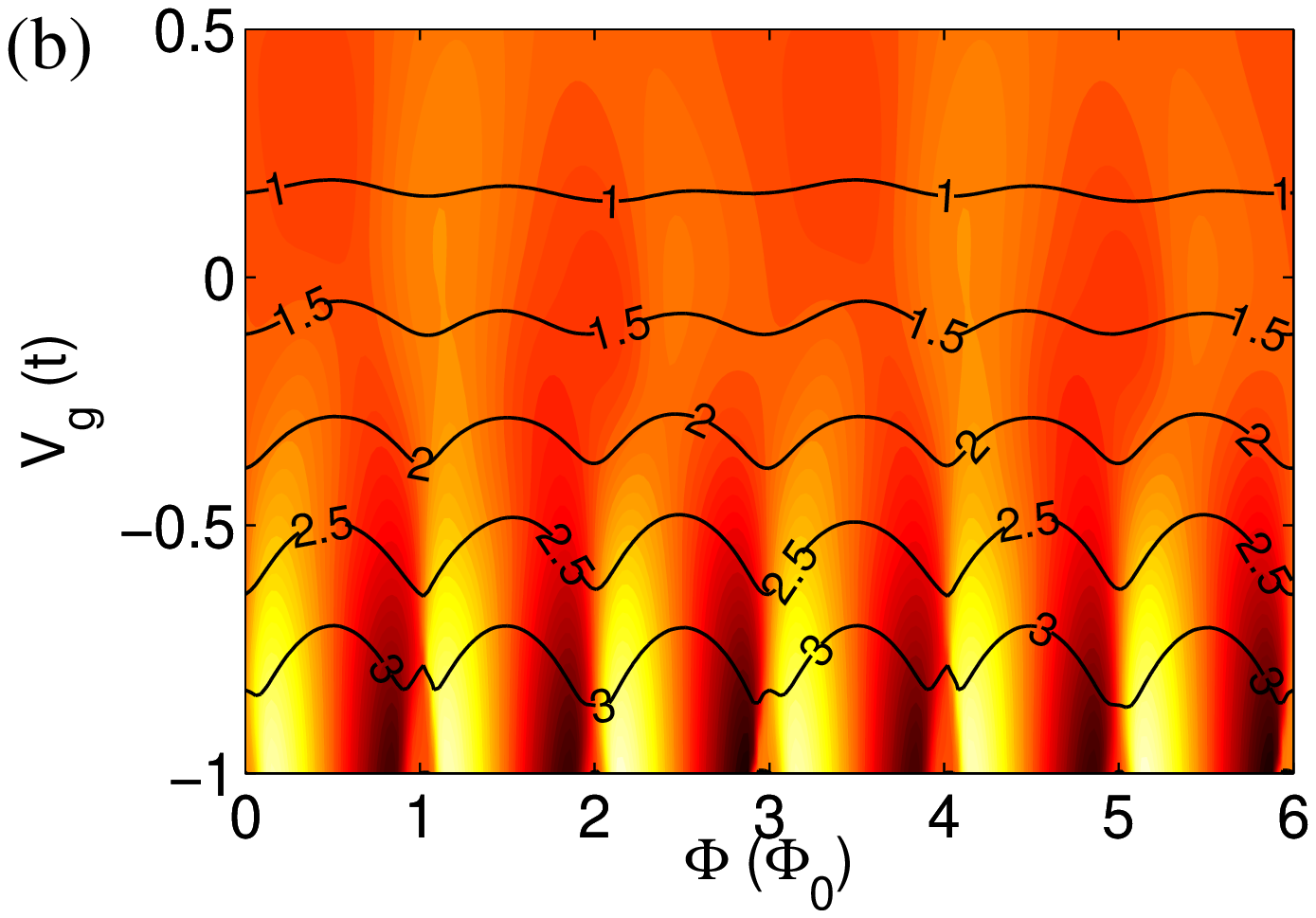} &
 \includegraphics[width=0.6\columnwidth]{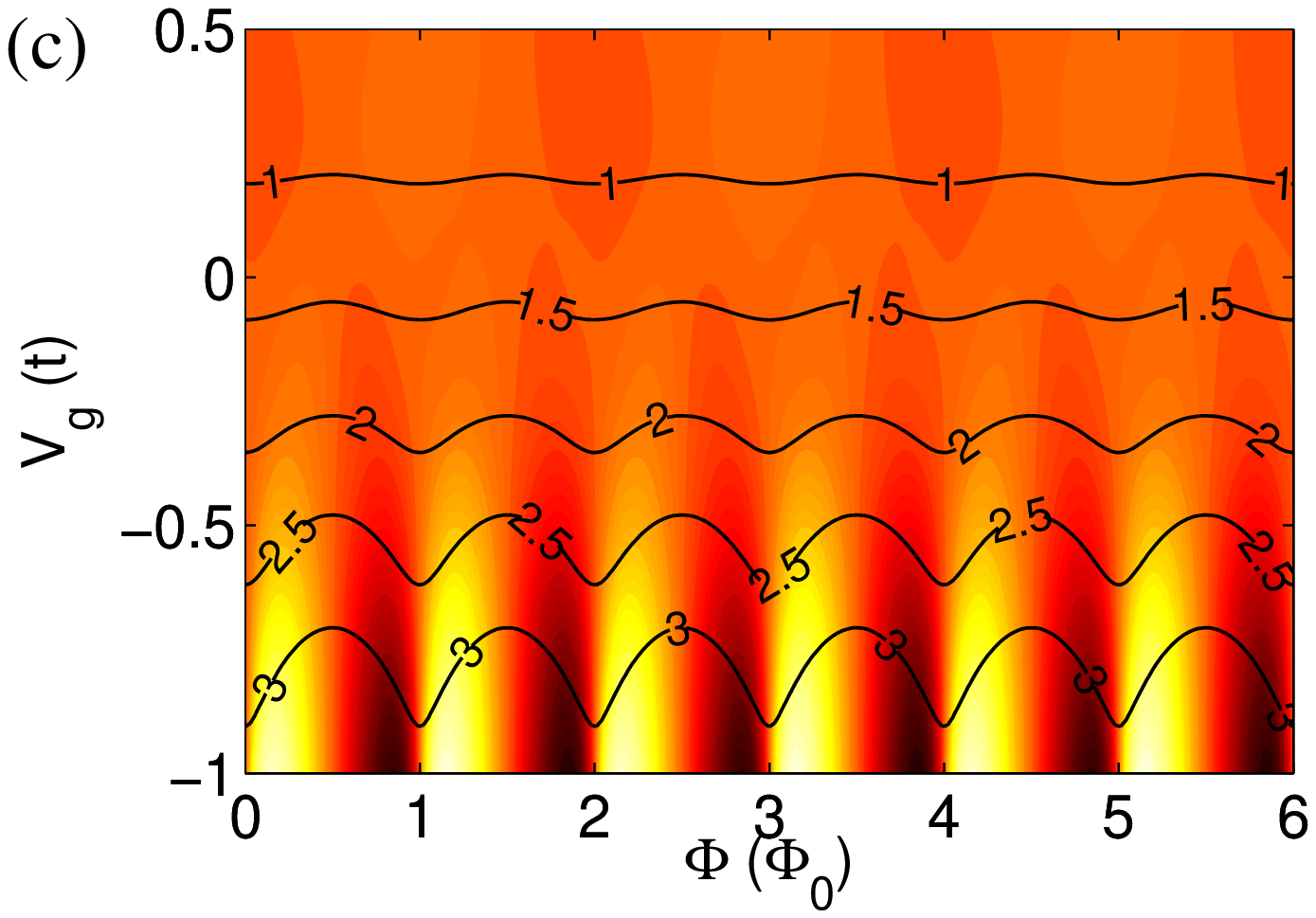}&
  \includegraphics[width=0.178\columnwidth]{cbar_J.eps}\\
\includegraphics[width=0.6\columnwidth]{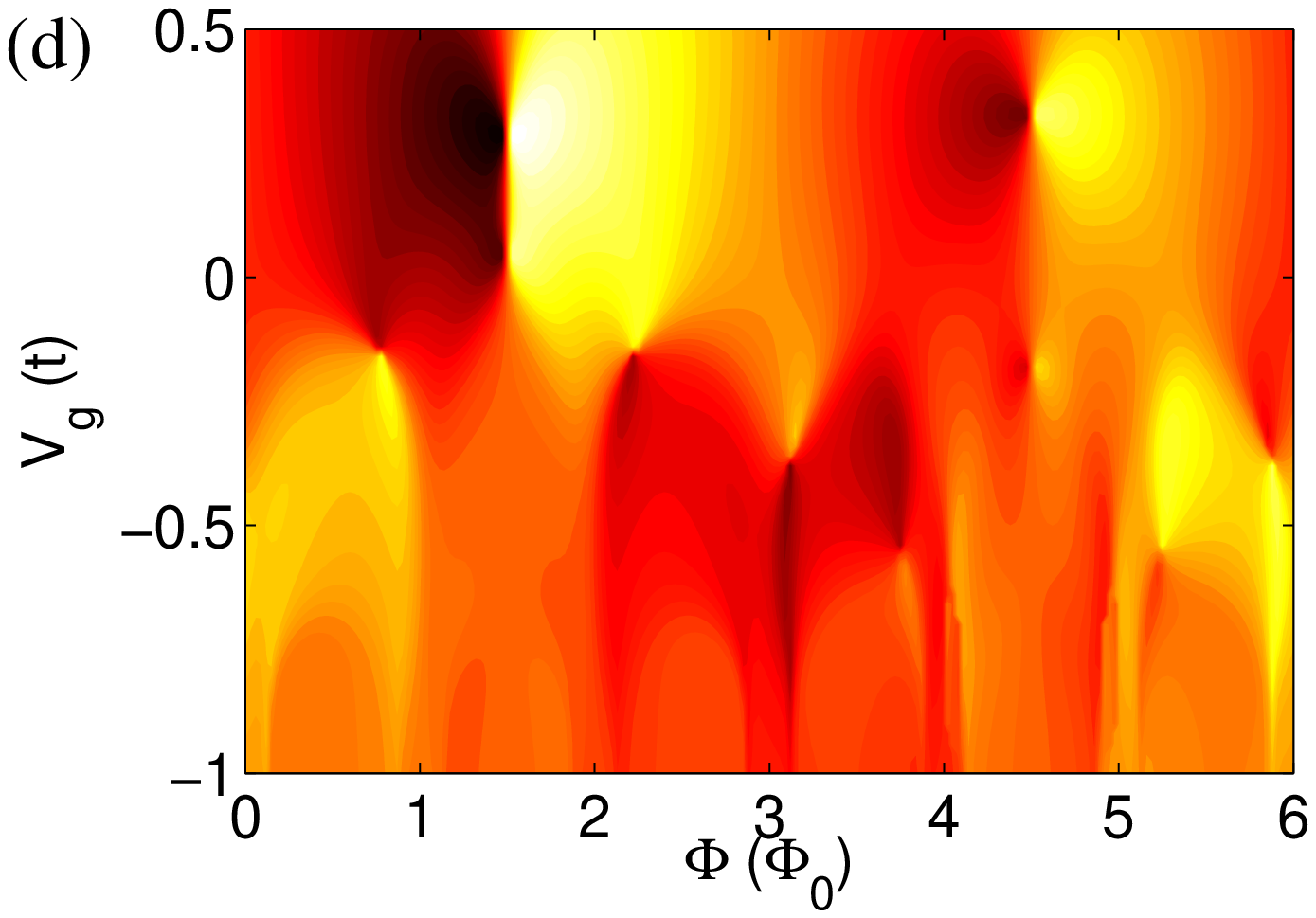}&
\includegraphics[width=0.6\columnwidth]{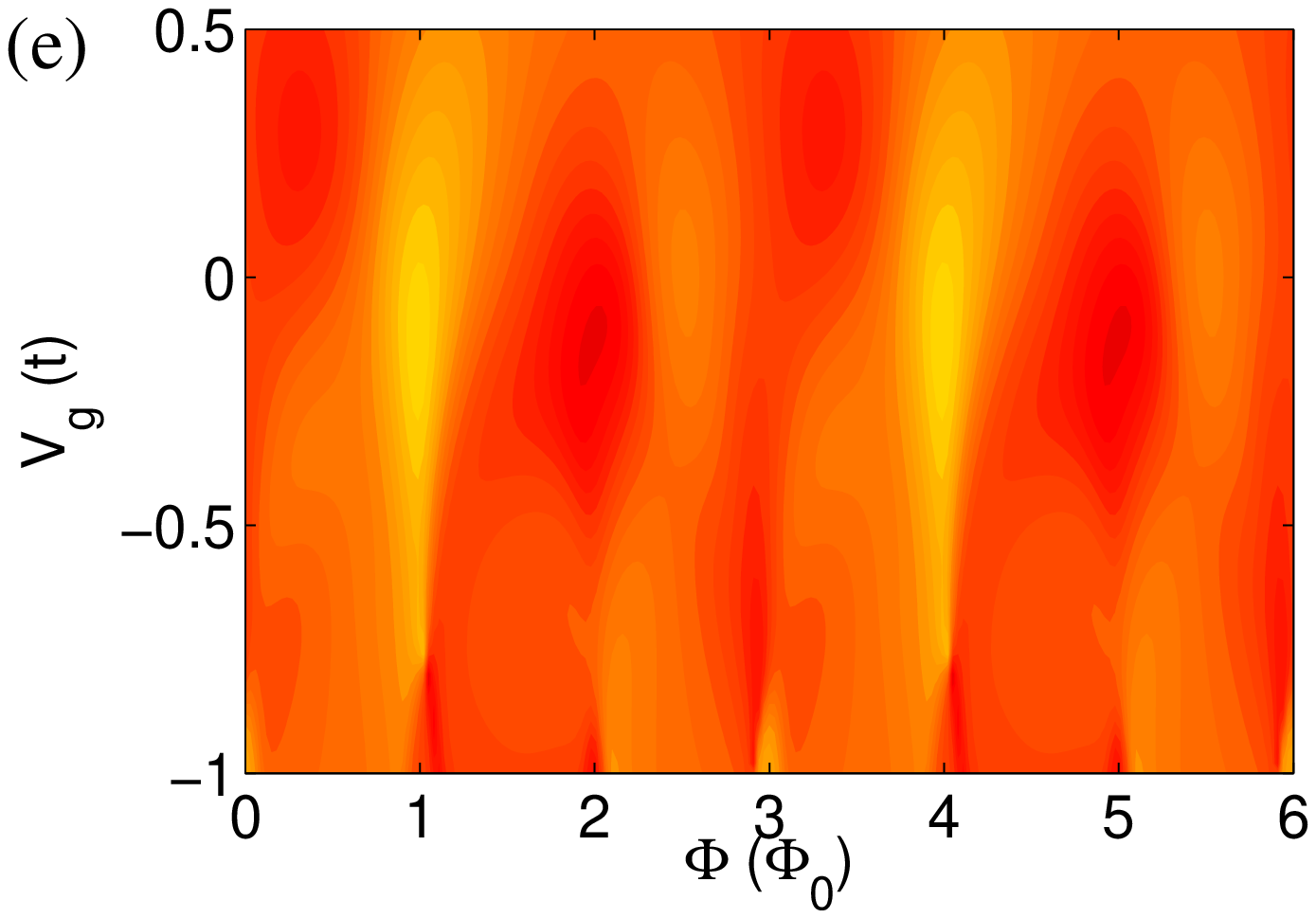} &
 \includegraphics[width=0.6\columnwidth]{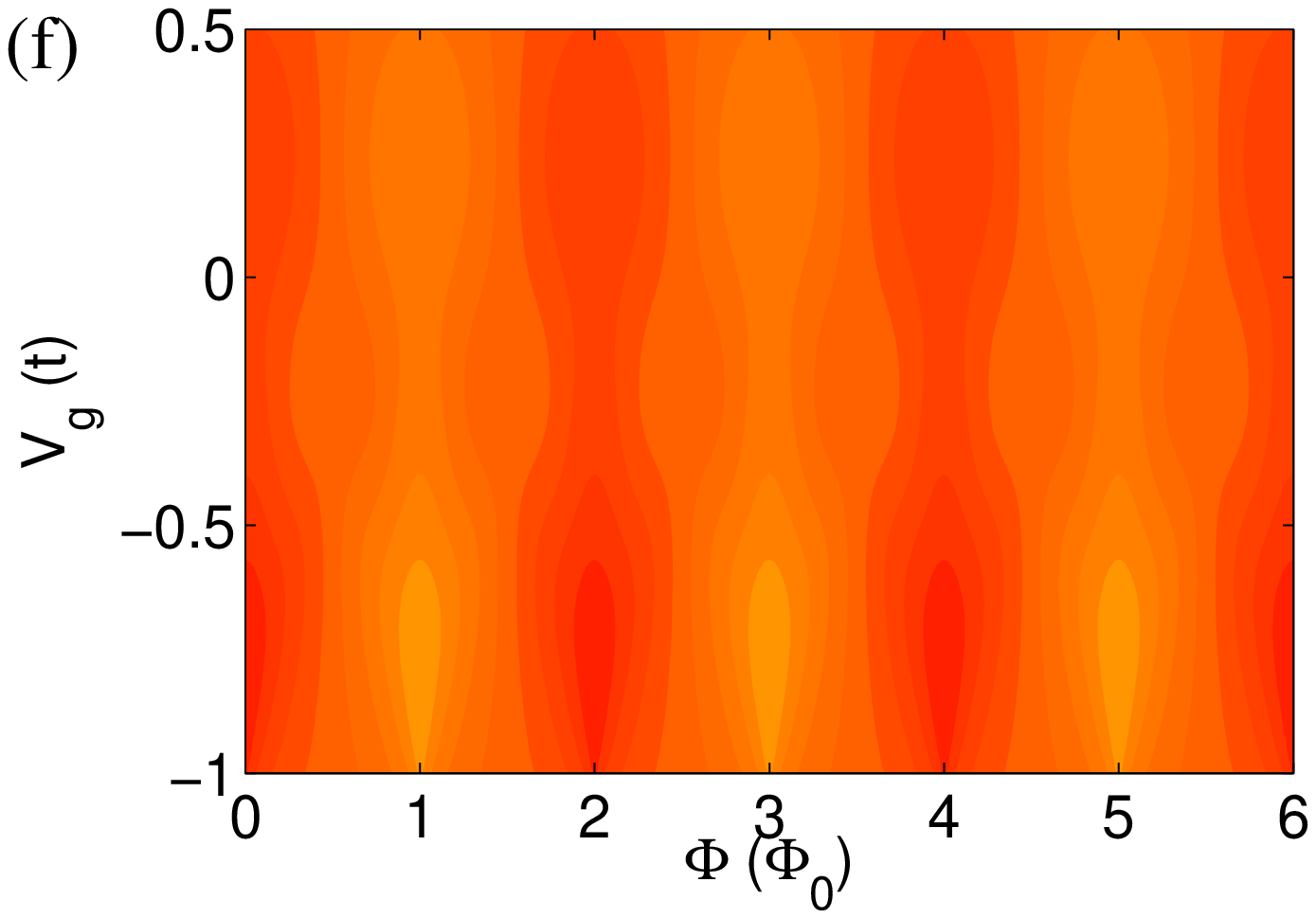}&
  \includegraphics[width=0.178\columnwidth]{cbar_J.eps}\\
\includegraphics[width=0.6\columnwidth]{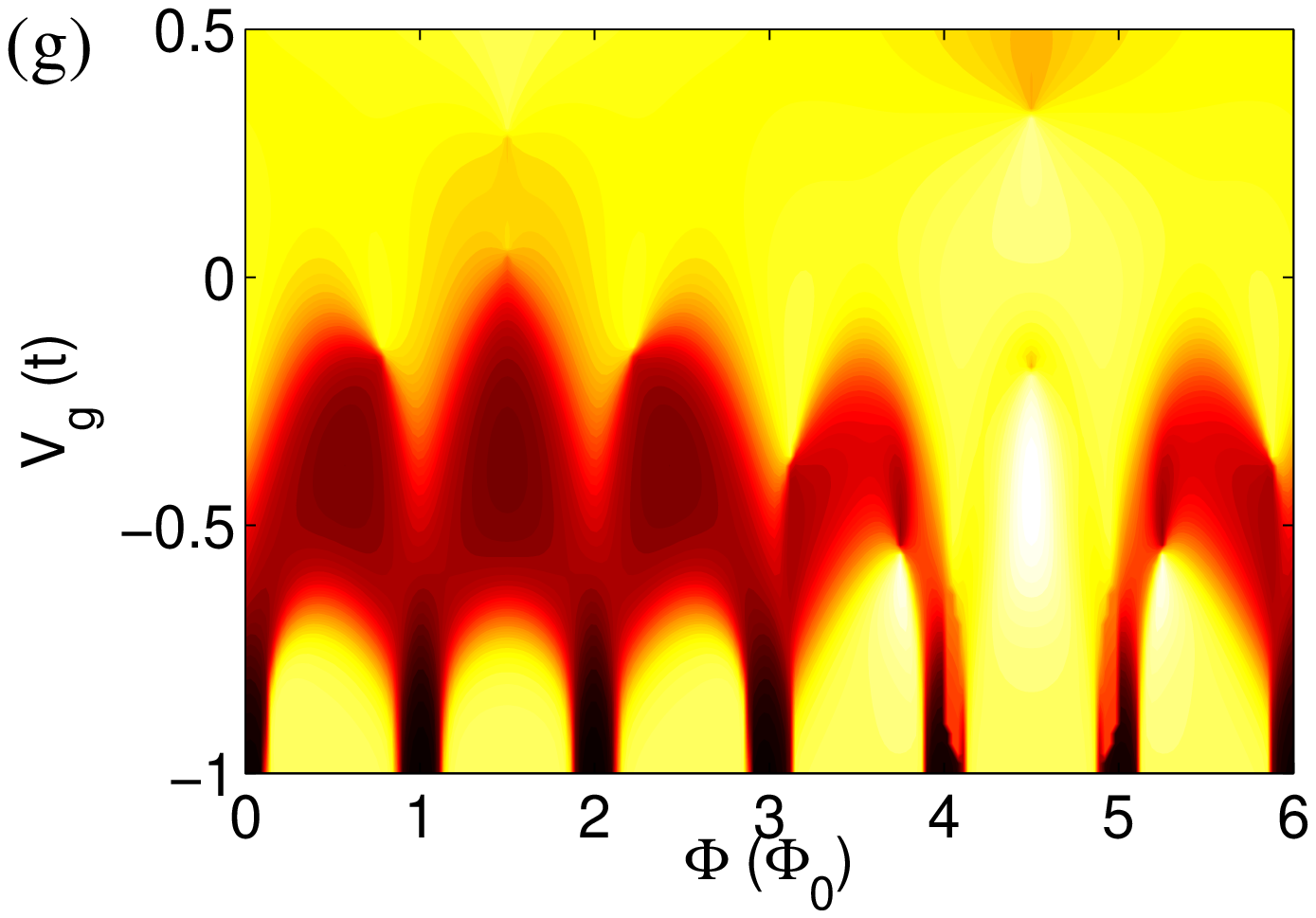}&
\includegraphics[width=0.6\columnwidth]{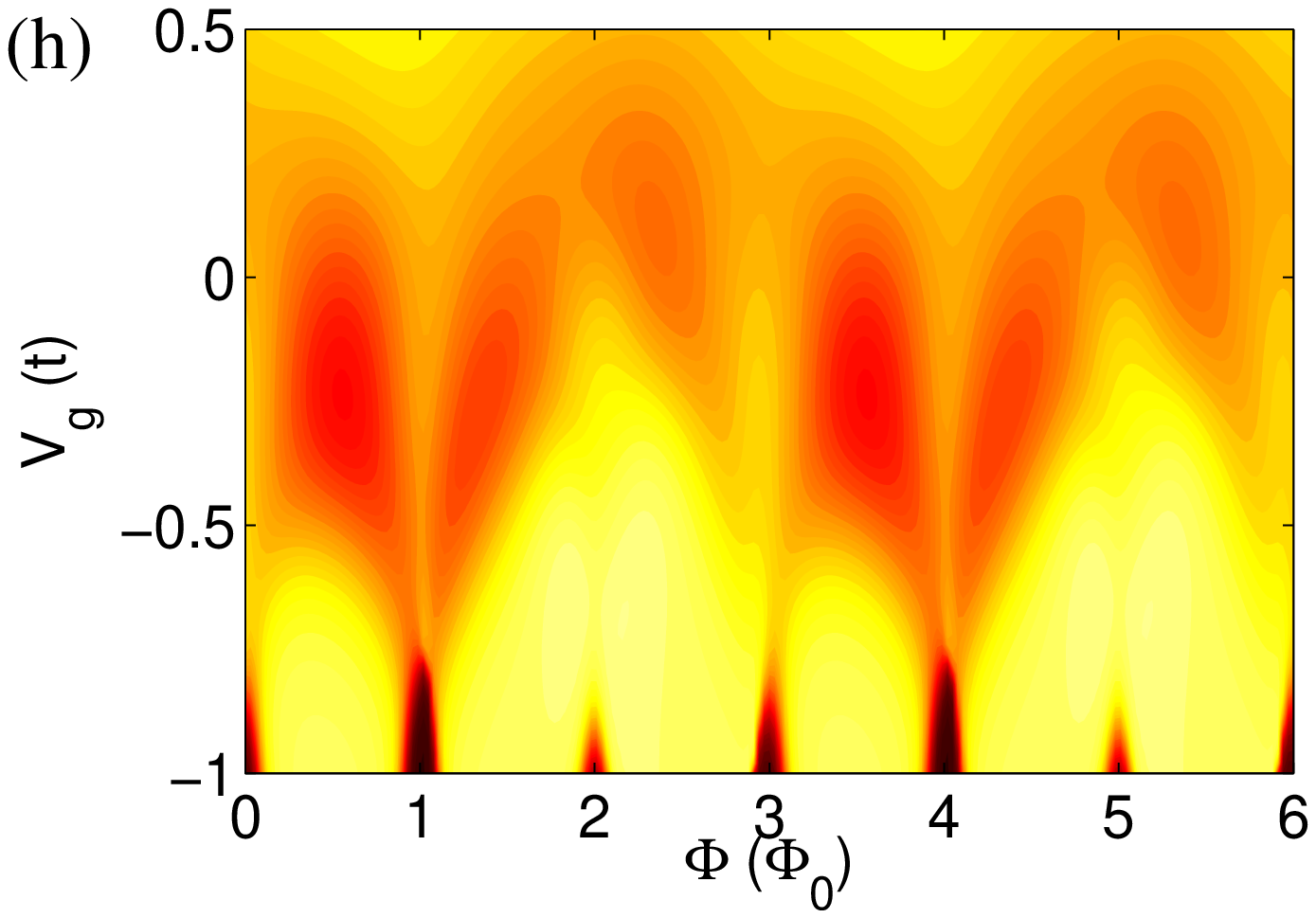} &
 \includegraphics[width=0.6\columnwidth]{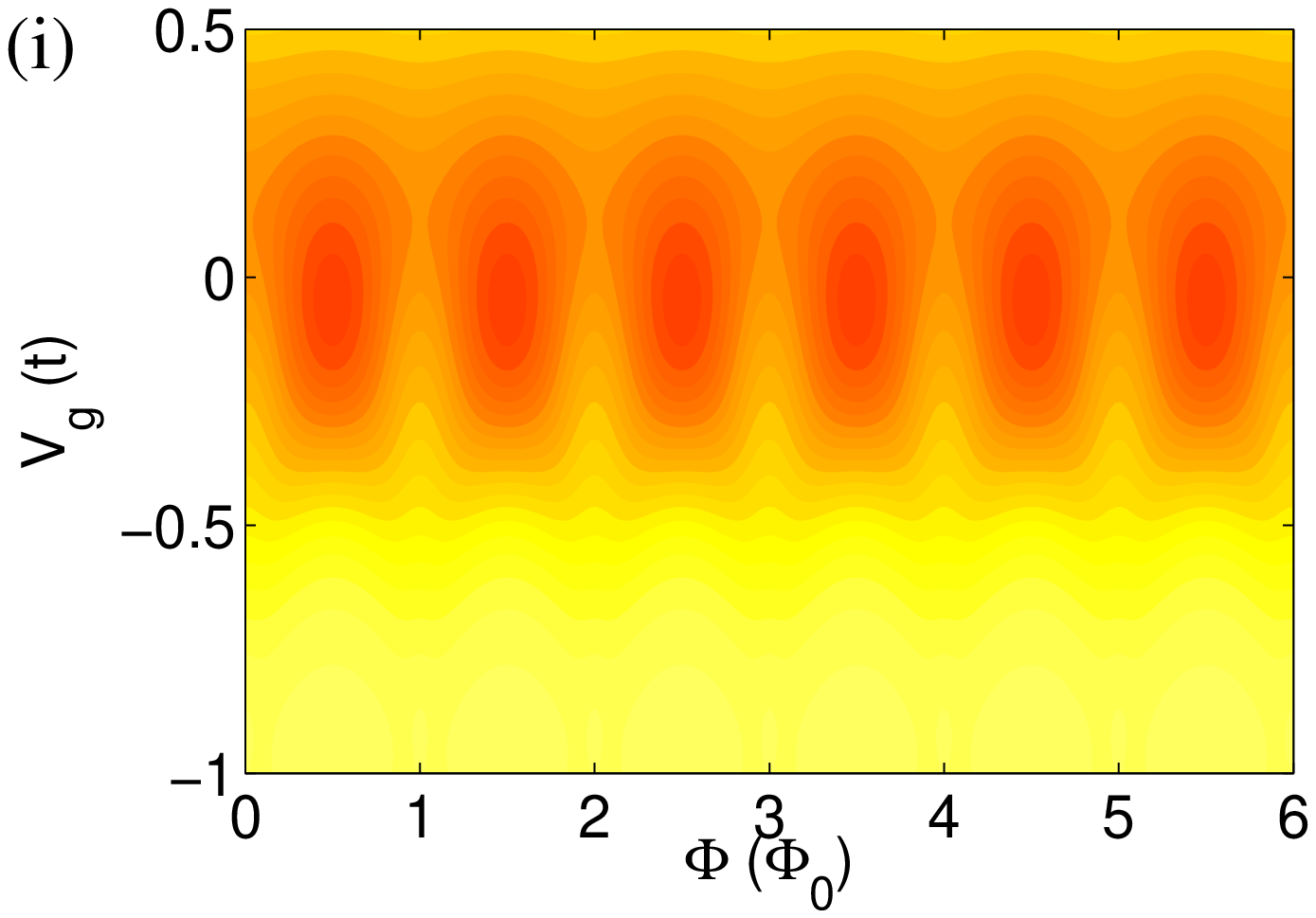}&
  \includegraphics[width=0.178\columnwidth]{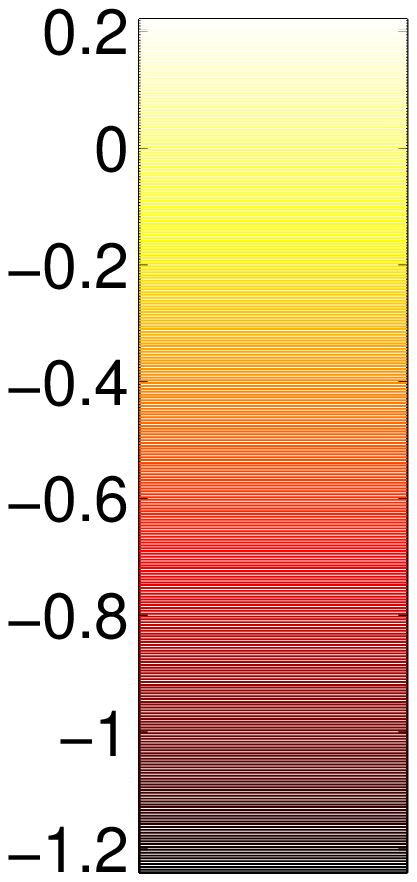}\\
\end{tabular}
 \caption{\label{fig:couplingsite_tc1} (Color online) The ring and lead currents [(a)-(c) and (d)-(f), respectively], as well as spin imbalance $n_{\uparrow}-n_{\downarrow}$ in the ring [(g)-(i)], in the ($\Phi$, $V^g$) space for the three different coupling sites for strong coupling between the lead and the ring.  (a),(d),(g): 1--2 coupling (b),(e),(h): 1--3 coupling (c),(f),(i): 1--4 coupling. In (a)-(c), the contours of integer and half-integer ring occupation $n_{\mathrm{ring}}$ are additionally drawn.  Other parameters: Six-site ring, 14-site lead, $t_c = -t$, $t' = 0.2t$, $N_{el}=5$, $U = 2t$. The color scales for each row are fixed to allow comparison in the magnitudes between different coupling positions. }
\end{figure*}

Studying the effect of the interaction strength $U$ on the Coulomb staircase provides additional insight into this effective decoupling. Fig.~\ref{fig:Uvar}(a) shows the effect of the interaction strength on the Coulomb steps. When the interaction is increased, shown in the different shades of gray, the number of steps increases. With an increasing electron-electron repulsion strength, the kinetic energy associated with the strong coupling becomes weaker relative to the interaction energy, and bound correlated states are formed in the ring part at sufficiently negative $V^g$ also for lower $n_{\mathrm{ring}}$.   The bound correlated states also support a ring current with fractional periodicity.  Indeed, we see in Fig.~\ref{fig:Uvar}(a) that at $U \geq 15t$, a Coulomb step with $n_{\mathrm{ring}} \approx 3$ has appeared, and the corresponding $n_{\mathrm{ring}}=3$ curves in Fig.~\ref{fig:murtoluku_tc1} show characteristics increasingly similar to the weakly coupled case above $U_c$ with increasing $U$. 

The interaction strength in the lead is actually less important. Figs.~\ref{fig:Uvar}(b) and ~\ref{fig:Uvar}(c) show the effect of increasing the interaction in the lead part of the system,  while the interaction in the ring part is $U_r = 5t$ and $U_r=20t$, respectively, for all curves. The highest-$V^g$ steps slightly shift when the interaction is increased but the overall shape of the curve is very similar regardless of the lead interaction strength. At weaker interaction in the ring ($U_r = 5t$ in Fig.~\ref{fig:Uvar}(b)) the presence of interaction in the lead has little effect, whereas of stronger ring interaction ($U_r = 20t$, Fig.~\ref{fig:Uvar}(c)),  the system is practically converged to the $U_l=20t$ case at $U_l = 5t$. 

\subsection{Asymmetry effects \label{sec:asymmetry}}

This far, we have only considered structures, in which the coupling between the lead and the ring is symmetric, both in strength and in the coupling position. In experiments, \cite{Leturcq} however, the coupling strength between the ring and the lead is typically controlled by gates and it is possible that the coupling strength is not equal for both contacts. Our approach allows us to study explicitly the effect of both types of asymmetry. Fig.~\ref{fig:unequal_tc} compares the Coulomb steps for a six-site ring 1-4 --coupled to a 14-site lead but with unequal coupling strengths. The product of the couplings, $t_{c1}t_{c2} = 1$ is fixed but the asymmetry, given by the fraction $t_{c1}/t_{c2}$, is varied. When the difference between $t_{c1}$ and $t_{c2}$ increases, the systems becomes effectively more weakly coupled, seen in the appearance of the Coulomb steps.Thus, a single weak link is sufficient for weakly coupled behavior.

The position of the lead-ring coupling can also be changed, the arms of the ring thus being of unequal length. Unequal-lenght arms appear also in the experiments.\cite{Leturcq} Fig.~\ref{fig:structure} illustrates the three different coupling positions available for the six-site ring. The effect of the coupling positions on the ring and lead currents is illustrated in Fig.~\ref{fig:couplingsite} in which the magnitude and sign of the ring current [(a)-(c)] and the lead current [(d)-(f)] is shown in the ($V^g$,$\Phi$) space for a weakly coupled ($t_c = -0.2t$) six-site ring with a 14-site lead ($U=2t$, $t'=0.2t$, $n_{\uparrow} = 3$, $n_{\downarrow}=2$), as well as the spin imbalance in the ring, $n_{\uparrow}-n_{\downarrow}$ [(g)-(i)]. The color scale is fixed so that in each row, the magnitudes for different coupling positions can be compared.  The contours overlaid with the ring current maps in Fig.~\ref{fig:couplingsite}(a)-(c) show the occupation of the ring, $n_{\mathrm{ring}}$.

It is easily understood that for weak coupling, the choice of the position has little effect on the behavior of the ring current and fractional periodicity, as seen in Fig.~\ref{fig:couplingsite}(a)-(c). The effect of the position is, however, seen in the current circulating in the lead part of the system. More precisely, an alternation in the magnitude of the lead current depending on the coupling position is clearly seen, as well as changes in the broadening of the lead current peaks. The high-current regions approximately parallel to the $\Phi$ axis are due to the electrons entering the ring, seen in the upper row as contours of half-integer $n_{\mathrm{ring}}$. The ridges parallel to the $V^g$ axis, on the other hand, are due to charge transfer changing the spin polarization $n_{\uparrow}-n_{\downarrow}$ within the ring at a constant ring occupation.  The density of one spin species increases and the other decreases abruptly, as seen in Fig.~\ref{fig:couplingsite}(g)-(i). 

In fact, the magnitude alternation in the lead current [Fig.~Ê\ref{fig:couplingsite}(d)-(f)] resembles the magnitude of persistent current in a ring, superimposed on the resonance peaks, only with a two, three or six times longer period. There is also a phase shift when the number of electrons in the lead changes, similar to the odd-even alternation in isolated, clean rings. Like already mentioned, the effect of increased period in the lead current is seen also in the symmetrically coupled case. In Figs.~\ref{fig:murtoluku} and~\ref{fig:murtoluku_tc1},  the periodicity of the lead current is twice that of the ring current, explained by a second ring formed by the lead and some ring sites that is pierced by half of the flux $\Phi$ in Section~\ref{sec:weak}. Also for the other coupling positions, we get the period of the lead current by comparing the flux piercing the ring, and a ring formed by the lead and some sites in the ring.  Thus, the 1--2 coupling results in a six times longer period in the lead current compared with the period of the ring, and the 1--3 coupling in a three times longer period. We also note that the maximal magnitude of the lead current in the symmetrically coupled case is approximately half of that for the other cases, shown in Fig.~\ref{fig:couplingsite}(c). This is most likely due to the appearance of destructive interferences when the ring arms are of equal length.

Fig.~\ref{fig:couplingsite_tc1} shows similar figures for the strongly coupled case. As there are no bound correlated states in the $V^g$ range considered and the electrons are delocalized both in the ring and the lead, the coupling position affects also the ring current. For larger negative $V^g$, regions with a well-defined constant $n_{\mathrm{ring}}$ similar to the weakly coupled case  would be seen also for strong coupling. The effect of the coupling position on the magnitude of the lead current is much larger than in the weakly coupled case, the maximal value of the lead current being six times larger at 1-2 --coupling compared to 1-4 --coupling. The diminishing magnitude of the lead current also makes the charge imbalance smaller in the 1-4 case, shown in the contours of the lower row in Fig.~\ref{fig:couplingsite_tc1}. The periodicities of the lead current, however, are the same as at weak coupling. 

\begin{figure}
 \includegraphics[width=0.99\columnwidth]{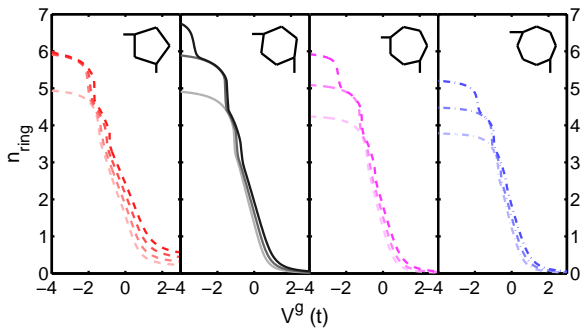}
\caption{\label{fig:ringsize} (Color online) The occupation in the ring, $n_{\mathrm{ring}}$ as a function of the gate voltage $V^g$ for  ring sizes $N_{\mathrm{ring}}=5-8$ for $N_{el}$~=~5,~ 6, and 7 (from lighter to darker curves). For the five-site ring, also $N_{el} = 8$ included. Other calculational parameters:  $U=2t$, $t'=0.2t$, $t_c=-t$, and $N_l = 14$. }
\end{figure}

Up to this point, we have only considered a six-site ring. We conclude by noting that our results on the six-site rings are representable also for other ring sizes. Fig.~\ref{fig:ringsize} compares the Coulomb steps at strong coupling for five- to eight-site rings with a 14-site lead. In all cases, the lowest-occupation steps are smoothened out, and for higher $n_{\mathrm{ring}}$ the bound states that are effectively weakly coupled to the lead appear. 

\section{Conclusions}

We have studied a Hubbard ring with second-nearest neighbor hoppings, connected to a ring-shaped lead using an embedding approach and solving the system Hamiltonian using exact diagonalization. In the case of a weak ring-lead coupling, the behavior of the persistent current is similar to that in isolated rings and the presence of the lead is only a weak perturbation. In addition,  the interaction strength required to observe fractional periodicity is increased with a higher electron occupation in the ring. 

In the strongly coupled case, on the contrary, fractional periodicity can not be observed for low electron occupations in the ring. For higher occupations corresponding to large negative gate voltages, fractional periodicity is observed due to the formation of bound, strongly correlated states. High interaction strength, however, are required. Our results provide additional insight to why the fractional periodicity is difficult to observe experimentally. 

\acknowledgments

M.I. acknowledges the financial support from the Finnish Doctoral Programme in Computational Sciences FICS and from V{\"{a}}is{\"{a}}l{\"{a}} foundation.

\bibliography{graah_final}

\begin{thebibliography}{52}%
\makeatletter
\providecommand \@ifxundefined [1]{%
 \@ifx{#1\undefined}
}%
\providecommand \@ifnum [1]{%
 \ifnum #1\expandafter \@firstoftwo
 \else \expandafter \@secondoftwo
 \fi
}%
\providecommand \@ifx [1]{%
 \ifx #1\expandafter \@firstoftwo
 \else \expandafter \@secondoftwo
 \fi
}%
\providecommand \natexlab [1]{#1}%
\providecommand \enquote  [1]{``#1''}%
\providecommand \bibnamefont  [1]{#1}%
\providecommand \bibfnamefont [1]{#1}%
\providecommand \citenamefont [1]{#1}%
\providecommand \href@noop [0]{\@secondoftwo}%
\providecommand \href [0]{\begingroup \@sanitize@url \@href}%
\providecommand \@href[1]{\@@startlink{#1}\@@href}%
\providecommand \@@href[1]{\endgroup#1\@@endlink}%
\providecommand \@sanitize@url [0]{\catcode `\\12\catcode `\$12\catcode
  `\&12\catcode `\#12\catcode `\^12\catcode `\_12\catcode `\%12\relax}%
\providecommand \@@startlink[1]{}%
\providecommand \@@endlink[0]{}%
\providecommand \url  [0]{\begingroup\@sanitize@url \@url }%
\providecommand \@url [1]{\endgroup\@href {#1}{\urlprefix }}%
\providecommand \urlprefix  [0]{URL }%
\providecommand \Eprint [0]{\href }%
\providecommand \doibase [0]{http://dx.doi.org/}%
\providecommand \selectlanguage [0]{\@gobble}%
\providecommand \bibinfo  [0]{\@secondoftwo}%
\providecommand \bibfield  [0]{\@secondoftwo}%
\providecommand \translation [1]{[#1]}%
\providecommand \BibitemOpen [0]{}%
\providecommand \bibitemStop [0]{}%
\providecommand \bibitemNoStop [0]{.\EOS\space}%
\providecommand \EOS [0]{\spacefactor3000\relax}%
\providecommand \BibitemShut  [1]{\csname bibitem#1\endcsname}%
\let\auto@bib@innerbib\@empty
\bibitem [{\citenamefont {Kotlyar}\ and\ \citenamefont
  {{Das~Sarma}}(1997)}]{Kotlyar-DasSarma}%
  \BibitemOpen
  \bibfield  {author} {\bibinfo {author} {\bibfnamefont {R.}~\bibnamefont
  {Kotlyar}}\ and\ \bibinfo {author} {\bibfnamefont {S.}~\bibnamefont
  {{Das~Sarma}}},\ }\href@noop {} {\bibfield  {journal} {\bibinfo  {journal}
  {Phys. Rev. B}\ }\textbf {\bibinfo {volume} {55}},\ \bibinfo {pages} {10205}
  (\bibinfo {year} {1997})}\BibitemShut {NoStop}%
\bibitem [{\citenamefont {Bleszynski-Jayich}\ \emph {et~al.}(2009)\citenamefont
  {Bleszynski-Jayich}, \citenamefont {Shanks}, \citenamefont {Peaudecerf},
  \citenamefont {Ginossar}, \citenamefont {von Oppen}, \citenamefont
  {Glazman},\ and\ \citenamefont {Harris}}]{Bleszynski-Jayich}%
  \BibitemOpen
  \bibfield  {author} {\bibinfo {author} {\bibfnamefont {A.~C.}\ \bibnamefont
  {Bleszynski-Jayich}}, \bibinfo {author} {\bibfnamefont {W.~E.}\ \bibnamefont
  {Shanks}}, \bibinfo {author} {\bibfnamefont {B.}~\bibnamefont {Peaudecerf}},
  \bibinfo {author} {\bibfnamefont {E.}~\bibnamefont {Ginossar}}, \bibinfo
  {author} {\bibfnamefont {F.}~\bibnamefont {von Oppen}}, \bibinfo {author}
  {\bibfnamefont {L.~I.}\ \bibnamefont {Glazman}}, \ and\ \bibinfo {author}
  {\bibfnamefont {J.~G.~E.}\ \bibnamefont {Harris}},\ }\href@noop {} {\bibfield
   {journal} {\bibinfo  {journal} {Science}\ }\textbf {\bibinfo {volume}
  {326}},\ \bibinfo {pages} {272} (\bibinfo {year} {2009})}\BibitemShut
  {NoStop}%
\bibitem [{\citenamefont {Bluhm}\ \emph {et~al.}(2009)\citenamefont {Bluhm},
  \citenamefont {Koshnick}, \citenamefont {Bert}, \citenamefont {Huber},\ and\
  \citenamefont {Moler}}]{Bluhm}%
  \BibitemOpen
  \bibfield  {author} {\bibinfo {author} {\bibfnamefont {H.}~\bibnamefont
  {Bluhm}}, \bibinfo {author} {\bibfnamefont {N.~C.}\ \bibnamefont {Koshnick}},
  \bibinfo {author} {\bibfnamefont {J.~A.}\ \bibnamefont {Bert}}, \bibinfo
  {author} {\bibfnamefont {M.~E.}\ \bibnamefont {Huber}}, \ and\ \bibinfo
  {author} {\bibfnamefont {K.~A.}\ \bibnamefont {Moler}},\ }\href@noop {}
  {\bibfield  {journal} {\bibinfo  {journal} {Phys. Rev. Lett.}\ }\textbf
  {\bibinfo {volume} {102}},\ \bibinfo {pages} {136802} (\bibinfo {year}
  {2009})}\BibitemShut {NoStop}%
\bibitem [{\citenamefont {Jariwala}\ \emph {et~al.}(2001)\citenamefont
  {Jariwala}, \citenamefont {Mohanty}, \citenamefont {Ketchen},\ and\
  \citenamefont {Webb}}]{Jariwala}%
  \BibitemOpen
  \bibfield  {author} {\bibinfo {author} {\bibfnamefont {E.~M.~Q.}\
  \bibnamefont {Jariwala}}, \bibinfo {author} {\bibfnamefont {P.}~\bibnamefont
  {Mohanty}}, \bibinfo {author} {\bibfnamefont {M.~B.}\ \bibnamefont
  {Ketchen}}, \ and\ \bibinfo {author} {\bibfnamefont {R.~A.}\ \bibnamefont
  {Webb}},\ }\href@noop {} {\bibfield  {journal} {\bibinfo  {journal} {Phys.
  Rev. Lett.}\ }\textbf {\bibinfo {volume} {86}},\ \bibinfo {pages} {1594}
  (\bibinfo {year} {2001})}\BibitemShut {NoStop}%
\bibitem [{\citenamefont {Deblock}\ \emph {et~al.}(2002)\citenamefont
  {Deblock}, \citenamefont {Bel}, \citenamefont {Reulet}, \citenamefont
  {Bouchiat},\ and\ \citenamefont {Mailly}}]{Deblock}%
  \BibitemOpen
  \bibfield  {author} {\bibinfo {author} {\bibfnamefont {R.}~\bibnamefont
  {Deblock}}, \bibinfo {author} {\bibfnamefont {R.}~\bibnamefont {Bel}},
  \bibinfo {author} {\bibfnamefont {B.}~\bibnamefont {Reulet}}, \bibinfo
  {author} {\bibfnamefont {H.}~\bibnamefont {Bouchiat}}, \ and\ \bibinfo
  {author} {\bibfnamefont {D.}~\bibnamefont {Mailly}},\ }\href@noop {}
  {\bibfield  {journal} {\bibinfo  {journal} {Phys. Rev. Lett.}\ }\textbf
  {\bibinfo {volume} {89}},\ \bibinfo {pages} {206803} (\bibinfo {year}
  {2002})}\BibitemShut {NoStop}%
\bibitem [{\citenamefont {Chandrasekhar}\ \emph {et~al.}(1991)\citenamefont
  {Chandrasekhar}, \citenamefont {Webb}, \citenamefont {Brady}, \citenamefont
  {Ketchen}, \citenamefont {Gallagher},\ and\ \citenamefont
  {Kleinsasser}}]{Chandrasekhar}%
  \BibitemOpen
  \bibfield  {author} {\bibinfo {author} {\bibfnamefont {V.}~\bibnamefont
  {Chandrasekhar}}, \bibinfo {author} {\bibfnamefont {R.~A.}\ \bibnamefont
  {Webb}}, \bibinfo {author} {\bibfnamefont {M.~J.}\ \bibnamefont {Brady}},
  \bibinfo {author} {\bibfnamefont {M.~B.}\ \bibnamefont {Ketchen}}, \bibinfo
  {author} {\bibfnamefont {W.~J.}\ \bibnamefont {Gallagher}}, \ and\ \bibinfo
  {author} {\bibfnamefont {A.}~\bibnamefont {Kleinsasser}},\ }\href@noop {}
  {\bibfield  {journal} {\bibinfo  {journal} {Phys. Rev. Lett.}\ }\textbf
  {\bibinfo {volume} {67}},\ \bibinfo {pages} {3578} (\bibinfo {year}
  {1991})}\BibitemShut {NoStop}%
\bibitem [{\citenamefont {Fuhrer}\ \emph {et~al.}(2001)\citenamefont {Fuhrer},
  \citenamefont {Ihn}, \citenamefont {Ensslin}, \citenamefont {Wegscheider},\
  and\ \citenamefont {Bichler}}]{Fuhrer}%
  \BibitemOpen
  \bibfield  {author} {\bibinfo {author} {\bibfnamefont {A.}~\bibnamefont
  {Fuhrer}}, \bibinfo {author} {\bibfnamefont {T.}~\bibnamefont {Ihn}},
  \bibinfo {author} {\bibfnamefont {K.}~\bibnamefont {Ensslin}}, \bibinfo
  {author} {\bibfnamefont {W.}~\bibnamefont {Wegscheider}}, \ and\ \bibinfo
  {author} {\bibfnamefont {M.}~\bibnamefont {Bichler}},\ }\href@noop {}
  {\bibfield  {journal} {\bibinfo  {journal} {Nature}\ }\textbf {\bibinfo
  {volume} {413}},\ \bibinfo {pages} {822} (\bibinfo {year}
  {2001})}\BibitemShut {NoStop}%
\bibitem [{\citenamefont {Mailly}\ \emph {et~al.}(1993)\citenamefont {Mailly},
  \citenamefont {Chapelier},\ and\ \citenamefont {Beno\^{\i}t}}]{Mailly}%
  \BibitemOpen
  \bibfield  {author} {\bibinfo {author} {\bibfnamefont {D.}~\bibnamefont
  {Mailly}}, \bibinfo {author} {\bibfnamefont {C.}~\bibnamefont {Chapelier}}, \
  and\ \bibinfo {author} {\bibfnamefont {A.}~\bibnamefont {Beno\^{\i}t}},\
  }\href@noop {} {\bibfield  {journal} {\bibinfo  {journal} {Phys. Rev. Lett.}\
  }\textbf {\bibinfo {volume} {70}},\ \bibinfo {pages} {2020} (\bibinfo {year}
  {1993})}\BibitemShut {NoStop}%
\bibitem [{\citenamefont {Morpurgo}\ \emph {et~al.}(1998)\citenamefont
  {Morpurgo}, \citenamefont {Heida}, \citenamefont {Klapwijk}, \citenamefont
  {van Wees},\ and\ \citenamefont {Borghs}}]{Morpurgo}%
  \BibitemOpen
  \bibfield  {author} {\bibinfo {author} {\bibfnamefont {A.~F.}\ \bibnamefont
  {Morpurgo}}, \bibinfo {author} {\bibfnamefont {J.~P.}\ \bibnamefont {Heida}},
  \bibinfo {author} {\bibfnamefont {T.~M.}\ \bibnamefont {Klapwijk}}, \bibinfo
  {author} {\bibfnamefont {B.~J.}\ \bibnamefont {van Wees}}, \ and\ \bibinfo
  {author} {\bibfnamefont {G.}~\bibnamefont {Borghs}},\ }\href@noop {}
  {\bibfield  {journal} {\bibinfo  {journal} {Phys. Rev. Lett.}\ }\textbf
  {\bibinfo {volume} {80}},\ \bibinfo {pages} {1050} (\bibinfo {year}
  {1998})}\BibitemShut {NoStop}%
\bibitem [{\citenamefont {Pedersen}\ \emph {et~al.}(2000)\citenamefont
  {Pedersen}, \citenamefont {Hansen}, \citenamefont {Kristensen}, \citenamefont
  {S\o{}rensen},\ and\ \citenamefont {Lindelof}}]{Pedersen}%
  \BibitemOpen
  \bibfield  {author} {\bibinfo {author} {\bibfnamefont {S.}~\bibnamefont
  {Pedersen}}, \bibinfo {author} {\bibfnamefont {A.~E.}\ \bibnamefont
  {Hansen}}, \bibinfo {author} {\bibfnamefont {A.}~\bibnamefont {Kristensen}},
  \bibinfo {author} {\bibfnamefont {C.~B.}\ \bibnamefont {S\o{}rensen}}, \ and\
  \bibinfo {author} {\bibfnamefont {P.~E.}\ \bibnamefont {Lindelof}},\
  }\href@noop {} {\bibfield  {journal} {\bibinfo  {journal} {Phys. Rev. B}\
  }\textbf {\bibinfo {volume} {61}},\ \bibinfo {pages} {5457} (\bibinfo {year}
  {2000})}\BibitemShut {NoStop}%
\bibitem [{\citenamefont {Giebers}\ \emph {et~al.}(2010)\citenamefont
  {Giebers}, \citenamefont {Zeitler}, \citenamefont {Katsnelson}, \citenamefont
  {Reuter}, \citenamefont {Wieck}, \citenamefont {Biasiol}, \citenamefont
  {Sorba},\ and\ \citenamefont {Maan}}]{Giebers}%
  \BibitemOpen
  \bibfield  {author} {\bibinfo {author} {\bibfnamefont {A.~J.~M.}\
  \bibnamefont {Giebers}}, \bibinfo {author} {\bibfnamefont {U.}~\bibnamefont
  {Zeitler}}, \bibinfo {author} {\bibfnamefont {M.~I.}\ \bibnamefont
  {Katsnelson}}, \bibinfo {author} {\bibfnamefont {D.}~\bibnamefont {Reuter}},
  \bibinfo {author} {\bibfnamefont {A.~D.}\ \bibnamefont {Wieck}}, \bibinfo
  {author} {\bibfnamefont {G.}~\bibnamefont {Biasiol}}, \bibinfo {author}
  {\bibfnamefont {L.}~\bibnamefont {Sorba}}, \ and\ \bibinfo {author}
  {\bibfnamefont {J.~C.}\ \bibnamefont {Maan}},\ }\href@noop {} {\bibfield
  {journal} {\bibinfo  {journal} {Nature Phys.}\ }\textbf {\bibinfo {volume}
  {6}},\ \bibinfo {pages} {173} (\bibinfo {year} {2010})}\BibitemShut {NoStop}%
\bibitem [{\citenamefont {B{\"{u}}ttiker}\ \emph {et~al.}(1983)\citenamefont
  {B{\"{u}}ttiker}, \citenamefont {Imry},\ and\ \citenamefont
  {Landauer}}]{Buttiker-Imry-Landauer}%
  \BibitemOpen
  \bibfield  {author} {\bibinfo {author} {\bibfnamefont {M.}~\bibnamefont
  {B{\"{u}}ttiker}}, \bibinfo {author} {\bibfnamefont {Y.}~\bibnamefont
  {Imry}}, \ and\ \bibinfo {author} {\bibfnamefont {R.}~\bibnamefont
  {Landauer}},\ }\href@noop {} {\bibfield  {journal} {\bibinfo  {journal}
  {Phys. Lett. A}\ }\textbf {\bibinfo {volume} {96}},\ \bibinfo {pages} {365}
  (\bibinfo {year} {1983})}\BibitemShut {NoStop}%
\bibitem [{\citenamefont {B{\"{u}}ttiker}(1985)}]{Buttiker}%
  \BibitemOpen
  \bibfield  {author} {\bibinfo {author} {\bibfnamefont {M.}~\bibnamefont
  {B{\"{u}}ttiker}},\ }\href@noop {} {\bibfield  {journal} {\bibinfo  {journal}
  {Phys. Rev. B}\ }\textbf {\bibinfo {volume} {32}},\ \bibinfo {pages} {1846}
  (\bibinfo {year} {1985})}\BibitemShut {NoStop}%
\bibitem [{\citenamefont {Cheung}\ \emph {et~al.}(1988)\citenamefont {Cheung},
  \citenamefont {Gefen}, \citenamefont {Riedel},\ and\ \citenamefont
  {Shih}}]{Cheung-Gefen}%
  \BibitemOpen
  \bibfield  {author} {\bibinfo {author} {\bibfnamefont {H.-F.}\ \bibnamefont
  {Cheung}}, \bibinfo {author} {\bibfnamefont {Y.}~\bibnamefont {Gefen}},
  \bibinfo {author} {\bibfnamefont {E.~K.}\ \bibnamefont {Riedel}}, \ and\
  \bibinfo {author} {\bibfnamefont {W.-H.}\ \bibnamefont {Shih}},\ }\href@noop
  {} {\bibfield  {journal} {\bibinfo  {journal} {Phys. Rev. B}\ }\textbf
  {\bibinfo {volume} {37}},\ \bibinfo {pages} {6050} (\bibinfo {year}
  {1988})}\BibitemShut {NoStop}%
\bibitem [{\citenamefont {Gogolin}\ and\ \citenamefont
  {Prokof'ev}(1994)}]{Gogolin-Prokofev}%
  \BibitemOpen
  \bibfield  {author} {\bibinfo {author} {\bibfnamefont {A.~O.}\ \bibnamefont
  {Gogolin}}\ and\ \bibinfo {author} {\bibfnamefont {N.~V.}\ \bibnamefont
  {Prokof'ev}},\ }\href@noop {} {\bibfield  {journal} {\bibinfo  {journal}
  {Phys. Rev. B}\ }\textbf {\bibinfo {volume} {50}},\ \bibinfo {pages} {4921}
  (\bibinfo {year} {1994})}\BibitemShut {NoStop}%
\bibitem [{\citenamefont {Yu}\ and\ \citenamefont {Fowler}(1992)}]{Yu-Fowler}%
  \BibitemOpen
  \bibfield  {author} {\bibinfo {author} {\bibfnamefont {N.}~\bibnamefont
  {Yu}}\ and\ \bibinfo {author} {\bibfnamefont {M.}~\bibnamefont {Fowler}},\
  }\href@noop {} {\bibfield  {journal} {\bibinfo  {journal} {Phys. Rev. B}\
  }\textbf {\bibinfo {volume} {45}},\ \bibinfo {pages} {11795} (\bibinfo {year}
  {1992})}\BibitemShut {NoStop}%
\bibitem [{\citenamefont {Loss}\ and\ \citenamefont
  {Goldbart}(1991)}]{Loss-Goldbart}%
  \BibitemOpen
  \bibfield  {author} {\bibinfo {author} {\bibfnamefont {D.}~\bibnamefont
  {Loss}}\ and\ \bibinfo {author} {\bibfnamefont {P.}~\bibnamefont
  {Goldbart}},\ }\href@noop {} {\bibfield  {journal} {\bibinfo  {journal}
  {Phys. Rev. B}\ }\textbf {\bibinfo {volume} {43}},\ \bibinfo {pages} {13762}
  (\bibinfo {year} {1991})}\BibitemShut {NoStop}%
\bibitem [{\citenamefont {Fye}\ \emph {et~al.}(1991)\citenamefont {Fye},
  \citenamefont {Martins}, \citenamefont {Scalapino}, \citenamefont {Wagner},\
  and\ \citenamefont {Hanke}}]{Fye}%
  \BibitemOpen
  \bibfield  {author} {\bibinfo {author} {\bibfnamefont {R.~M.}\ \bibnamefont
  {Fye}}, \bibinfo {author} {\bibfnamefont {M.~J.}\ \bibnamefont {Martins}},
  \bibinfo {author} {\bibfnamefont {D.~J.}\ \bibnamefont {Scalapino}}, \bibinfo
  {author} {\bibfnamefont {J.}~\bibnamefont {Wagner}}, \ and\ \bibinfo {author}
  {\bibfnamefont {W.}~\bibnamefont {Hanke}},\ }\href@noop {} {\bibfield
  {journal} {\bibinfo  {journal} {Phys. Rev. B}\ }\textbf {\bibinfo {volume}
  {44}},\ \bibinfo {pages} {6909} (\bibinfo {year} {1991})}\BibitemShut
  {NoStop}%
\bibitem [{\citenamefont {Kusmartsev}(1991)}]{Kusmartsev_1991}%
  \BibitemOpen
  \bibfield  {author} {\bibinfo {author} {\bibfnamefont {F.~V.}\ \bibnamefont
  {Kusmartsev}},\ }\href@noop {} {\bibfield  {journal} {\bibinfo  {journal} {J.
  Phys.: Condens. Matter}\ }\textbf {\bibinfo {volume} {3}},\ \bibinfo {pages}
  {3199} (\bibinfo {year} {1991})}\BibitemShut {NoStop}%
\bibitem [{\citenamefont {Weisz}\ \emph {et~al.}(1994)\citenamefont {Weisz},
  \citenamefont {Kishore},\ and\ \citenamefont
  {Kusmartsev}}]{Weisz-Kishore-Kusmartsev}%
  \BibitemOpen
  \bibfield  {author} {\bibinfo {author} {\bibfnamefont {J.~F.}\ \bibnamefont
  {Weisz}}, \bibinfo {author} {\bibfnamefont {R.}~\bibnamefont {Kishore}}, \
  and\ \bibinfo {author} {\bibfnamefont {F.~V.}\ \bibnamefont {Kusmartsev}},\
  }\href@noop {} {\bibfield  {journal} {\bibinfo  {journal} {Phys. Rev. B}\
  }\textbf {\bibinfo {volume} {49}},\ \bibinfo {pages} {8126} (\bibinfo {year}
  {1994})}\BibitemShut {NoStop}%
\bibitem [{\citenamefont {Kusmartsev}\ \emph {et~al.}(1994)\citenamefont
  {Kusmartsev}, \citenamefont {Weisz}, \citenamefont {Kishore},\ and\
  \citenamefont {Takahashi}}]{Kusmartsev-Weisz}%
  \BibitemOpen
  \bibfield  {author} {\bibinfo {author} {\bibfnamefont {F.~V.}\ \bibnamefont
  {Kusmartsev}}, \bibinfo {author} {\bibfnamefont {J.~F.}\ \bibnamefont
  {Weisz}}, \bibinfo {author} {\bibfnamefont {R.}~\bibnamefont {Kishore}}, \
  and\ \bibinfo {author} {\bibfnamefont {M.}~\bibnamefont {Takahashi}},\
  }\href@noop {} {\bibfield  {journal} {\bibinfo  {journal} {Phys. Rev. B}\
  }\textbf {\bibinfo {volume} {49}},\ \bibinfo {pages} {16234} (\bibinfo {year}
  {1994})}\BibitemShut {NoStop}%
\bibitem [{\citenamefont {Kusmartsev}(1995)}]{Kusmartsev}%
  \BibitemOpen
  \bibfield  {author} {\bibinfo {author} {\bibfnamefont {F.~V.}\ \bibnamefont
  {Kusmartsev}},\ }\href@noop {} {\bibfield  {journal} {\bibinfo  {journal}
  {Phys. Rev. B}\ }\textbf {\bibinfo {volume} {52}},\ \bibinfo {pages} {14445}
  (\bibinfo {year} {1995})}\BibitemShut {NoStop}%
\bibitem [{\citenamefont {Kusmartsev}(1997)}]{Kusmartsev_1997}%
  \BibitemOpen
  \bibfield  {author} {\bibinfo {author} {\bibfnamefont {F.~V.}\ \bibnamefont
  {Kusmartsev}},\ }\href@noop {} {\bibfield  {journal} {\bibinfo  {journal}
  {Phys. Lett. A}\ }\textbf {\bibinfo {volume} {232}},\ \bibinfo {pages} {135}
  (\bibinfo {year} {1997})}\BibitemShut {NoStop}%
\bibitem [{\citenamefont {Kusmartsev}(1999)}]{Kusmartsev_1999}%
  \BibitemOpen
  \bibfield  {author} {\bibinfo {author} {\bibfnamefont {F.~V.}\ \bibnamefont
  {Kusmartsev}},\ }\href@noop {} {\bibfield  {journal} {\bibinfo  {journal}
  {Phys. Lett. A}\ }\textbf {\bibinfo {volume} {251}},\ \bibinfo {pages} {143}
  (\bibinfo {year} {1999})}\BibitemShut {NoStop}%
\bibitem [{\citenamefont {Maiti}(2006)}]{Maiti_PhysicaE}%
  \BibitemOpen
  \bibfield  {author} {\bibinfo {author} {\bibfnamefont {S.~K.}\ \bibnamefont
  {Maiti}},\ }\href@noop {} {\bibfield  {journal} {\bibinfo  {journal} {Physica
  E}\ }\textbf {\bibinfo {volume} {31}},\ \bibinfo {pages} {117} (\bibinfo
  {year} {2006})}\BibitemShut {NoStop}%
\bibitem [{\citenamefont {Cardamone}\ \emph {et~al.}(2006)\citenamefont
  {Cardamone}, \citenamefont {Stafford},\ and\ \citenamefont
  {Mazumdar}}]{Cardamone}%
  \BibitemOpen
  \bibfield  {author} {\bibinfo {author} {\bibfnamefont {D.~M.}\ \bibnamefont
  {Cardamone}}, \bibinfo {author} {\bibfnamefont {C.~A.}\ \bibnamefont
  {Stafford}}, \ and\ \bibinfo {author} {\bibfnamefont {S.}~\bibnamefont
  {Mazumdar}},\ }\href@noop {} {\bibfield  {journal} {\bibinfo  {journal} {Nano
  Lett.}\ }\textbf {\bibinfo {volume} {6}},\ \bibinfo {pages} {2422} (\bibinfo
  {year} {2006})}\BibitemShut {NoStop}%
\bibitem [{\citenamefont {Ke}\ \emph {et~al.}(2008)\citenamefont {Ke},
  \citenamefont {Yang},\ and\ \citenamefont {Baranger}}]{Ke}%
  \BibitemOpen
  \bibfield  {author} {\bibinfo {author} {\bibfnamefont {S.-H.}\ \bibnamefont
  {Ke}}, \bibinfo {author} {\bibfnamefont {W.}~\bibnamefont {Yang}}, \ and\
  \bibinfo {author} {\bibfnamefont {H.~U.}\ \bibnamefont {Baranger}},\
  }\href@noop {} {\bibfield  {journal} {\bibinfo  {journal} {Nano Lett.}\
  }\textbf {\bibinfo {volume} {8}},\ \bibinfo {pages} {3257} (\bibinfo {year}
  {2008})}\BibitemShut {NoStop}%
\bibitem [{\citenamefont {Solomon}\ \emph {et~al.}(2008)\citenamefont
  {Solomon}, \citenamefont {Andrews}, \citenamefont {Hansen}, \citenamefont
  {Goldsmith}, \citenamefont {Wasielewski}, \citenamefont {Duyne},\ and\
  \citenamefont {Ratner}}]{Solomon}%
  \BibitemOpen
  \bibfield  {author} {\bibinfo {author} {\bibfnamefont {G.~C.}\ \bibnamefont
  {Solomon}}, \bibinfo {author} {\bibfnamefont {D.~Q.}\ \bibnamefont
  {Andrews}}, \bibinfo {author} {\bibfnamefont {T.}~\bibnamefont {Hansen}},
  \bibinfo {author} {\bibfnamefont {R.~H.}\ \bibnamefont {Goldsmith}}, \bibinfo
  {author} {\bibfnamefont {M.~R.}\ \bibnamefont {Wasielewski}}, \bibinfo
  {author} {\bibfnamefont {R.~P.~V.}\ \bibnamefont {Duyne}}, \ and\ \bibinfo
  {author} {\bibfnamefont {M.~A.}\ \bibnamefont {Ratner}},\ }\href@noop {}
  {\bibfield  {journal} {\bibinfo  {journal} {J. Chem. Phys.}\ }\textbf
  {\bibinfo {volume} {129}},\ \bibinfo {pages} {054701} (\bibinfo {year}
  {2008})}\BibitemShut {NoStop}%
\bibitem [{\citenamefont {Charlier}\ \emph {et~al.}(2007)\citenamefont
  {Charlier}, \citenamefont {Blase},\ and\ \citenamefont {Roche}}]{Charlier}%
  \BibitemOpen
  \bibfield  {author} {\bibinfo {author} {\bibfnamefont {J.-C.}\ \bibnamefont
  {Charlier}}, \bibinfo {author} {\bibfnamefont {X.}~\bibnamefont {Blase}}, \
  and\ \bibinfo {author} {\bibfnamefont {S.}~\bibnamefont {Roche}},\
  }\href@noop {} {\bibfield  {journal} {\bibinfo  {journal} {Rev. Mod. Phys.}\
  }\textbf {\bibinfo {volume} {79}},\ \bibinfo {pages} {677} (\bibinfo {year}
  {2007})}\BibitemShut {NoStop}%
\bibitem [{\citenamefont {Sangalli}\ and\ \citenamefont
  {Marini}(2011)}]{Sangalli-Marini}%
  \BibitemOpen
  \bibfield  {author} {\bibinfo {author} {\bibfnamefont {D.}~\bibnamefont
  {Sangalli}}\ and\ \bibinfo {author} {\bibfnamefont {A.}~\bibnamefont
  {Marini}},\ }\href {\doibase 10.1021/nl200871v} {\bibfield  {journal}
  {\bibinfo  {journal} {Nano Lett.}\ }\textbf {\bibinfo {volume} {11}},\
  \bibinfo {pages} {4052} (\bibinfo {year} {2011})}\BibitemShut {NoStop}%
\bibitem [{\citenamefont {L\'{e}vy}\ \emph {et~al.}(1990)\citenamefont
  {L\'{e}vy}, \citenamefont {Dolan}, \citenamefont {Dunsmuir},\ and\
  \citenamefont {Bouchiat}}]{Levy}%
  \BibitemOpen
  \bibfield  {author} {\bibinfo {author} {\bibfnamefont {L.~P.}\ \bibnamefont
  {L\'{e}vy}}, \bibinfo {author} {\bibfnamefont {G.}~\bibnamefont {Dolan}},
  \bibinfo {author} {\bibfnamefont {J.}~\bibnamefont {Dunsmuir}}, \ and\
  \bibinfo {author} {\bibfnamefont {H.}~\bibnamefont {Bouchiat}},\ }\href@noop
  {} {\bibfield  {journal} {\bibinfo  {journal} {Phys. Rev. Lett.}\ }\textbf
  {\bibinfo {volume} {64}},\ \bibinfo {pages} {2074} (\bibinfo {year}
  {1990})}\BibitemShut {NoStop}%
\bibitem [{\citenamefont {Bouchiat}\ and\ \citenamefont
  {Montambaux}(1989)}]{Bouchiat-Montambaux}%
  \BibitemOpen
  \bibfield  {author} {\bibinfo {author} {\bibfnamefont {H.}~\bibnamefont
  {Bouchiat}}\ and\ \bibinfo {author} {\bibfnamefont {G.}~\bibnamefont
  {Montambaux}},\ }\href@noop {} {\bibfield  {journal} {\bibinfo  {journal} {J.
  Phys. France}\ }\textbf {\bibinfo {volume} {50}},\ \bibinfo {pages} {2695}
  (\bibinfo {year} {1989})}\BibitemShut {NoStop}%
\bibitem [{\citenamefont {Keyser}\ \emph {et~al.}(2003)\citenamefont {Keyser},
  \citenamefont {F{\"{u}}hner}, \citenamefont {Borck}, \citenamefont {Haug},
  \citenamefont {Bichler}, \citenamefont {Abstreiter},\ and\ \citenamefont
  {Wegscheider}}]{Keyser}%
  \BibitemOpen
  \bibfield  {author} {\bibinfo {author} {\bibfnamefont {U.~F.}\ \bibnamefont
  {Keyser}}, \bibinfo {author} {\bibfnamefont {C.}~\bibnamefont
  {F{\"{u}}hner}}, \bibinfo {author} {\bibfnamefont {S.}~\bibnamefont {Borck}},
  \bibinfo {author} {\bibfnamefont {R.~J.}\ \bibnamefont {Haug}}, \bibinfo
  {author} {\bibfnamefont {M.}~\bibnamefont {Bichler}}, \bibinfo {author}
  {\bibfnamefont {G.}~\bibnamefont {Abstreiter}}, \ and\ \bibinfo {author}
  {\bibfnamefont {W.}~\bibnamefont {Wegscheider}},\ }\href@noop {} {\bibfield
  {journal} {\bibinfo  {journal} {Phys. Rev. Lett.}\ }\textbf {\bibinfo
  {volume} {90}},\ \bibinfo {pages} {196601} (\bibinfo {year}
  {2003})}\BibitemShut {NoStop}%
\bibitem [{\citenamefont {Hernandez}\ \emph {et~al.}(2011)\citenamefont
  {Hernandez}, \citenamefont {Gusev}, \citenamefont {Kvon},\ and\ \citenamefont
  {Portal}}]{Hernandez-Gusev}%
  \BibitemOpen
  \bibfield  {author} {\bibinfo {author} {\bibfnamefont {F.~G.~G.}\
  \bibnamefont {Hernandez}}, \bibinfo {author} {\bibfnamefont {G.~M.}\
  \bibnamefont {Gusev}}, \bibinfo {author} {\bibfnamefont {Z.~D.}\ \bibnamefont
  {Kvon}}, \ and\ \bibinfo {author} {\bibfnamefont {J.~C.}\ \bibnamefont
  {Portal}},\ }\href@noop {} {\bibfield  {journal} {\bibinfo  {journal} {Phys.
  Rev. B}\ }\textbf {\bibinfo {volume} {84}},\ \bibinfo {pages} {075332}
  (\bibinfo {year} {2011})}\BibitemShut {NoStop}%
\bibitem [{\citenamefont {Fuhrer}\ \emph {et~al.}(2004)\citenamefont {Fuhrer},
  \citenamefont {Ihn}, \citenamefont {Ensslin}, \citenamefont {Wegscheider},\
  and\ \citenamefont {Bichler}}]{Fuhrer_prl}%
  \BibitemOpen
  \bibfield  {author} {\bibinfo {author} {\bibfnamefont {A.}~\bibnamefont
  {Fuhrer}}, \bibinfo {author} {\bibfnamefont {T.}~\bibnamefont {Ihn}},
  \bibinfo {author} {\bibfnamefont {K.}~\bibnamefont {Ensslin}}, \bibinfo
  {author} {\bibfnamefont {W.}~\bibnamefont {Wegscheider}}, \ and\ \bibinfo
  {author} {\bibfnamefont {M.}~\bibnamefont {Bichler}},\ }\href@noop {}
  {\bibfield  {journal} {\bibinfo  {journal} {Phys. Rev. Lett.}\ }\textbf
  {\bibinfo {volume} {93}},\ \bibinfo {pages} {176803} (\bibinfo {year}
  {2004})}\BibitemShut {NoStop}%
\bibitem [{\citenamefont {B{\"{u}}ttiker}(1986)}]{Buttiker_PRL}%
  \BibitemOpen
  \bibfield  {author} {\bibinfo {author} {\bibfnamefont {M.}~\bibnamefont
  {B{\"{u}}ttiker}},\ }\href@noop {} {\bibfield  {journal} {\bibinfo  {journal}
  {Phys. Rev. Lett.}\ }\textbf {\bibinfo {volume} {57}},\ \bibinfo {pages}
  {1761} (\bibinfo {year} {1986})}\BibitemShut {NoStop}%
\bibitem [{\citenamefont {Meir}\ \emph {et~al.}(1993)\citenamefont {Meir},
  \citenamefont {Wingreen},\ and\ \citenamefont {Lee}}]{Meir-Wingreen-Lee}%
  \BibitemOpen
  \bibfield  {author} {\bibinfo {author} {\bibfnamefont {Y.}~\bibnamefont
  {Meir}}, \bibinfo {author} {\bibfnamefont {N.~S.}\ \bibnamefont {Wingreen}},
  \ and\ \bibinfo {author} {\bibfnamefont {P.~A.}\ \bibnamefont {Lee}},\
  }\href@noop {} {\bibfield  {journal} {\bibinfo  {journal} {Phys. Rev. Lett.}\
  }\textbf {\bibinfo {volume} {70}},\ \bibinfo {pages} {2601} (\bibinfo {year}
  {1993})}\BibitemShut {NoStop}%
\bibitem [{\citenamefont {Jagla}\ and\ \citenamefont
  {Balseiro}(1993)}]{Jagla-Balseiro}%
  \BibitemOpen
  \bibfield  {author} {\bibinfo {author} {\bibfnamefont {E.~A.}\ \bibnamefont
  {Jagla}}\ and\ \bibinfo {author} {\bibfnamefont {C.~A.}\ \bibnamefont
  {Balseiro}},\ }\href@noop {} {\bibfield  {journal} {\bibinfo  {journal}
  {Phys. Rev. Lett.}\ }\textbf {\bibinfo {volume} {70}},\ \bibinfo {pages}
  {639} (\bibinfo {year} {1993})}\BibitemShut {NoStop}%
\bibitem [{\citenamefont {Hallberg}\ \emph {et~al.}(2004)\citenamefont
  {Hallberg}, \citenamefont {Aligia}, \citenamefont {Kampf},\ and\
  \citenamefont {Normand}}]{Hallberg-Aligia}%
  \BibitemOpen
  \bibfield  {author} {\bibinfo {author} {\bibfnamefont {K.}~\bibnamefont
  {Hallberg}}, \bibinfo {author} {\bibfnamefont {A.~A.}\ \bibnamefont
  {Aligia}}, \bibinfo {author} {\bibfnamefont {A.~P.}\ \bibnamefont {Kampf}}, \
  and\ \bibinfo {author} {\bibfnamefont {B.}~\bibnamefont {Normand}},\
  }\href@noop {} {\bibfield  {journal} {\bibinfo  {journal} {Phys. Rev. Lett.}\
  }\textbf {\bibinfo {volume} {93}},\ \bibinfo {pages} {067203} (\bibinfo
  {year} {2004})}\BibitemShut {NoStop}%
\bibitem [{\citenamefont {Rinc\'{o}n}\ \emph {et~al.}(2009)\citenamefont
  {Rinc\'{o}n}, \citenamefont {Aligia},\ and\ \citenamefont
  {Hallberg}}]{Rincon-Aligia-Hallberg}%
  \BibitemOpen
  \bibfield  {author} {\bibinfo {author} {\bibfnamefont {J.}~\bibnamefont
  {Rinc\'{o}n}}, \bibinfo {author} {\bibfnamefont {A.~A.}\ \bibnamefont
  {Aligia}}, \ and\ \bibinfo {author} {\bibfnamefont {K.}~\bibnamefont
  {Hallberg}},\ }\href@noop {} {\bibfield  {journal} {\bibinfo  {journal}
  {Phys. Rev. B}\ }\textbf {\bibinfo {volume} {79}},\ \bibinfo {pages} {035112}
  (\bibinfo {year} {2009})}\BibitemShut {NoStop}%
\bibitem [{\citenamefont {Friederich}\ and\ \citenamefont
  {Meden}(2008)}]{Friederich-Meden}%
  \BibitemOpen
  \bibfield  {author} {\bibinfo {author} {\bibfnamefont {S.}~\bibnamefont
  {Friederich}}\ and\ \bibinfo {author} {\bibfnamefont {V.}~\bibnamefont
  {Meden}},\ }\href@noop {} {\bibfield  {journal} {\bibinfo  {journal} {Phys.
  Rev. B}\ }\textbf {\bibinfo {volume} {77}},\ \bibinfo {pages} {195122}
  (\bibinfo {year} {2008})}\BibitemShut {NoStop}%
\bibitem [{\citenamefont {Meden}\ and\ \citenamefont
  {Schollw{\"{o}}ck}(2003)}]{Meden-Schollwock_rings}%
  \BibitemOpen
  \bibfield  {author} {\bibinfo {author} {\bibfnamefont {V.}~\bibnamefont
  {Meden}}\ and\ \bibinfo {author} {\bibfnamefont {U.}~\bibnamefont
  {Schollw{\"{o}}ck}},\ }\href@noop {} {\bibfield  {journal} {\bibinfo
  {journal} {Phys. Rev. B}\ }\textbf {\bibinfo {volume} {67}},\ \bibinfo
  {pages} {035106} (\bibinfo {year} {2003})}\BibitemShut {NoStop}%
\bibitem [{\citenamefont {Rejec}\ and\ \citenamefont
  {Ram\v{s}ak}(2003{\natexlab{a}})}]{Rejec-Ramsak}%
  \BibitemOpen
  \bibfield  {author} {\bibinfo {author} {\bibfnamefont {T.}~\bibnamefont
  {Rejec}}\ and\ \bibinfo {author} {\bibfnamefont {A.}~\bibnamefont
  {Ram\v{s}ak}},\ }\href@noop {} {\bibfield  {journal} {\bibinfo  {journal}
  {Phys. Rev. B}\ }\textbf {\bibinfo {volume} {68}},\ \bibinfo {pages} {035342}
  (\bibinfo {year} {2003}{\natexlab{a}})}\BibitemShut {NoStop}%
\bibitem [{\citenamefont {Rejec}\ and\ \citenamefont
  {Ram\v{s}ak}(2003{\natexlab{b}})}]{Rejec-Ramsak_AB}%
  \BibitemOpen
  \bibfield  {author} {\bibinfo {author} {\bibfnamefont {T.}~\bibnamefont
  {Rejec}}\ and\ \bibinfo {author} {\bibfnamefont {A.}~\bibnamefont
  {Ram\v{s}ak}},\ }\href@noop {} {\bibfield  {journal} {\bibinfo  {journal}
  {Phys. Rev. B}\ }\textbf {\bibinfo {volume} {68}},\ \bibinfo {pages} {033306}
  (\bibinfo {year} {2003}{\natexlab{b}})}\BibitemShut {NoStop}%
\bibitem [{\citenamefont {Maiti}\ \emph {et~al.}(2011)\citenamefont {Maiti},
  \citenamefont {Saha},\ and\ \citenamefont {Karmakar}}]{Maiti-EPJB}%
  \BibitemOpen
  \bibfield  {author} {\bibinfo {author} {\bibfnamefont {S.~K.}\ \bibnamefont
  {Maiti}}, \bibinfo {author} {\bibfnamefont {S.}~\bibnamefont {Saha}}, \ and\
  \bibinfo {author} {\bibfnamefont {S.~N.}\ \bibnamefont {Karmakar}},\
  }\href@noop {} {\bibfield  {journal} {\bibinfo  {journal} {Eur. Phys. J. B}\
  }\textbf {\bibinfo {volume} {79}},\ \bibinfo {pages} {209} (\bibinfo {year}
  {2011})}\BibitemShut {NoStop}%
\bibitem [{\citenamefont {Viefers}\ \emph {et~al.}(2004)\citenamefont
  {Viefers}, \citenamefont {Koskinen}, \citenamefont {Deo},\ and\ \citenamefont
  {Manninen}}]{Viefers}%
  \BibitemOpen
  \bibfield  {author} {\bibinfo {author} {\bibfnamefont {S.}~\bibnamefont
  {Viefers}}, \bibinfo {author} {\bibfnamefont {P.}~\bibnamefont {Koskinen}},
  \bibinfo {author} {\bibfnamefont {P.~S.}\ \bibnamefont {Deo}}, \ and\
  \bibinfo {author} {\bibfnamefont {M.}~\bibnamefont {Manninen}},\ }\href@noop
  {} {\bibfield  {journal} {\bibinfo  {journal} {Physica E}\ }\textbf {\bibinfo
  {volume} {21}},\ \bibinfo {pages} {1} (\bibinfo {year} {2004})}\BibitemShut
  {NoStop}%
\bibitem [{\citenamefont {Hancock}\ \emph {et~al.}(2008)\citenamefont
  {Hancock}, \citenamefont {Suorsa}, \citenamefont {T{\"{o}}l{\"{o}}},\ and\
  \citenamefont {Harju}}]{Hancock}%
  \BibitemOpen
  \bibfield  {author} {\bibinfo {author} {\bibfnamefont {Y.}~\bibnamefont
  {Hancock}}, \bibinfo {author} {\bibfnamefont {J.}~\bibnamefont {Suorsa}},
  \bibinfo {author} {\bibfnamefont {E.}~\bibnamefont {T{\"{o}}l{\"{o}}}}, \
  and\ \bibinfo {author} {\bibfnamefont {A.}~\bibnamefont {Harju}},\
  }\href@noop {} {\bibfield  {journal} {\bibinfo  {journal} {Phys. Rev. B}\
  }\textbf {\bibinfo {volume} {77}},\ \bibinfo {pages} {155103} (\bibinfo
  {year} {2008})}\BibitemShut {NoStop}%
\bibitem [{\citenamefont {Sushkov}(2001)}]{Sushkov}%
  \BibitemOpen
  \bibfield  {author} {\bibinfo {author} {\bibfnamefont {O.~P.}\ \bibnamefont
  {Sushkov}},\ }\href@noop {} {\bibfield  {journal} {\bibinfo  {journal} {Phys.
  Rev. B}\ }\textbf {\bibinfo {volume} {64}},\ \bibinfo {pages} {155319}
  (\bibinfo {year} {2001})}\BibitemShut {NoStop}%
\bibitem [{\citenamefont {Dagotto}(1994)}]{Dagotto}%
  \BibitemOpen
  \bibfield  {author} {\bibinfo {author} {\bibfnamefont {E.}~\bibnamefont
  {Dagotto}},\ }\href@noop {} {\bibfield  {journal} {\bibinfo  {journal} {Rev.
  Mod. Phys.}\ }\textbf {\bibinfo {volume} {66}},\ \bibinfo {pages} {763}
  (\bibinfo {year} {1994})}\BibitemShut {NoStop}%
\bibitem [{\citenamefont {Koskinen}\ and\ \citenamefont
  {Manninen}(2003)}]{Koskinen-Manninen}%
  \BibitemOpen
  \bibfield  {author} {\bibinfo {author} {\bibfnamefont {P.}~\bibnamefont
  {Koskinen}}\ and\ \bibinfo {author} {\bibfnamefont {M.}~\bibnamefont
  {Manninen}},\ }\href@noop {} {\bibfield  {journal} {\bibinfo  {journal}
  {Phys. Rev. B}\ }\textbf {\bibinfo {volume} {68}},\ \bibinfo {pages} {195304}
  (\bibinfo {year} {2003})}\BibitemShut {NoStop}%
\bibitem [{\citenamefont {Ferrari}\ \emph {et~al.}(1999)\citenamefont
  {Ferrari}, \citenamefont {Chiappe}, \citenamefont {Anda},\ and\ \citenamefont
  {Davidovich}}]{Ferrari}%
  \BibitemOpen
  \bibfield  {author} {\bibinfo {author} {\bibfnamefont {V.}~\bibnamefont
  {Ferrari}}, \bibinfo {author} {\bibfnamefont {G.}~\bibnamefont {Chiappe}},
  \bibinfo {author} {\bibfnamefont {E.~V.}\ \bibnamefont {Anda}}, \ and\
  \bibinfo {author} {\bibfnamefont {M.~A.}\ \bibnamefont {Davidovich}},\
  }\href@noop {} {\bibfield  {journal} {\bibinfo  {journal} {Phys. Rev. Lett.}\
  }\textbf {\bibinfo {volume} {82}},\ \bibinfo {pages} {5088} (\bibinfo {year}
  {1999})}\BibitemShut {NoStop}%
\bibitem [{\citenamefont {Leturcq}\ \emph {et~al.}(2006)\citenamefont
  {Leturcq}, \citenamefont {S{\'{a}}nchez}, \citenamefont {G{\"{o}}tz},
  \citenamefont {Ihn}, \citenamefont {Ensslin}, \citenamefont {Driscoll},\ and\
  \citenamefont {Gossard}}]{Leturcq}%
  \BibitemOpen
  \bibfield  {author} {\bibinfo {author} {\bibfnamefont {R.}~\bibnamefont
  {Leturcq}}, \bibinfo {author} {\bibfnamefont {D.}~\bibnamefont
  {S{\'{a}}nchez}}, \bibinfo {author} {\bibfnamefont {G.}~\bibnamefont
  {G{\"{o}}tz}}, \bibinfo {author} {\bibfnamefont {T.}~\bibnamefont {Ihn}},
  \bibinfo {author} {\bibfnamefont {K.}~\bibnamefont {Ensslin}}, \bibinfo
  {author} {\bibfnamefont {D.~C.}\ \bibnamefont {Driscoll}}, \ and\ \bibinfo
  {author} {\bibfnamefont {A.~C.}\ \bibnamefont {Gossard}},\ }\href@noop {}
  {\bibfield  {journal} {\bibinfo  {journal} {Phys. Rev. Lett.}\ }\textbf
  {\bibinfo {volume} {96}},\ \bibinfo {pages} {126801} (\bibinfo {year}
  {2006})}\BibitemShut {NoStop}%
\end{thebibliography}%

\end{document}